\documentclass[fleqn,usenatbib]{mnras}

\usepackage[T1]{fontenc}
\usepackage{ae,aecompl}

\usepackage{graphicx}	% Including figure files
\graphicspath{{figures_select/}} %Setting the graphicspath
\usepackage{amsmath}	% Advanced maths commands
\usepackage{amssymb}	% Extra maths symbols
\usepackage{caption}    % for subfigures

\usepackage{hyperref}
\usepackage{xcolor}
\definecolor{citationblue}{HTML}{00657E}

\hypersetup{colorlinks,
            citecolor=citationblue,
            urlcolor=citationblue,
            linkcolor=citationblue}
\usepackage[symbol]{footmisc}
\usepackage{fp}
\usepackage{multirow}
\usepackage{footnote}
\usepackage{caption}
\usepackage{subcaption}
\usepackage{subfloat}
\usepackage{float}
\newcommand{\gcc}{\,g\,s$^{-3}$} %
\newcommand{\Mearth}{\,M$_\oplus$} %
\newcommand{\Rearth}{\,R$_\oplus$} %
\newcommand{\Msun}{\,M$_\odot$} %
\title[Stellar activity and the radius gap]{The influence of host star activity evolution on the population of super-Earths and mini-Neptunes}

\author[L. Ketzer et al.]{
L. Ketzer$^{1, 2}$\thanks{E-mail: lketzer@aip.de}
and K. Poppenhaeger$^{1, 2}$
\\
$^{1}$Leibniz Institute for Astrophysics Potsdam (AIP), An der Sternwarte 16, 14482 Potsdam, Germany\\
$^{2}$Universit\"at Potsdam,  Institut f\"ur Physik und Astronomie,  Karl-Liebknecht-Stra\ss e 24/25, 14476 Potsdam, Germany}

\date{Accepted 2022 September 12. Received 2022 July 15; in original form 2022 June 2.}

\pubyear{2022}

\begin{document}
\label{firstpage}
\pagerange{\pageref{firstpage}--\pageref{lastpage}}
\maketitle

\begin{abstract}

\noindent The detected exoplanet population displays a dearth of planets with sizes of about two Earth radii, the so-called radius gap. This is interpreted as an evolutionary effect driven by a variety of possible atmospheric mass loss processes of exoplanets. For mass loss driven by an exoplanet's irradiation by stellar X-ray and extreme-UV photons, the time evolution of the stellar magnetic activity is important. It is known from observations of open stellar clusters that stars of the same age and mass do not all follow the same time evolution of activity-induced X-ray and extreme-UV luminosities. Here we explore how a realistic spread of different stellar activity tracks influences the mass loss and radius evolution of a simulated population of small exoplanets and the observable properties of the radius gap. Our results show qualitatively that different saturation time scales, i.e.\ the young age at which stellar high-energy emission starts to decline, and different activity decay tracks over moderate stellar ages can cause changes in the population density of planets in the gap, as well as in the observable width of the gap. We also find that while the first 100 million years of mass loss are highly important to shape the radius gap, significant evolution of the gap properties is expected to take place for at least the first 500-600 million years, i.e.\ the age of the Hyades cluster. Observations of exoplanet populations with defined ages will be able to shed more light on the radius gap evolution.

\end{abstract}

 % additional keywords, sth about the models, simulation..

%%%%%%%%%%%%%%%%%%%%%%%%%%%%%%%%%%%%%%%%%%%%%%%%%%

%%%%%%%%%%%%%%%%% BODY OF PAPER %%%%%%%%%%%%%%%%%%

\section{Introduction}

One of the surprising discoveries of the Kepler mission is not only the abundance of planets below the size of Neptune \citep{2010Borucki}, but even more so the bimodal-structure of the radius distribution of these small planets. The observed population seems to be divided into two distinct groups, the so-called super-Earths and sub-Neptunes \citep{2015Rogers}, which are separated by a deficit in planets of around $2\, \mathrm{R}_\oplus$. This gap-like feature or valley has been shown to have a dependence on the planetary orbital period or irradiation, supporting the interpretation that atmospheric mass loss is the main cause of the substructure in the observed exoplanet population. This had been predicted by theoretical studies \citep{2013Owen, Lopez2013} before being observed \citep{Fulton2017, VanEylen2018b, 2021David}.

The prevailing explanation for this gap-like feature is atmospheric erosion caused by the high-energy X-ray and ultraviolet (together: XUV) irradiation from the host star, also known as photoevaporation \citep[e.g.][]{Lopez2013, 2013Owen, 2017Owen}; but also mass-loss driven by the internal luminosity of the cooling planetary core is able to reproduce the observed feature \citep[e.g][]{2018Ginzburg, 2019Gupta}. In both scenarios it is assumed that planets are born with rocky cores, generally having masses of a few $\mathrm{M}_\oplus$, which are surrounded by primordial hydrogen/helium (H/He) atmospheres. Over the course of a planet's life, the strength of the atmospheric mass loss determines whether it is completely stripped of its atmosphere, leaving behind only the bare rocky core (super-Earths), or if it can hold on to some fraction of its envelope, retaining a radius which places it above the radius gap (sub-Neptunes).

In the photoevaporation scenario, the stellar environment plays a crucial role in the evolution of the planetary atmosphere over time. Stellar high-energy photons, ranging from UV to X-rays, shape a planet's atmosphere by dissociating molecules and photoionizing atoms. They alter the chemistry in the atmosphere and cause significant heating, which, in turn, can drive a hydrodynamic outflow, leading to the loss of the atmosphere over time \citep{1981Watson, 2003Lammer, 2004Lecavelier}. To properly take into account this externally-driven mass-loss process, it is necessary to understand the evolution of the host-star activity over time. Stellar activity is an umbrella term used to describe phenomena related to the presence of magnetic fields in cool stars, like the chromospheric and coronal high energy emission, or stellar flares and coronal mass ejections.  As stars age, stellar rotation and the magnetic activity, which is driven by the stellar dynamo, decrease hand in hand. The amount of XUV radiation a star produces depends on several factors like age, activity level and rotation, as well as spectral type \citep{2005Preibisch, 2011Wright, 2012Jackson, 2014Reiners, 2015Tu, 2021Johnstone}.

The first few hundred million years are thought to sculpt planetary atmospheres most strongly, since this is the timescale over which stars maintain high activity levels and thus planets receive the highest XUV flux. Additionally, in the early stages of planetary evolution, planets still host extended atmospheres because they have not had enough time too cool and contract. The interplay between these factors leads to the general assumption that the most significant photoevaporative mass loss is taking place early on \citep[e.g.][]{2016Owen}. Core-powered mass loss on the other hand is a more gradual process, sculpting planets on gigayear timescales.
Recent observational studies investigating the age dependence and time evolution of the radius gap hint that the fraction of super-Earths to sub-Neptunes increases with stellar age, going from young planets less than 1-2 Gyr to older ones \citep{2020Berger, 2021Sandoval}. Additionally, the radius gap appears to fill in over gigayear timescales \citep{2021David}. While this could be seen as an indication of core-powered mass loss, the results, due to the very small number of planets with ages less than a few 100 Myr, do not yet constrain the evolution at the earliest times, when photoevaporation is thought to dominate the mass loss \citep[e.g.][]{2017Owen}. On top of that, the exact time evolution of the stellar X-ray and EUV output can extend the timescale where photoevaporative mass loss is important to Gyrs \citep{2021King}.

An additional complication arises from stars of similar mass experiencing a non-uniform stellar spin down. One cause may be the different initial rotation rates that stars are born with, or differences in the magnetic field complexity \citep{2015Tu, Garraffo2016}. As a consequence, a stars' spin-down behavior varies and with it the age at which the spin-down, and thus the activity decay, sets in. If differences in the initial rotation rate are the cause, the timescale for the onset of the activity decay can range from a few Myr to a few 100 Myr, for slow and fast rotators, respectively. This leads to a spread of about an order-of-magnitude in the activity levels of young stars with similar masses, comparable to what has been observed \citep{2011Wright, 2015Tu, 2021Johnstone}. The stellar activity evolution in the first several 100 Myrs is thus another factor that needs to be taken into account when estimating the future planetary mass and radius evolution of young systems over time \citep[e.g.][]{2021Kubyshkina}. For individual systems with mature ages of several Gyrs where the planets have retained some fraction of their primordial atmosphere, it is already possible to put constraints on the past rotational history of the host star using the measurable present day planet properties. This approach exploits the fact that the observed properties like the planetary radius depend strongly on the mass-loss history of the planet, which is driven by the amount of high enerergy flux received by the planet during its lifetime; and the emitted XUV flux, in turn, is determined by the rotational evolution of the host star \citep{2019Kubyshkina_b, 2019Kubyshkina_a, 2021Bonfanti}. While this approach is promising for well constrained, individual systems, the rotational histories of most exoplanet host stars known to date remain unconstrained.

In this work, we study the effect of various stellar activity tracks on the radius distribution of an observationally motivated population of exoplanets. We compare an energy- and radiation-recombination-limited model with hydro-based mass-loss rates, and highlight how sensitively photoevaporation calculations depend on the assumed X-ray saturation timescale and luminosity, as well as the estimation of the EUV contribution to the total high-energy flux. Details on the input physics of the planetary structure model and the mass-loss calculations are given in Section~\ref{sec:evo_framework}. A detailed description of the host star activity model, i.e. the stellar XUV evolutionary tracks used for this work, are provided in Section~\ref{sec:activity_decay}. In Section~\ref{sec:planetary_sample}, we describe the input planet population and provide details on the simulation setup. We summarize our findings regarding the influence of the activity tracks on the 1D and 2D radius distribution of our sample population in Section~\ref{sec:results}.

\section{Planetary Mass Loss}
\label{sec:evo_framework}

We use a coupled thermal evolution and atmospheric photoevaporation model to investigate the mass-loss and subsequent radius evolution of a large population of planets \citep[similar to e.g.][]{Lopez2013, LopezFortney2014, 2013Owen, 2017Owen}. We consider here the atmospheric mass-loss driven by stellar XUV photons, and investigate how the inclusion of a distribution of low, medium and high activity tracks for the host star - as opposed to assuming only one track for all stars in the population - affects the final planetary radius distribution at mature system ages. Our calculations are performed with the publicly available python code PLATYPOS\footnote{\url{https://github.com/lketzer/platypos/}}, for which a description of the code's fuctionality was presented in \citet{2022Ketzer}. In the following, we describe the relevant inputs for this work in more detail.

\subsection{Planetary Models}
\label{subsec:planet_models}

PLATYPOS gives the choice to select either the planetary structure models by \cite{LopezFortney2014} or the MESA-based models by \cite{ChenRogers2016} (hereafter: LoFo14 and ChRo16). The models assume the planetary cores to be of an Earth-like mixture of silicates and iron, with the surrounding envelope composed of predominantly hydrogen and helium. \cite{LopezFortney2014} and \cite{ChenRogers2016} compute planetary radii for a large grid of planetary parameters, and provide fitting formulae to estimate the radius at a specified age, core mass, envelope mass fraction, $f_{\mathrm{env}}$, and bolometric incident flux. For specific details of these models, we refer to the original publications. We use their fitted mass-radius-age relations for low-mass gaseous sub-Neptune-sized planets to estimate their radius at any given time over the course of the simulation. For this paper, we choose to use the MESA-based ChRo16 model because the resulting radius gap qualitatively best matches the observations. This comparison is based on the assumption the planet radius from the evolutionary models is equal to the optical white light transit radius. Transit radii in narrow wavelength bands can be different due to additional absorption in the thin atmosphere; howoever, this is outside the scope of this work. We discuss differences with the LoFo14 model and motivate our choice for the ChRo16 model in Appendix~\ref{sec:app_planet_model}.

A caveat of both models is that the fitting formulae should only be used within a certain range of planetary parameters, with age and envelope mass fraction being of particular importance for our calculation. In general, the first 100 Myr of planetary evolution are influenced by the assumed initial conditions.  Both the LoFo14 and ChRo16 models assume the so-called hot-start scenario, where planets form with large initial entropy and thus present enlarged radii at a young age \citep{2007Marley, 2007Fortney}. Those authors note, however, that due to the short cooling timescale for the enlarged, low-mass hot-start planets, the rapid initial cooling and subsequent contraction erases any differences resulting from the choice of initial entropy by $\sim$ 10 Myr. \cite{LopezFortney2014} explicitly showed that the modeled planetary radii can be reasonably backwards extrapolated to ages of 10 Myr (see their Figure 2), and we therefore chose to extrapolate both models to 10 Myr as the starting age of our simulations.

Both model-fits are also only valid for planets with an atmospheric mass fraction greater than $0.01\%$, but planets in our sample undergo mass-loss, and thus might enter a regime with an envelope less than this lower $f_{\mathrm{env}}$ limit. \citet{2021Rogers_a} note that below an atmospheric mass fraction of 0.01\% the planetary radius becomes indistinguishable from the core radius in transit observations, making it reasonable to assume complete stripping below this limit. We performed two test simulations to assess the effect of extrapolating the models to atmospheric mass fractions below $0.01\%$. Once, we allow for extrapolation and continue decreasing the envelope mass until no envelope is remaining, or the simulation reaches the final age. For comparison, once a planet reaches the lower limit, we keep $f_{\mathrm{env}}$ fixed at $0.01\%$ for the remainder of the simulation run, and continue the mass-loss calculation with a constant envelope mass. We find that in general, if a planet reaches an envelope mass fraction of $f_{\mathrm{env}}=0.01\%$, it cannot hold on to its atmosphere regardless of the radius estimation in the final stages.

\subsection{Mass-loss rate calculation}
\label{subsec:evaporation_calcs}

Several regimes of hydrodynamic escape of hydrogen-dominated atmospheres have been identified in theoretical studies \citep[e.g.][]{2003Lammer, MurrayClay2009, OwenJackson2012, 2016Owen}. This includes energy-limited, radiation-recombination limited and photon-limited escape. Recent observational evidence of giant planets support the presence of these regimes \citep{2021Lampon}, which differ in terms of the underlying physics of the escaping planetary wind. Important is the consideration of how neutral hydrogen is produced and lost, but also the processes of converting the absorbed stellar radiation into work, which ultimately drives the evaporative outflow, has to be taken into account.

We built different evaporation schemes into PLATYPOS, which allows us to compare them against each other. Photoevaprative mass-loss can be estimated using an energy-limited approximation only, including the radiation-recombination limited regime, or via a hydro-based approximation, which takes into account not only the contribution from high-energy radiation but also the planetary intrinsic thermal energy and surface gravity \citep{Kubyshkina2018b}. The different evaporation scenarios and their respective parameterized mass-loss rate calculations are explained below.

\subsubsection{Energy-limited mass loss}
\label{subsec:Elim}

The stellar high-energy radiation impinging on the planet photoionizes and heats the gas in the upper atmosphere. Mass loss occurs in the energy-limited regime when a (significant) fraction of the external energy supply is efficiently used to do PdV work expanding the atmosphere and lifting material out of the planet's gravitational well, while little energy is lost to radiation and internal energy changes \citep{MurrayClay2009}. We adopt the commonly used energy-limited hydrodynamic escape model \cite[see e.g.][]{OwenJackson2012, Lopez2012}, which assumes that the mass-loss rates are limited by the stellar radiative energy deposition and scale linearly with the high energy incident flux:
\begin{equation}
\dot{M}_{\mathrm{Elim}} = - \epsilon \frac{(\pi R_{\mathrm{XUV}}^2) F_{\mathrm{XUV}}}{K G M_{\mathrm{pl}}/R_{\mathrm{pl}}} = - \epsilon \frac{3 \beta^2 F_{\mathrm{XUV}}}{4 G K \rho_{\mathrm{pl}}}\,,
\label{eq1}
\end{equation}

where $\epsilon$ is the efficiency of the atmospheric escape, $M_{\mathrm{pl}}$ the mass and $\rho_{\mathrm{pl}}$ the density of the planet, $F_{\mathrm{XUV}}$ the high-energy flux recieved by the planet, and $R_{\mathrm{pl}}$ and $R_{\mathrm{XUV}}$ the planetary radii at optical and XUV wavelengths, respectively; we use $\beta = R_{\mathrm{XUV}}/R_{\mathrm{pl}}$ as a shorthand in the following. The factor $K$ encompasses the impact of Roche lobe overflow \citep{Erkaev2007}, and can take on values of 1 for no Roche lobe influence and $<\,1$ for planets filling significant fractions of their Roche lobes. The values for the efficiency parameter, $\epsilon$, reported in the literature range from as high as 0.4 \citep{Lalitha2018} down to 0.01 and lower for Jupiter-mass planets \citep{MurrayClay2009, Salz2016b, 2021Lampon}. Planets in the mass regime of sub-Neptunes are predicted to have heating efficiencies between 10 and 30\% \citep[e.g][]{2013Owen, Salz2016b}.

\subsubsection{Radiation/recombination-limited mass loss}
\label{subsec:RRlim}

At high UV fluxes, the temperature of the planetary wind and the ionization fraction become so high that the flow enters a state of radiation-recombination equilibrium. Recombination of hydrogen becomes so efficient that a considerable fraction of the absorbed UV energy is efficiently lost to the subsequent cooling radiation, particularly through the hydrogen Lyman-alpha (Ly$\alpha$) line.
The losses regulate the gas to a near-constant temperature of $\sim10^4$K, which leads to a shallower dependence of the mass-loss rates on the incoming high energy flux ($\dot{M} \propto \sqrt{\mathrm{F}_{XUV}}$) \citep[e.g.][]{MurrayClay2009, Salz2016b}. Recent calculations using a hydrodynamics code for exoplanetary atmospheres and explicitly solving the photoionozation equilibrium give a more detailed view on the relation between high-energy heating and radiative cooling of individual modelled exoplanets \citep{Caldiroli2021, Caldiroli2022}.

Following \citet{ChenRogers2016, 2017Lopez, 2018Lopez}, we use a modified prescription for the mass-loss rate, which takes into account the influence of significant radiative cooling for highly irradiated planets \citep{MurrayClay2009}. The mass-loss rate is given by:

\begin{multline}
\dot{M}_{\mathrm{RRlim}} = - 4\pi c_{\mathrm{s}} R_{\mathrm{s}}^2 \mu_{\mathrm{+,wind}} m_{\mathrm{H}} \left( \frac{F_{\mathrm{XUV}} G M_{\mathrm{pl}}}{h \nu_{0} \alpha_{\mathrm{rec,B}} c_{\mathrm{s}}^2 R_{\mathrm{XUV}}^2} \right)^{1/2} \\ \times \mathrm{exp} \left[ \frac{G M_{\mathrm{pl}}}{R_{\mathrm{XUV}} c_{\mathrm{s}}^2} \left( \frac{R_{\mathrm{XUV}}}{R_{\mathrm{s}}} - 1 \right) \right]\,,
\end{multline}
\label{eq2}

where $R_{\mathrm{s}}$ is the radius of the sonic point and ${c_{\mathrm{s}} = (k_{\mathrm{B}} T_{\mathrm{wind}} / \mu_{\mathrm{wind}} M_{H})^{1/2}}$ the isothermal sound speed of a fully ionized wind at the sonic point. For H/He envelopes, the mean molecular weight of the wind is set to ${\mu_{\mathrm{wind}} = 0.62}$, taking into account that most of the hydrogen is ionized. The case B radiative recombination coefficient for hydrogen at $10^4$\,K is ${\alpha_{\mathrm{rec,B}} = 2.70\times10^{-13}\,\mathrm{cm}^3 \mathrm{s}^{-1}}$, and the mean molecular weight of ions at the base of the wind is given by ${\mu_{\mathrm{+,wind}} = 1.3}$.
The formula given above is only correct when $R_{\mathrm{XUV}}$ does not exceed $R_{\mathrm{s}}$, which may not be true for many low-mass planets early in their evolution. In case of ${R_{\mathrm{XUV}} > R_{\mathrm{s}}}$ at any given point in the simulation, we impose that ${R_{\mathrm{XUV}} = R_{\mathrm{s}}}$.
To greatly simplify the calculation, the spectral dependence of the incoming photons is neglected, and instead, following \citet{MurrayClay2009}, we assume that the ionizing radiation has a typical photon energy of $20\,\mathrm{eV}$. For a more detailed explanation on the calculations of this mass-loss rate, see Section 2.3 in \citet{2017Lopez}.

For each time step of our calculation, we evaluate both the radiation/recombination-limited and energy-limited mass-loss rate and adopt the lesser of the two. This is to ensure that the mass-loss rates for highly irradiated planets, particularly at young ages when the host star is still very active, are not overpredicted.

\subsubsection{Hydrodynamic-based approximation}
\label{subsec:HBA}

\cite{Kubyshkina2018a} argue that the energy-limited mass-loss rates can be severly underestimated for highly irradiated low-density planets or overestimated for planets with hydrostatic atmospheres. They compute a large model grid of hydrodynamic upper atmosphere models \citep{Kubyshkina2018b} and present, based on the grid results, an analytical expression for the mass-loss rates as a function of the system parameters. This "hydro-based approximation" assumes an efficiency of ${\epsilon = 0.15}$, takes into account Roche-lobe effects, and self-consistently accounts for the effective absorption radius (i.e. $\beta$, which is given by the ratio of the XUV to the optical planetary radius). It is important to note that in their upper-atmosphere simulations, atmospheric escape can, in addition to being XUV-driven, also occur due to a favorable combination of planetary intrinsic thermal energy and low surface gravity. This is a major difference to the energy- and/or radiation/recombination limited mass-loss rate calculations, where only XUV-driven escape is taken into account. Their results show that mass-loss rates for planets in a so-called "boil-off" regime are several orders of magnitude, up to $10^8$, larger than compared to energy-limited mass-loss rates. For planets with higher masses and/or lower incident fluxes, on the other hand, the energy-limited mass-loss rates can be overestimated by up to a factor 50 \citep{Kubyshkina2018a}.

For this paper, we choose to calculate the mass-loss rates based on the energy-limited approximation, but with the inclusion of a radiation/recombination-limit as described in Section~\ref{subsec:RRlim}. Due to recent cautions \citep[see][]{2021Krenn} about energy-limited mass loss, we also perform our simulations using hydro-based mass-loss rates for comparison. The main differences and implications for our simulations are discussed in Appendix~\ref{sec:app_evaporation}, where we show that hydro-based mass loss evaporates most mini-Neptunes in our simulations, leaving behind almost no planets above the gap for the initial planet population used in this work.

\subsection{Effective absorption radius}
\label{subsec:XUV_radius}

Planetary radii can have vastly different sizes when observed at X-ray, EUV or optical wavelengths \citep{2013Poppenhaeger,2014Kulow,2015Ehrenreich}. Optical photons can penetrate much deeper into the atmosphere, while high-energy photons are readily absorbed at higher altitudes. In addition, planetary parameters like the gravitational potential affect the height at which XUV photons are absorbed, with some lower-mass planets hosting extended atmospheres a few times the size of the optial radius \citep[e.g.][]{Salz2016b, 2021Lampon}. That planets appear significantly larger at shorter wavelengths is further supported by UV and X-ray observations of the gas giant HD\,189733\,b, which show an enhanced transit depth and thus indicate an enlarged XUV absorption radius \citep{2013Poppenhaeger}.

The effective absorption radius of the high-energy radiation, $R_{\mathrm{XUV}}$, is only weakly constrained by existing observations. To obtain reasonable mass-loss rate estimates, $R_{\mathrm{XUV}}$ (i.e $\beta$) needs to be estimated. The two available methods are the relation by \citet{Salz2016b} (from here on called "Salz-$\beta$"), as well as the relation presented in \citet{ChenRogers2016} and \citet{2017Lopez}, which follows the arguments presented in \citet{MurrayClay2009} (we label this "Lopez-$\beta$"). The "Salz-$\beta$" calculation is motivated by results from detailed numerical simulations, while the "Lopez-$\beta$" is derived from analytical calculations. In our simulations, we choose to use the "Lopez"-$\beta$". We also introduce "Salz"-$\beta$" in Appendix~\ref{sec:app_beta}, and then compare and discuss the two methods. For both methods, we impose that the effective absorption radius, $R_{\mathrm{XUV}}$, cannot be larger than the Roche lobe radius of the planet. If $R_{\mathrm{XUV}}$ exceeds the Roche lobe at any point in the simulation, we set $R_{\mathrm{XUV}}$ equal to the Roche lobe radius, $R_{\mathrm{Rl}}$.
It is given by ${R_{\mathrm{Rl}} = a (M_{\mathrm{pl}} / (3 M_{\mathrm{pl}} + M_{*}))^{1/3}}$, where $a$ denotes the semi-major axis of the planet. In general, we find that young, low mass planets which are most prone to fulfilling this criterion in their early evolution, do not stand a chance of retaining an atmosphere in our simulations.

\subsubsection{"Lopez-$\beta$" calculation}
\label{subsec:Lopezbeta}

Following \citet{ChenRogers2016} and \citet{2017Lopez}, we also implement a theoretical approximation to estimate the radius of the XUV photosphere, which is based on simplifying assumptions on the structure of the atmosphere. The photoionization base of the evaporative wind is where the atmosphere becomes optically thick to XUV photons, and the difference between the XUV and optical photosphere can be estimated as follows
\begin{equation}
R_{\mathrm{XUV}} - R_{\mathrm{pl}} \approx H_{\mathrm{below}} \times \mathrm{ln} \left(\frac{P_{\mathrm{photo}}}{P_{\mathrm{XUV}}}\right)\,,
\end{equation}

with ${H_{\mathrm{below}}=(k_{\mathrm{B}} T_{\mathrm{eq}}) / (\mu_{\mathrm{below}} m_{\mathrm{H}} g)}$ approximating the scale height of the atmosphere between the optical and the XUV photosphere. This layer is taken to be close to isothermal at the equilibrium temperature of the visible photosphere, $T_{\mathrm{eq}}$, and having a mean molecular weight of ${\mu_{\mathrm{below}} = 2.5}$ due to H/He being in molecular form below the photoionization base. At the visible photosphere, the pressure is set to ${P_{\mathrm{photo}} \simeq 20\,\mathrm{mbar}}$ \citep{2007Fortney}, while at the XUV photosphere it is approximated by ${P_{\mathrm{XUV}} \approx (m_{\mathrm{H}} G M_{\mathrm{pl}})/(\sigma_{\nu_{0}} R_{\mathrm{pl}}^{2})}$, using ${\sigma_{\nu_0} = 6\,\times\,10^{-18}(h\nu_0/13.6\mathrm{eV})^{-3} \mathrm{cm}^2}$.
Instead of taking the whole spectrum into account, we assume that the typical energy of an ionizing photon is ${h\nu_0 = 20\,\mathrm{eV}}$ (${\sim60\,\mathrm{nm}}$) \citep{MurrayClay2009}.

\section{Host star activity evolution}
\label{sec:activity_decay}

Once the protoplanetary disk dissolves, a planet orbiting a young star becomes exposed to high energy radiation from the host star. For orbital periods less than 100 days, this is expected to drive the erosion of the planetary atmosphere. As stars age, rotational spin-down causes a decrease in magnetic activity and with it their high-energy X-ray and ultraviolet emission \citep[e.g.][]{1997Guedel, Ribas2005, Booth2017}. Young stars with ages less than a Gyr emit XUV radiation that can be several orders of magnitude higher than for the present-day Sun. For planets, the XUV irradiation level thus is much higher in the early stages compared to more mature ages.
As a result, their atmospheres are hotter, more expanded and therefore more susceptible to mass loss. When investigating the atmospheric erosion of planetary atmospheres it is thus important to account for changes in the mass loss rates induced by changes in the received XUV flux.

In this study, we take into account the observed spread in X-ray luminosities of stars in open clusters with ages below a gigayear by modelling the host star activity evolution with a range of spin-down ages \citep{2015Tu, 2021Johnstone}. By spin-down or saturation time, we mean the age until which a star emits intense XUV flux at a constant level before decreasing. This is in contrast to most previous population studies \citep[e.g][]{2017Owen, 2017Lopez, 2021Rogers_a}, which assume only one track with a saturation timescale of 100 Myr for all planet hosting stars. As has been pointed out by \citet{2021King}, the EUV contribution to the total high energy flux received by a planet can also significantly influence the mass-loss history of a planet, in particular after the X-ray dominated saturated phase due to the shallower decay of the stellar EUV emission compared to X-rays. To estimate the stellar EUV flux for a wide range of ages and activity levels, we make use of an X-ray and EUV surface flux relation, instead of assuming X-rays and EUVs decline at the same rate \citep{2015Chadney, 2021Johnstone}.

\subsection{Stellar activity decay}
\label{subsec:Xray_decay}

% rotational spin-down & importance of rotation rate on saturation time
Magnetic activity and rotation are closely linked via the stellar dynamo, and decrease together as stars age. Observations of young cluster stars with similar masses and ages show that there is a large spread in rotation rates before $\sim$ 500 Myr \citep{2012Jackson, 2015Tu}. Stars at such young ages tend to cluster either on the fast or the slow rotational branch \citep{Barnes2003}. As mentioned previously, models for the stellar rotation evolution predict a relatively rapid spin-down at a wide range of ages, or saturation times, where the long-term rotational evolution is determined primarily by stellar mass and initial rotation rate \citep{2011Wright, 2015Matt, 2015Tu, 2018Gondoin, 2021Johnstone}. Magnetic field complexity has also been considered as a factor impacting the spin-down \citep{2018Garraffo}, and the underlying cause for the spin-down is still not fully explained. What has been observed is that earlier-type stars staying in the saturated regime longer than later-type stars, and that stars with similar stellar mass undergo a non-uniform spin-down. Some are able to maintain their high activity level for prolonged periods of time, while others start their spin-down and thus decrease in activity around the time the disk dissipates.

\begin{figure}
\centering
\includegraphics[width=0.49\textwidth]{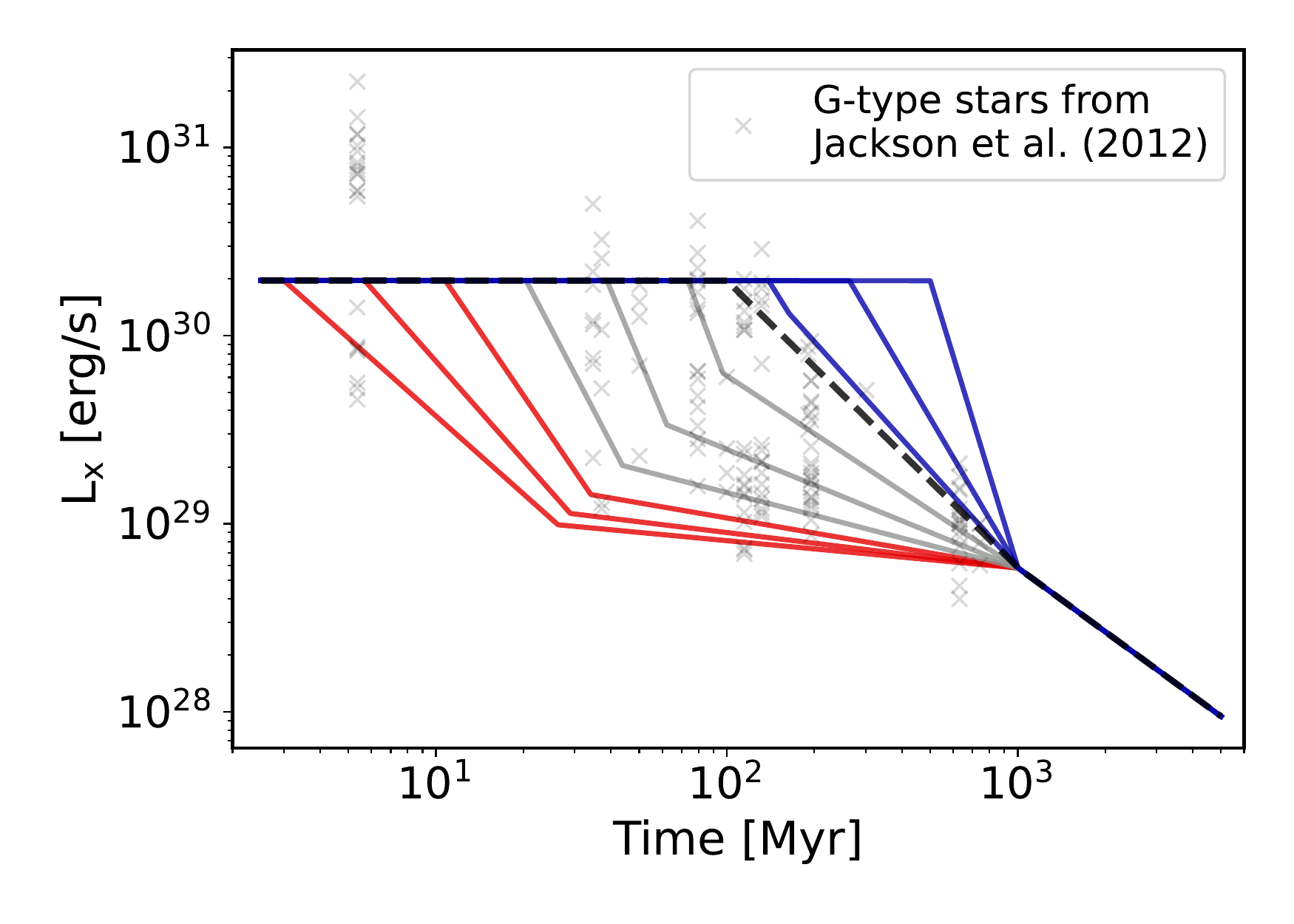}
\caption{X-ray evolutionary tracks for a solar-mass star used in this work. Our tracks, which are motivated by the models from \citep{2015Tu, 2021Johnstone}, encompass the observed spread in observed X-ray luminosities for stars with ages younger than a Gyr. The X-ray measurements for G-type stars in the sample from \citet{2012Jackson} are shown as grey X's. In red, grey and blue colors we show low, intermediate and high activity tracks. For comparison, we show the widely used "median track" with a saturation time of 100 Myr as the black dashed line.}
\label{fig:Lx_tracks}
\end{figure}

The dearth of stars with intermediate rotation periods in young open cluster \citep[e.g][]{2021aFritzewski} implies that the spin-down of an individual star, once it has started, happens relatively quickly. Therefore, we assume a spread of onset ages for rotational spin-down, ranging from a few Myr to a few hundred Myr, followed by a rapid decay phase. At ages of around $1$ Gyr, stars seem to converge and continue a similar rotation and X-ray luminosity evolution. This ultimately translates into a wide distribution of evolutionary tracks for the X-ray luminosity at ages younger than $\sim$ 500 Myr (see Fig.~\ref{fig:Lx_tracks}).

% Tu track approximation and generation of track distribution
Motivated by the rotational evolution model for solar-like stars ($0.9-1.1\, \mathrm{M}_{\odot}$) by \citet{2015Tu} and the updated models by \citet{2021Johnstone}, we approximate the different stellar activity tracks by broken power-law models; as opposed to assuming a saturated phase followed by a single power-law decline only \cite[see e.g.][]{2017Owen, 2020Mordasini}. Each activity track corresponds to a specific saturation time, $t_{\mathrm{sat}}$, until which the star stays at a constant X-ray saturation luminosity. When the star falls out of the saturated regime, the X-ray decay can proceed with a one- or two-piece power-law until the age of $1$ Gyr, where all tracks are set to converge. Tracks with a two-piece drop in X-ray luminosity are set to decay within $\sim\,$25 Myr. For the convergence point, we choose an X-ray luminosity that is motivated by the median value for all G-type stars in the oldest cluster in the sample by \citet{2012Jackson}. Beyond 1 Gyr, the X-ray decrease proceeds with a power-law index which is set to $-1.13$, a value typical for Sun-like stars \citep{2012Jackson}. This X-ray decay slope is also in good agreement with the activity evolution models by \citet{2021Johnstone}, but shallower than the slope used in \citet{2015Tu} ($\sim -1.6$), which was calibrated to match the current solar X-ray luminosity \citep{2015Tu}. We note that the slope used for the X-ray and EUV decay in many other evaporation studies is often quite steep, with values up to $-1.5$ \citep[e.g.][]{2017Owen, 2021Rogers_a}. A power-law decline with such a steep slope implies that the total high-energy emission is dominated mostly by the saturated phase.

In Figure~\ref{fig:Lx_tracks} we show the set of nine activity tracks for a one solar-mass star used in this work, together with a 100 Myr "median track" for comparison. The saturation timescales are evenly spaced in log-space, ranging from 3 to 500 Myr. Since we want to investigate the effect of a mixture of stellar activity tracks on the final radius distribution of a sample of exoplanets, we need to assign each track with a certain occurrence probability.
Informed by the spin-down-age dependence on stellar mass from \citet{2021Johnstone} (see their Figure 10), we choose a mean saturation time of 40, 25 and 15 Myr for K, G and F stars, respectively. From 0.9-1.1 M$_\odot$ we count a star as G-type, while lower masses are taken as K, and higher masses as F spectral type. For each of the three mass bins, we construct a lognormal probability distribution for the saturation time, with 40, 25 and 15 Myr as the mean and a standard deviation of 0.45.
With this in hand, we can estimate the probability that a random star, with a given mass, will evolve along one of the nine tracks. For a G-type star, 60\% of stars will follow one of the grey intermediate tracks, while 30\% follow one of the low activity, red tracks and 10\% one of the blue high activity ones. A consequence of the mass dependence is that more K stars will follow the higher activity grey and blue tracks in Figure~\ref{fig:Lx_tracks}, while most F stars fall out of the saturation regime early on and will mainly follow the red and grey tracks.

Another important parameter of a realistic stellar activity track is the value of the X-ray luminosity, $L_{\mathrm{X}_\mathrm{sat}}$, in the saturated phase. We use the relation by \citet{2011Wright}, given by ${\log(L_{X_{\mathrm{sat}}}/L_{\mathrm{bol}}) = 10^{-3.13}}$, to estimate $L_{\mathrm{X}_\mathrm{sat}}$. This relation is almost independent of spectral type, which means we can use it for all the K, G and F stars in our sample. To take into account that we start our calculations at very early ages when some stars are still on the pre-main sequence, we use the bolometric luminosity right when a star reaches the Zero Age Main Sequence (ZAMS) to estimate the X-ray saturation luminosity. We find that this approach gives more reasonable saturation values for stars with masses higher than the Sun, when compared to using a main-sequence mass-luminosity relation \citep{2013Pecaut}. The bolometric luminosities at the ZAMS for stars of a given mass are extracted from the pre-MS MESA models by T. Steindl (priv. comm.).

\subsection{Estimating the stellar EUV luminosity}
\label{sec:EUV_estimation}

While X-rays ($\sim 10-100$ \AA) can contribute to atmospheric heating, in particular for young active stars \citep{OwenJackson2012}, it is the stellar EUV emission ($100-912$ \AA) that provides the majority of radiation power to ionize hydrogen in planetary atmospheres \citep[e.g.][]{Murray-Clay2009, Sanz-Forcada2011, Wang2018}. Atmospheric mass loss ultimately depends on the combined XUV input, since both X-rays and EUVs heat the upper atmosphere and thus drive the escaping wind. While X-ray measurements are readily available with current space-based telescopes, observing the UV emission from stars other than the Sun is challenging. Between $\sim 400$ and $912\,$\AA\,(the H ioniyation threshold) the interstellar medium absorbs all radiation, even for nearby stars. Currently, no spaced-based telescopes for the observable portion of the EUV spectrum ($\lesssim 400$\,\AA) are in operation, and archival data is only available for a handful of stars. This complicates the estimation of the important EUV content for planet hosting stars.

A simplification that is often made in photoevaporation studies is to assume that the EUV and X-ray irradiation of planetary atmospheres decline at the same rate \citep[e.g.][]{OwenJackson2012, 2017Owen, 2020Mordasini, 2021Rogers_a}. Studies of the Sun's high energy emission over the course of the solar activity cycle have shown, however, that the EUV emission remains rather strong as X-ray surface flux decreases \citep{2015Chadney, 2018King}. \citet{2015Chadney} further demonstrated that simply scaling the entire solar spectrum based on measured stellar X-ray fluxes is not appropriate for stars with spectral types and/or activity levels other that the Sun. They show that, instead, a power-law relation between EUV and X-ray \textit{surface fluxes} provides a more effective way to estimate the unobservable stellar EUV emission. This relation prevents a significant overestimation of the stellar EUV output for young and active stars, but also implies that the stellar EUV output falls off less steeply in time than X-rays. This is also in agreement with e.g. \citet{Ribas2005, 2012Claire, Shkolnik2014, 2021King} who showed that for sun-like stars, the X-ray emission decays faster than the EUV one. The use of a single relation between X-ray and EUV surface fluxes for all spectral types is further supported by \citet{2021Johnstone}, who note that the relation seems to hold true also for young and active planet-hosting stars on the pre-main sequence.

As a consequence of the shallower EUV decline, the total XUV radiation received by an exoplanet at Gyr-timescales can still be high enough to cause considerable mass-loss at later times well past the saturated phase \citep{2021King}. To take this into account in our mass-loss simulations, we assume than X-ray and EUV luminosities evolve hand in hand (see Section~\ref{subsec:Xray_decay}), but decaying at different rates. The power-law slopes of the X-ray decline are given by our model for the stellar activity evolution, while the EUV emission is estimated via the updated surface-flux scaling relation by \citet{2021Johnstone} (short: Jo21). We use the main-sequence mass-radius relation by \citet{2018Eker} to convert the surface fluxes to luminosities, and then take the total high energy output of the star as the combined X-ray and EUV luminosity.

\begin{figure}
\includegraphics[width=0.49\textwidth]{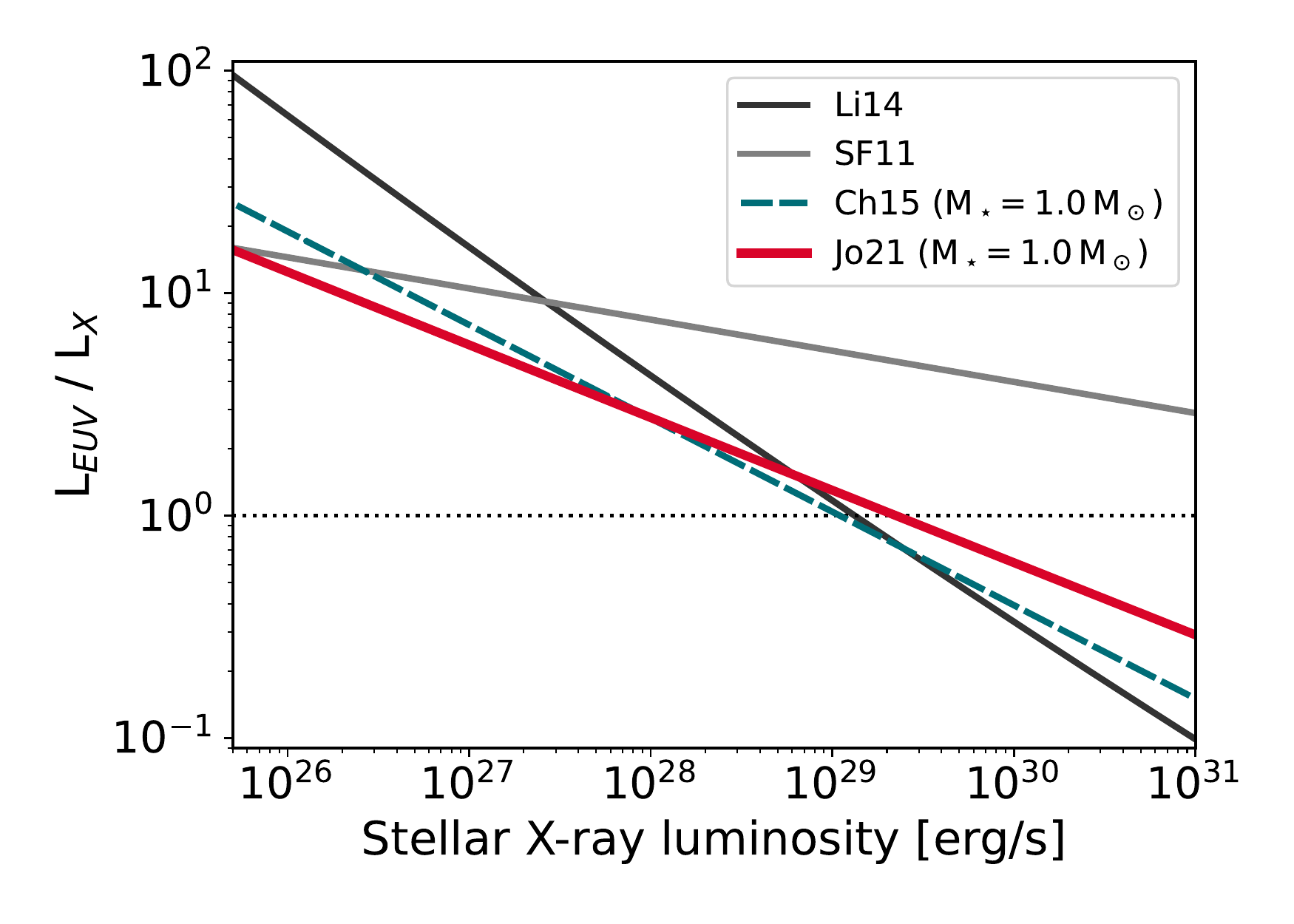}
\caption{Estimates of stellar EUV emission for a range of X-ray luminosities based on four different methods. The relation by \citet{Sanz-Forcada2011} (SF11) is shown in grey, the one by \citet{Linsky2014} (Li14) in black, and the surface-flux relations by \citet{2015Chadney} (Ch15) and \citet{2021Johnstone} (Jo21) in blue and red, respectively. The surface-flux relations have a slight mass-dependence and we only show the $1\,\mathrm{M}_\odot$ case.}
\label{fig:EUV_rel}
\end{figure}

In Figure~\ref{fig:EUV_rel} we compare several EUV estimation methods for a $1\,\mathrm{M}_\odot$ star. In addition to the X-ray and EUV surface flux relation by \citet{2015Chadney} and \citet{2021Johnstone}, we show two other EUV estimation methods widely used in the literature. One is the empirical scaling relation between X-ray and EUV energy bands for late-type stars based on synthetic XUV spectra by \citet{Sanz-Forcada2011}, while the other makes use of the tight correlation between X-ray and Ly$\alpha$ luminosities for a sample of K to F dwarfs \citep{Linsky2013}, which in turn can be used to estimate EUV luminosities from Ly$\alpha$ measurements \citep{Linsky2014} (short: Li14). In this work, we have chosen the method by \citet{2021Johnstone}, Jo21, as our baseline, but the other options can be calculated with PLATYPOS as well (see Appendix~\ref{sec:EUV_est}). Figure~\ref{fig:EUV_rel} shows that for highly active stars the relation by \citet{Sanz-Forcada2011} (short: SF11) predicts EUV luminosities which can be more than an order of magnitude higher than the estimates from Ly$\alpha$ or the surface flux relations.

\subsection{Integrated XUV emission}
\label{subsec:XUV_integrated}

\begin{figure}
\centering
\includegraphics[width=0.49\textwidth]{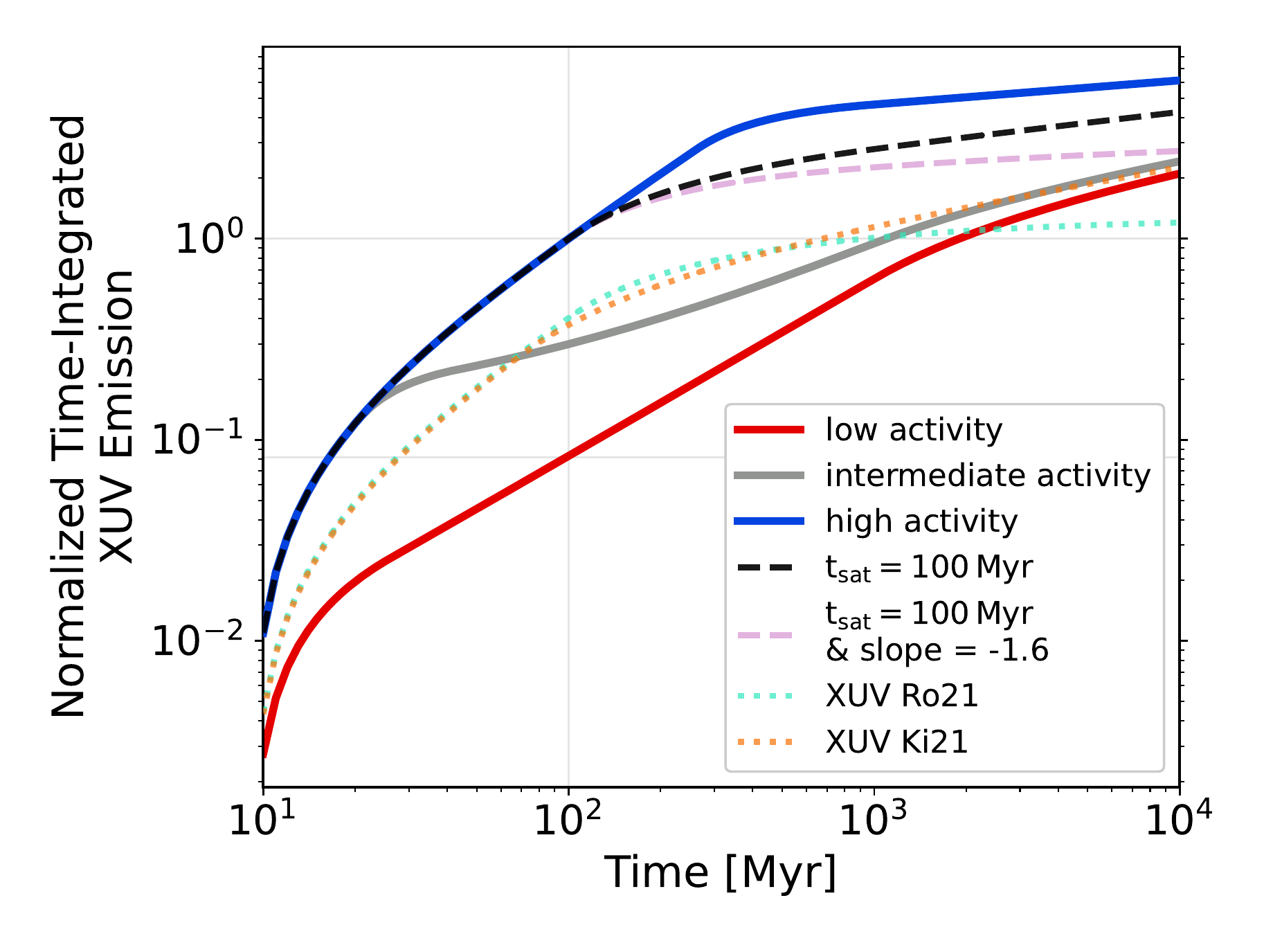}
\caption{Cumulative XUV emission from 10 Myr to 10 Gyr normalized to the cumulative XUV emission of the reference track  at 100 Myr (t$_{sat}=$100 Myr; black dashed line). A low, intermediate and high activity track for a one solar mass star are shown in red, grey and blue, respectively. The large difference in emitted XUVs below 1 Gyr is clearly visible. For comparison, we also show the cumulative XUV emission used in \citet{2021Rogers_a} (Ro21) and \citet{2021King} (Ki21) as green and orange dotted lines, as well a track with a saturation time of 100 Myr, but with a steeper X-ray slope of -1.6, in pink.}
\label{fig:integrated_XUV}
\end{figure}

Figure~\ref{fig:integrated_XUV} shows the time-integrated XUV emission for a low, intermediate and high activity track (red, grey, blue) of a solar mass star, together with a commonly used 100 Myr track (black). These tracks are constructed as described in Sec.~\ref{subsec:Xray_decay}, with the EUV content estimated according to the power law realtion between X-ray surface fluxes and EUVs by \citet{2021Johnstone} (Jo21). For comparison, we also show the XUV tracks from \citet{2021King} (Ki21) and \citet{2021Rogers_a} (Ro21). For easier comparability, we normalize everything to the cumulative XUV emission of the 100 Myr track at 100 Myr. We show this comparison for two reasons. One is to make the reader aware of how much differences in the choice of X-ray and EUV saturation luminosity and decay slope can impact the emitted stellar XUV radiation over time, and the other is to show how different saturation timescales, i.e. a low, intermediate and high stellar activity track, reflect on the total XUV luminosity around 100 Myr and at Gyr ages.

The first thing we want to highlight is the difference in XUV saturation luminosity bewtween our tracks with the Jo21 EUV estimation and the Ro21 tracks. Due to the lower saturation luminosity (factor 2.4 lower than ours) and the steeper XUV decline of the Ro21 track, their integrated XUV emission at $10\,$Gyr is comparable with the emitted radiation in the first 100 Myr of a star which follows one of our higher activity tracks. So overall, the planets in our sample receive significantly larger integrated XUV fluxes when compared to the works by e.g. \citet{2021Rogers_a} or \citet{2017Owen}. Due to the similar X-ray saturation luminosity between the Ki21 and Ro21 tracks, the Ki21 tracks give a XUV emission similar to the Ro21 tracks in the first 100 Myr, but their shallower EUV decline causes the EUV emission beyond one Gyr to still contribute significantly to the total XUV flux.

The second thing we want to point out is the difference in the time-integrated XUV emission before $\sim 1 \mathrm{Gyr}$ between our low, intermediate, and high activity track with the Jo21 EUV estimation. If a star drops out of the saturated regime early on, the total XUV flux received by a planet by the age of 100 Myr can be a factor 10 lower than for a planet around a star that stays saturated longer. So while the cumulative emission at 10 Gyr is comparable within a factor 4 between the low activity and reference track, a planet around a low activity star receives much less flux early on. For a plant whose mass is dominated by its core, the amount of XUV exposure early in its life can be crucial because the atmosphere is still warm and inflated and material can be most easily lifted out of the planet's gravitational well.

\section{Evolution of the planetary sample}
\label{sec:planetary_sample}

Our aim is to investigate how the inclusion of a stellar activity track distribution affects the planetary radius distribution at Gyr ages. \citet{2021Kubyshkina} recently performed a comparative study of sub-Neptune-like planets orbiting a range of stellar masses and including different evolutionary histories. They show the impact of a slow and fast rotating star on a fixed grid of exoplanets. %exo/-pla/-nets
In contrast to their study, we construct an exoplanet population, which is motivated on the one hand by the properties of the observed super-Earth and sub-Neptune planets discovered with \emph{Kepler}, and on the other hand by numerical simulation results from planet formation studies predicting the primordial planet composition after the dispersal of the protoplanetary disk.

We do not aim to fit the observed bimodal radius distribution, but instead, using a reasonable primordial planet population, show the relative influence of a distribution of stellar activity tracks on the final radius distribution, and the slope, width and location of the radius gap.

\subsection{Input planet population}
\label{subsec:population_details}
% details on core-mass, orb. sep, host star mass distro, initial fenv

We construct a planet population consisting of short-period exoplanets \mbox{($P \leq 100$ days)} with Earth-like rocky cores underneath gaseous H/He envelopes. To generate a planet population with properties similar to the observed \emph{Kepler} population, we choose an orbital period distribution which has been obtained by fitting the planets in the \emph{Kepler} sample and correcting for transit probabilities \citep{2017Owen, 2018Ginzburg}:
\begin{equation}
\frac{\mathrm{d}N}{\mathrm{d log}P} \propto
    \begin{cases}
    P^2,& \text{if } P < 8 \text{ days} \\
    \text{constant},   & \text{otherwise}\,.
    \end{cases}
\end{equation}

The core masses range from from $1$ to $25$\,\Mearth, and are distributed according to a broken-power law, with cores below a certain threshold mass, $M_{c}$, being described by a Rayleigh distribution, and more massive cores by an inverse square tail:
\begin{equation}
\frac{\mathrm{d}N}{\mathrm{d}M_{c}} \propto
    \begin{cases}
    M_{c}\,\mathrm{exp}\,\Big(-M_{c}^2/(2 \sigma_{M}^2)\Big),& \text{if } M_{c}\geq 5\,M_{\odot}\\
    M_{c}^{-2},              & \text{otherwise}\,.
    \end{cases}
\end{equation}

\noindent \citet{2018Ginzburg} showed that such distribution is in good agreement with the observed high-mass tail from radial velocity measurements. Similar to \citet{2019Gupta}, we set the dividing mass, $M_{c} = 5$\,\Mearth, and use $\sigma_{M} = 3$\,\Mearth\,for the Rayleigh distribution \citep{2017Owen}.

The planets in our population are set to orbit sun-like stars with masses similar to the ones found in the \emph{Kepler} sample \citep{Fulton2017}. The stellar mass distribution can be described by a Gaussian centered at $0.97$\,\Msun\,and a standard deviation of $0.14$\,\Msun. To be able to better disentangle the influence of varying XUV saturation levels due to different stellar masses and the effect of different stellar activity tracks, we also conduct simulation runs with one solar mass host stars, only.

As stated before, we do not attempt to fit the observed bimodal radius distribution, but rather reproduce the general shape by making reasonable assumptions about the primordial planet population and assuming photoevaporation as the main mechanism shaping the envelopes of the planets over time. We assume the primary atmosphere of the planets has been accreted during their formation inside the protoplanetary disk, and consists mainly of hydrogen and helium. To estimate the post-formation envelope mass fraction, $f_{env}$, which is defined as the fraction between the mass of the envelope and the total mass of the planet, we rely on the results from two numerical simulations. \citet{2020Mordasini} (short: M20) predict that post-formation envelopes are positively correlated with core mass, but with an inverse dependence on orbital separation.
For our sample, this corresponds to a median initial envelope of around 4\% in mass. \citet{2019Gupta} (short: Gu19) on the other hand, based on a study on accretion and mass loss during the disk dispersal phase \citep{2016Ginzburg}, report a positive dependence on core mass only. This predicts the planets in our sample to be born with more massive envelopes, with a median mass fraction of 10\%. We choose the M20 study to predict the primordial envelope masses of the planets in our sample, and show a comparison to the Gu19 envelope mass fractions in Appendix~\ref{sec:app_fenv_init}.

By 10 Myr, a typical protoplanetary disk has dispersed \citep[e.g.][]{2001Haisch, 2011Williams, 2016Pecaut, 2017Venuti}, and we set this as the starting age for all planets in our sample. We tested different starting ages but find that at Gyr-ages, the exact starting time does not have a significant influence on the final planetary radii.
Additionally, we make the simplifying assumption that planetary orbits are circular, planets undergo negligible migration beyond the disk dissipation age, and that the change in stellar bolometric luminisity from pre- to post-MS has a negligible effect on the thermal contraction of the planets. 
For works covering these effects we did not include, see \citep[e.g.][]{LopezFortney2014}.

To avoid having planets with unphysical properties in the unevolved population, we follow \citet{Kubyshkina2018b} and further remove any planets with a bulk density lower than $0.03$ \gcc, where the Roche lobe is closer than $0.5$ \Rearth from the planetary surface, and that have a Jeans escape parameter greater than $80$, which indicates that the outflow not in the hydrodynamic regime \citep{Fossati2017}.

We run our simulations from 10 Myr to 10 Gyr, and in general, show the final population at one single chosen age. Since in reality, the observed population includes a distribution of star-planet system ages, we also include the option of generating an age distributed exoplanet population. Similar to \citet{2020Modirrousta}, we fit a truncated Gaussian to the ages of a selected sample of observed exoplanets. We downloaded the catalog from exoplanet.eu on March 9th, 2022, and select only confirmed planets with periods $\leq$ 300 days, and age errors less than 50\%. We obtain a mean of $\sim$ 3.1 Gyr, standard deviation of $\sim$3.8 Gyr, setting the minimum and maximum age for the fit to 10 Myr and 13.8 Gyr. We create snapshots of our simulation run at 20 different log-spaced ages from 10 Myr to 10 Gyr. From the fit to the observed ages, we estimate the probability of a planet having one of the given snapshot ages. To construct the age-distributed population, for each planet in the sample we randomly pick one of the 20 ages based on the estimated probabilities, and obtain with it the corresponding planet parameters at the given age. Our findings show that for the chosen age distribution, where the majority of star-planet systems have ages of several Gyr, and only a small fraction ($\leq 30\%$) has ages younger than 2 Gyr or older than 7 Gyr, the 1D radius distribution is not significantly different from a single intermediate age of several Gyrs. For the sample with a single intermediate stellar acivity track, less than 2\% of the planets in the age-distributed sample end up either above or below the gap compared to the 5 Gyr single-age sample. We further discuss the effect of an age distribution in Section~\ref{subsec:age_distro}.

\subsection{Chosen parameters for the simulations}
\label{subsec:simulation}

For the simulation results shown in Section~\ref{sec:results}, we use the ChRo16 planet models for a population with a core mass distribution peaked at 3 M$_\oplus$. The primordial envelope mass fractions are estimated according to M20. Regarding the host stars, we either assume a single host star mass of $1\,\mathrm{M}_\odot$, or a distribution of stellar masses ranging from K to G to F stars.

To estimate the mass loss rates, we combine energy- and radiation\slash recombination-limited mass loss with the Lopez-$\beta$ estimation (unless stated otherwise). Previous works have demonstrated that the evaporation efficiency for close-in exoplanets can vary based on planetary properties, in particular the gravitational potential, and irradiation levels \citep[e.g.][]{Murray-Clay2009, OwenJackson2012, Salz2016b}. Hydrodynamic simulations predict values between $\sim 0.1-0.3$ for the planetary masses present in our sample. We perform all runs with a fixed constant evaporation efficiency of 10\%. All simulation outcomes are shown at a single chosen age.

We focus our Results and Discussion sections on the impact of the stellar activity evolution on the predicted planetary radii at Gyr ages. The Appendix shows the impact of other simulation inputs like the planetary structure model, effective absorption radius, primordial gas envelope mass fraction, or core-mass distribution. Some limitations of our assumptions and model details are discussed in Section~\ref{subsec:limitations}.

Besides the input planet population and the evaporation model, the strength and evolution of the host star XUV emission has to be chosen. Planetary XUV fluxes change according to the host star activity track, directly affecting the photoevaporative mass-loss rates and thus the fate of the planets. For each planet in our sample, we calculate the temporal mass and radius evolution for the nine tracks introduced in Section~\ref{sec:activity_decay}, and choose the Jo21 method to estimate the important EUV contribution. A detailed description of the modeled stellar activity decay is given in Section~\ref{subsec:Xray_decay}, and Appendix~\ref{sec:EUV_est} discusses the impact of the EUV estimation and the slope of the X-ray decay.

Combining all these inputs, we compute the momentary mass-loss rate at each time step from the starting age to the final age of the simulation run. For each planet, we use the latest radius, envelope-mass fraction, XUV absorption radius and stellar XUV flux to calculate the mass-loss rate at the age of the simulation run. If the XUV absorption radius exceeds a planet's Roche lobe, it is set to the Roche lobe value. We then compute the mass lost in a given time step, and adjust the time step if the radius change is too drastic (> 0.5\%) or too little (< 0.02\%). We allow for time steps to range from 0.01 to 10 Myr, and reduce or increase the time step by a factor of 10 if the radius change is too extreme. We then update the planet mass and use ChRo16 planetary models to calculate the new radius with the reduced gaseous envelope. Next, we update the XUV flux based on the specified stellar evolution track and then continue this cyclic procedure until the planetary radius has reached the core radius and no atmosphere is remaining, or the end of the simulation is reached. This allows us to trace the temporal mass and radius evolution induced by atmospheric photoevaporation and planetary cooling for a range of stellar activity tracks (see Fig.~\ref{fig:Lx_tracks}), and compare them against each other.

\section{Results}
\label{sec:results}

We study the effect of an observationally-motivated distribution of stellar activity tracks (see Section~\ref{sec:activity_decay}) on the 1D and 2D radius distribution of a reasonable population of exoplanets.
First, we describe the impact of different individual activity tracks, and then how a distribution of activity tracks impacts the radius gap.

Photoevaporation simulations are complex, multidimensional problems due to the large number of required assumptions and inputs.
We try to disentangle and visualize individual influences on the radius distribution of our planetary sample in Appendix~\ref{sec:app_planet_model}-\ref{sec:EUV_est}, and focus here only on the impact of different activity tracks for a single set of simulation assumptions (see Sec.~\ref{subsec:simulation}). We do not try to reproduce a radius gap which matches the observed radius valley slope and location, but instead investigate the relative changes in the radius distribution caused by the host star activity evolution. Our simulations qualitatively reproduce the bimodal radius distribution, and show that the duration a star spends in the saturated regime shifts the location of the radius valley. In addition, a spread in saturation times can cause the borders of the radius gap to become fuzzier.

%%%%%%%%%%%%%%%%%%%%%%%%%%%%%%%%%%%%%%%%%%%%%%%%%%%%%%%%%%%%%%%%%%%%%%%%%%%%%%%%%
%%%%%%%%%%%%%%%%%%%%%%%%%%%%%%%%%%%%%%%%%%%%%%%%%%%%%%%%%%%%%%%%%%%%%%%%%%%%%%%%%
\subsection{Single stellar activity track vs. distribution of activity tracks}
\label{subsec:1Dresults}

\begin{figure}
\centering
\includegraphics[width=0.45\textwidth]{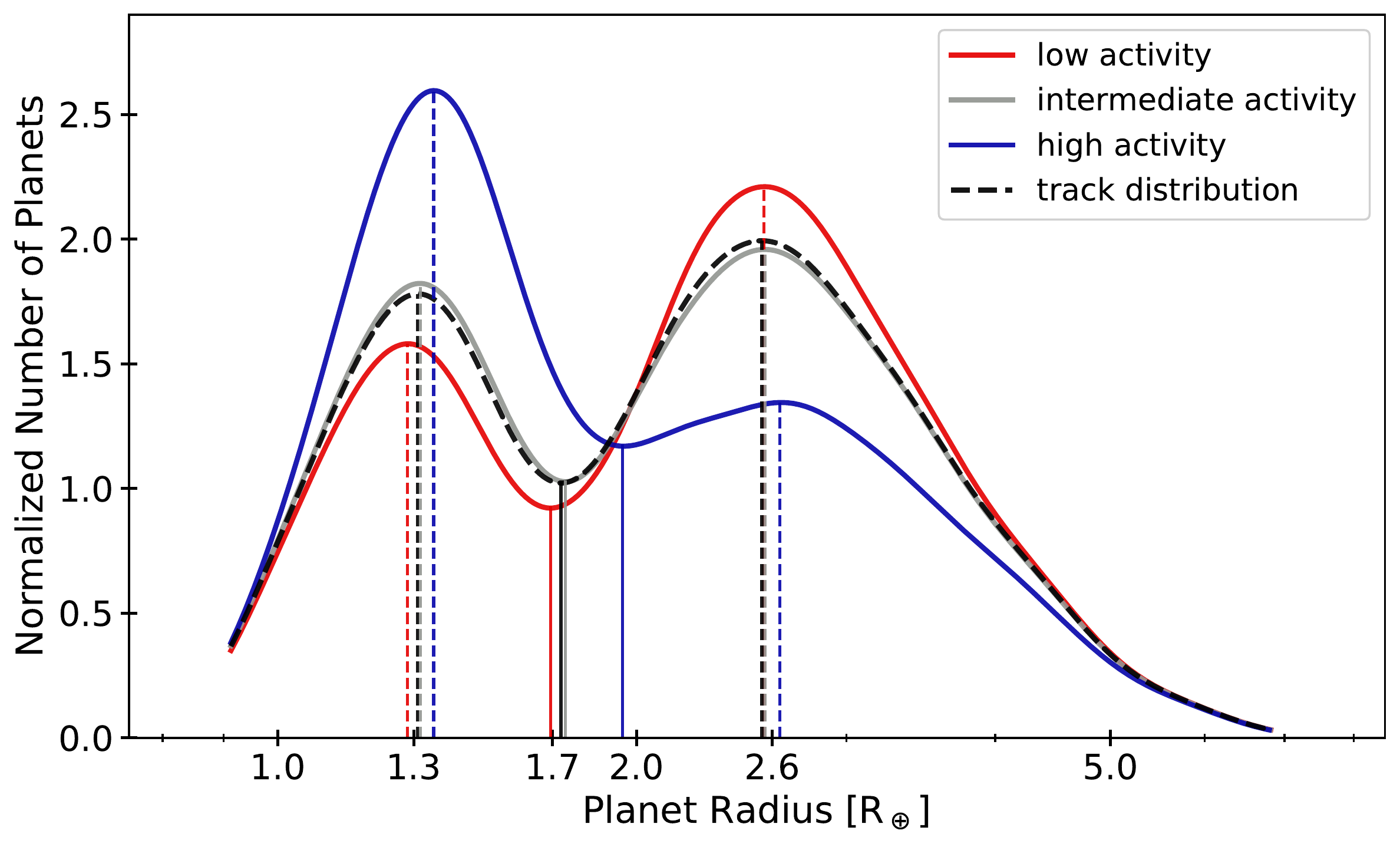}
\caption{Influence of the stellar activity track on the 1D radius distribution for a single stellar mass. We compare the 1D KDEs of the radius distribution at 5 Gyr for three individual stellar activity tracks, ranging from stars which start their spin down after a few Myr and thus are considered to have very low activity (red), to stars with a saturation time around 40 Myr and thus an intermediate activity (grey), to highly active stars which remain saturated for several hundred Myr (blue). The bimodal shape is visible, but the second peak is very diminished for the high activity scenario, due to the prolonged intense XUV irradiation and thus high mass loss rates in the first Gyr. This also causes the first peak, and in particular the radius gap minimum to shift to larger radii, because planets with more massive and thus larger bare cores get completely stripped and fall below the gap. We also show the same planet sample but with a distribution of stellar activity tracks as the black dashed line. The shape of the 1D radius distribution and the location of the gap minimum is still set by the large number of intermediate tracks in the sample, and is not significantly changed by the inclusion of a low and high activity tail.}
\label{fig:1D_FGK_1_5_10}
\end{figure}

In Figure~\ref{fig:1D_FGK_1_5_10}, we show several Gaussian kernel density estimates (KDEs) of the radius distribution of our planet population after ongoing evaporation around one solar mass host stars at 5 Gyr. We compare the radius distribution for three individual activity tracks to a mixture of stellar activity tracks. The simulations with a single stellar activity track are chosen to cover the extremes, from very low, to intermediate, to very high activity (red, grey and blue, respectively). Note that we do not simply scale the intermediate track up or down as e.g. in \citet{2020Mordasini}, but in our case, low, intermediate and high refers to different saturation times (see Section~\ref{subsec:Xray_decay}). As a reminder, in the sample with a distribution stellar activity tracks, the bulk of G-type planets will follow an intermediate track (60\%), while 30\% follow one of the lower activity, and about 10\% a high activity ones.

As expected, in the simulation with the highest activity track, the largest number of planets completely lose their atmosphere and become bare rocky cores. In the first few hundred Myr, when planets are still cooling and contracting, they are most susceptible to mass loss. Thus, the longer a star spends in the saturated phase, constantly emitting large amounts of XUV photons, the more detrimental the mass loss for the planet. As a consequence, more planets with small cores further out and planets with more massive cores closer in are completely stripped of their atmosphere, compared to stars with shorter saturation timescales. For the simulation run shown in Figures~\ref{fig:1D_FGK_1_5_10} and \ref{fig:2D_1M_FGK_1_5_10}, this translates to a fraction of bare cores below the gap of 30, 35 and 51\% of the entire planet sample for the low, intermediate and high activity track, and about 34\% for the mixed track sample.

In the 1D radius distribution in Figure~\ref{fig:1D_FGK_1_5_10}, the bimodal shape is present across the whole range of activity tracks, although the second peak is quite diminished for the highest activity track. In this case, the overall evaporation is so strong that the large number of lower mass planets in our sample with core masses below 5 M$_\oplus$ stand almost no chance of retaining an envelope by the age of 5 Gyr. The slight shift of the first peak to larger radii for the highest activity track is caused by more massive ($\sim$ 5-10 M$_\oplus$) and thus larger bare cores. The location of the gap minimum is comparable for low and intermediate tracks in this run, and only shifts to larger radii for the extreme high activity track. Again, this shift is caused by more massive, and thus larger bare cores populating the first radius peak and thus pushing the gap to larger radii. In the population with a distribution of stellar activity tracks, the gap minimum is not significantly affected by the inclusion of very low and very high stellar activity tracks. The shape of the radius distribution in 1D is mostly dominated by the large number of star-planet systems with intermediate activity tracks.

While we investigate here the effect of varying saturation times, or activity tracks, the result also highlights how the overall XUV exposure - for which the first several 100 Myr contribute significantly - can alter the radius distribution, in particular the location of the gap minimum and the relative height of the peaks. It is thus very important to pay attention to the estimation of the X-ray saturation luminosity and the EUV contribution. We show the impact of other simulation inputs like the planetary core mass distribution or the effective absorption radius on the radius distribution in the appendix, but our simulations suggest that the inclusion of an observationally-motivated stellar activity track distribution compared to a single intermediate track does not significantly affect the shape of the radius distribution in 1D. Other simulation parameters, like for example the EUV estimation method, the estimation of the effective absorption radius or the mass loss rate calculation, seem to have a stronger impact on the location of the gap minimum.

\subsection{Location and slope of the gap}

\subsubsection{2D structure of the gap in period space}

We show in Figure~\ref{fig:2D_1M_FGK_1_5_10} and \ref{fig:2D_smass_FGK_1_5_10} the 2D radius distribution at 5 Gyr for a single stellar mass and a distribution of stellar masses - on the left for a distribution of stellar activity tracks, and on the right for three single activity tracks ranging from low, to intermediate to high. The main features - two populations of planets separated by a less populated gap - are clearly visible in all of our simulation runs, but the number of planets below the gap increases significantly going from the low to the high activity track. The lower edge of the gap is populated by the heaviest cores that can be stripped, and thus a shift of the gap to larger radii can be observed, going from low to high activity. The slope of the gap is consistent across all individual tracks, and the inclusion of a mixture of activity tracks with only a small fraction of very extreme low or high activity tracks also does not change the slope of the gap in period or flux space (typically, the gap slope in period space is dlog$R_\mathrm{p}$/dlog$P\approx-0.19\pm0.01$ for a single stellar mass and $\approx-0.16\pm0.01$ for a mix of stellar masses).

To quantify the slope and width of the gap in our simulations, we derived a gap fitting method suitable for those respective planet distributions. In line with previous radius-gap determinations, we assume a linear relationship between the radius gap and the planetary period or the insolation in log-log space \citep[e.g.][]{VanEylen2018b, 2019Martinez, 2020Loyd}.

We found that many of the existing gap-fitting approaches struggle either with capturing the gap properties in the case where a significant number of planets exist inside the gap - which we call here ''fuzziness'' of the gap, or in the case when the gap is completely empty \citep[e.g.][]{VanEylen2018b, 2019Martinez, 2020Loyd, 2022Petigura}. In particular, \citet{VanEylen2018b} fitted the gap position and slope in their data set by starting out with the usual likelihood function for fitting a straight line to a set of data. However, since the desired output is a fit to an absence of data, they inverted the likelihood function, so that the likelihood was maximized when the fitted line was placed away from the data at both edges of the gap. We note here that this approach can only work for a completely empty gap, because their inverted likelihood is ill-defined and goes towards infinity whenever the line happens to go through a data point. If there are planets located inside the gap, as is the case for several of our scenarios, it happens quite often that an MCMC sampling of the line fit parameters reaches those ill-defined points in the parameter space, making the \citet{VanEylen2018b} fit method not applicable for the simulations which yield a fuzzy gap. \citet{VanEylen2018b} and \citet{2021David} also used an approach based on support vector machines (SVMs) to find the line which maximizes the boarders between two distinct classes of planets in the period-radius or insolation-radius plane. \citet{2022Petigura} note that this approach struggles for a sample where the gap is not completely devoid of planets.

There are other approaches in the literature which are based on kernel density estimates (KDEs) of the planet population, for example in the period-radius plane. \citet{2019Martinez} and \citet{2022Petigura} apply a KDE to get the number density of detected planets, find the minimum density along a number of vertical lines, or 1D projections through the chosen plane, and fit a powerlaw to the train of minima. The 'gapfit' code by \cite{2020Loyd} subtracts off a trial gap relationship and evaluates the 1D KDE of the residuals. These approaches are significantly influenced by the chosen smoothing parameter, which determines how much weight the planets near the gap edges have on the fit compared to the planets further away.

\begin{figure}
\centering
\includegraphics[width=0.4\textwidth]{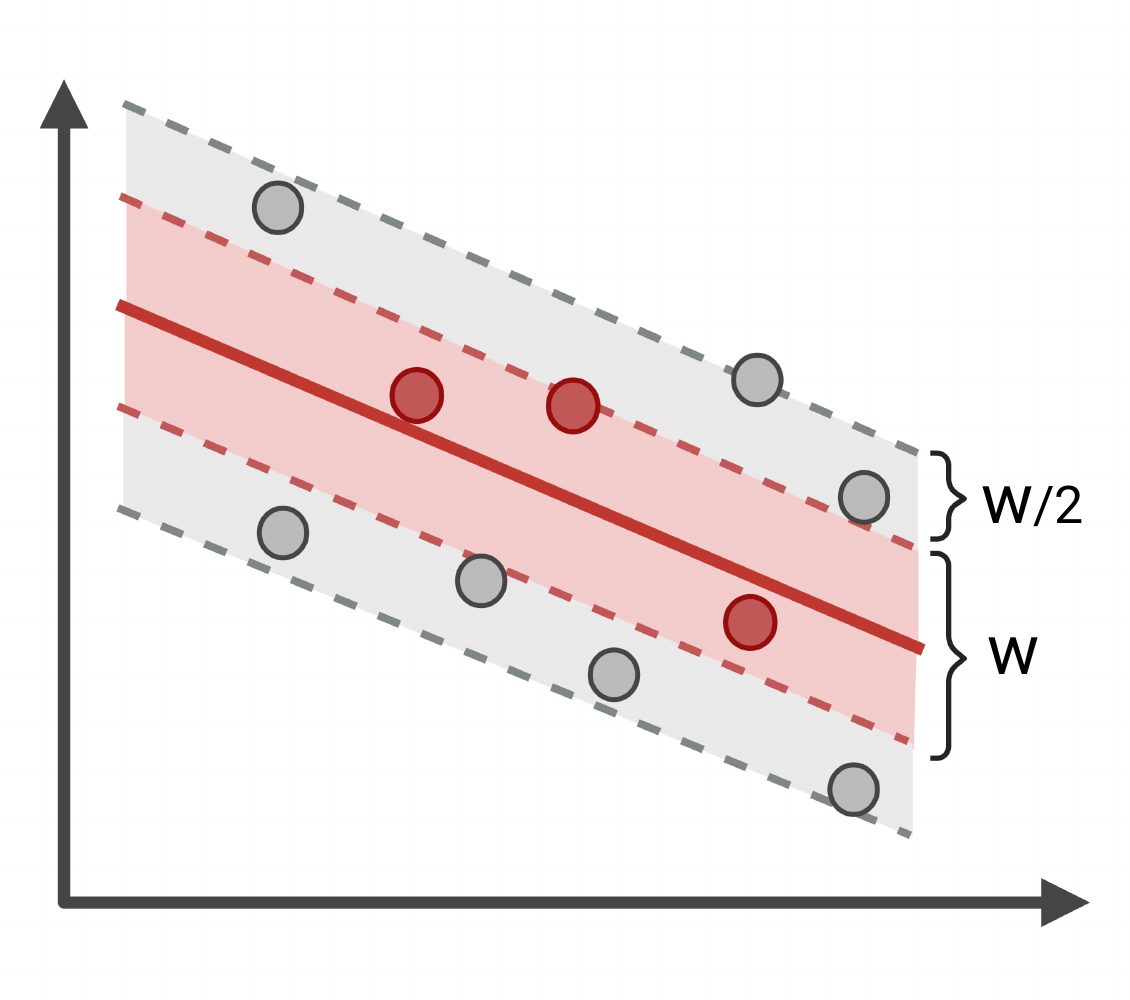}
\caption{Schematic of the used fitting routine. We determine, for a given strip with, $w$, how many planets lie in the central strip around the empirical location of the gap, and how many lie in the two comparison strips of width $w/2$ next to it. We then minimize this ratio of planets inside and outside the strip across a grid of appropriately chosen parameters for the slope, y-intercept and width, to find the line and strip width that best fits the radius gap.}
\label{fig:gapfit}
\end{figure}

\begin{figure*}
\includegraphics[width=0.95\textwidth]{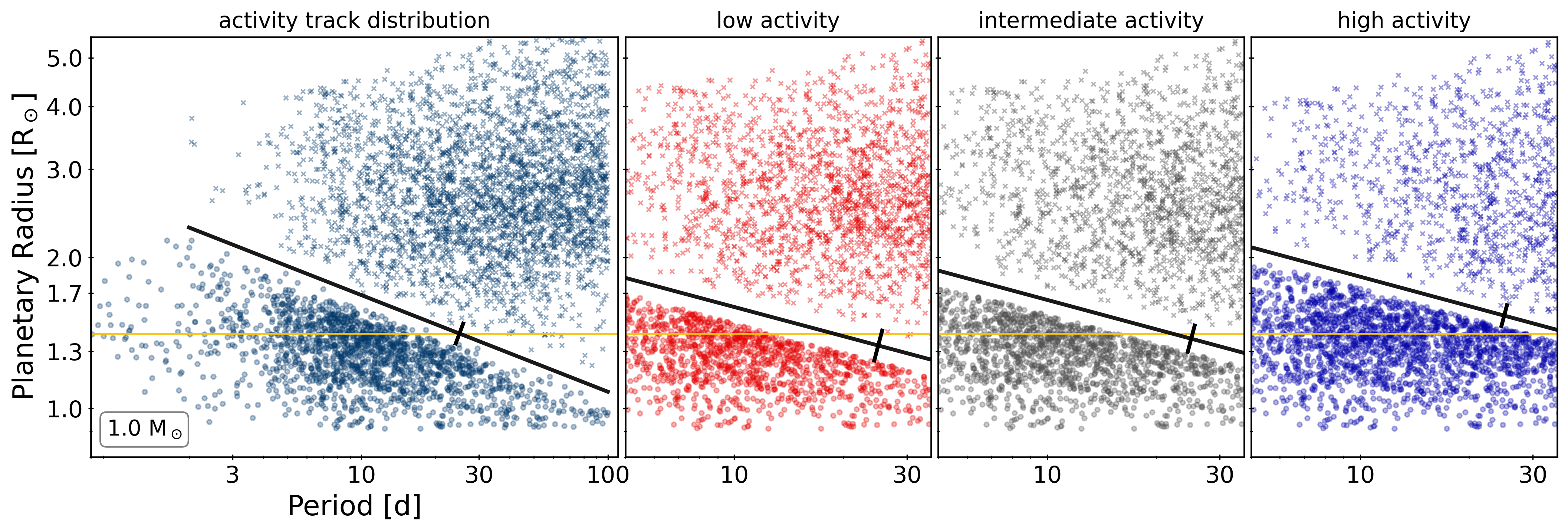}
\caption{Influence of the stellar activity track on the exoplanet radius gap, for a single stellar mass. We show from left to right the same planet population at 5 Gyr for the mixed-track sample, and a single low, intermediate and high activity track. Two distinct populations separated by a band with almost no planets is visible in all panels. The orthogonal distance fit of the gap and the corresponding gap width are shown in black. The gap slope is consistent across all four samples, but going from low to high stellar activity, the gap shifts upwards due to more massive and larger cores below the gap. The fitted gap width decreases with increasing stellar activity from  0.06, 0.05, to 0.04 dex for the low, intermediate and high activity track. Higher levels of evaporation lift the bare core boundary and narrow the gap. For the mixed track sample, the width of the gapfit is even smaller, with 0.03 dex. The gap is less pronounced and fuzzier due to the superposition of planets shaped by the different stellar activity tracks. The horizontal dashed line is drawn to highlight the shift of the gap between individual tracks and the mixed-track sample.}
\label{fig:2D_1M_FGK_1_5_10}
\end{figure*}

\begin{figure*}
\includegraphics[width=0.95\textwidth]{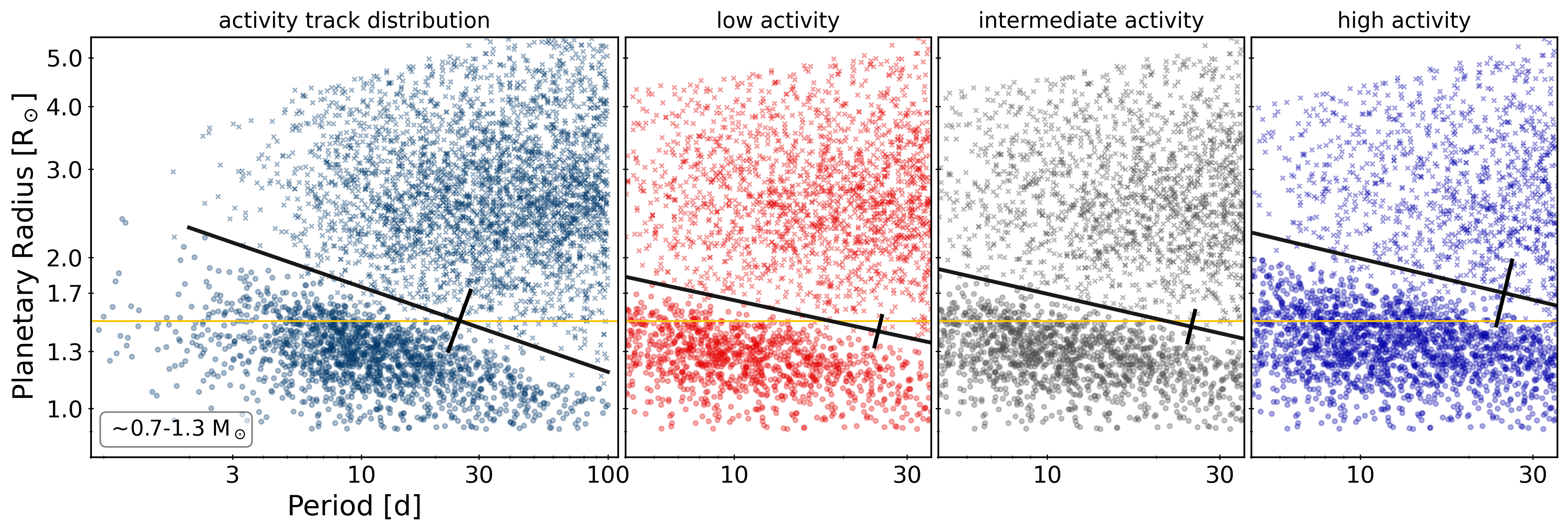}
\caption{Influence of the stellar activity track on the exoplanet radius gap, for a sample with mixed stellar masses. Same as Figure~\ref{fig:2D_1M_FGK_1_5_10}, but for the planet population with a distribution of stellar masses of F, G and K spectral type. Although less pronounced, a deficit of planets separating the planets with envelope from the bare cores is still visible. For the highest activity track, the gap becomes hard to discern.  Due to the extreme mass loss, very few planets remain above the gap. Here, the width of the gap is largest in the mixed-track sample on the left. The superposition of planets with various activity tracks makes the already filled-in gap in the stellar mass sample appear wider. Planets inside the gap are either bare cores or planets that can hold on to very thin envelopes of $\leq 1\%$ in total planet mass in our simulation setup.}
\label{fig:2D_smass_FGK_1_5_10}
\end{figure*}

We instead try a different approach to fit the gap by defining a test strip of width $w$, surrounded by two comparison strips of width $w/2$ next to it (see Fig.~\ref{fig:gapfit}). A test strip that captures the empirical location of the gap (in terms of slope, y-intercept and width) will produce a low number of planets inside the test strip compared to the number of planets in the comparison strips. We therefore calculate the ratio of the counted planets in the test and comparison strips for a grid of reasonable gap widths and slopes\footnote{This method only works if the test and comparison strips actually capture the relevant parts of the planet distribution and manage to capture a decent number of planets in the strips. We therefore restricted our test grid to gap positions motivated by the 1D population, and restricted the minimum gap width to avoid combinations where zero planets were captured in the strips.} and select the parameters yielding the lowest ratio as the representation of the gap. In the case of fuzzy gaps, our fit results agree very well with the gap-fitting methods which are based on KDEs \citep{2019Martinez, 2020Loyd, 2022Petigura}; those KDE approaches,  however, do not yield width estimates. For empty gaps, we find that those methods only yield results similar to our method when decreasing the kernel bandwidth to better capture the sharp boundaries of the gap. While the slope of the gap is always in agreement, the gap fit is strongly drawn towards the cloud of planets above the gap, if the smoothing parameter is too large. We also find that the gap fits from the SVM approach are in good agreement with our orthogonal distance approach for both a non-empty, or noisy gap, as well as a well-defined empty gap. In our approach, we have extra information on whether the planets have some remaining envelope or not since our planets are simulated, and can therefore label them accordingly for SVMs -- however, such information is not readily available for the observed exoplanet population. This labelling allows the SVMs to also work well on our data with a fuzzy gap. These agreements show that our test strip method captures the parameters of the gap well, while adding a gap width estimate to the toolbox.

\subsubsection{Differences of the 2D gap in period or flux space}

While the effect of a mixture of stellar activity tracks on the 1D radius distribution seems negligible (see Sec.~\ref{subsec:1Dresults}), in planetary radius vs. period or planetary irradiation space, some influences on the gap are visible. For the single stellar mass run in Figure~\ref{fig:2Dplots}, panels (a)-(d), going from an individual intermediate track to a mix of tracks, the gap becomes slightly narrower and fuzzier in period and flux space. This arises from the fact that the planet population with a stellar activity track distribution is a superposition of the populations with a single stellar activity track. Around stars with high activity, for the same orbital separation, planets with more massive cores can be stripped compared to stars with low or intermediate activity. This causes the bare core boundary, or lower edge of the gap, in the mixed track sample to be composed mostly of bare cores around stars which stay active for prolonged periods. At 10 (50) days, the bare core mass increases from 5.4 (4.0), to 5.6 (4.1) and 6.3 (4.7) M$_\oplus$ going from low to high stellar activity, and with 5.8 (4.3) M$_\oplus$ is slightly higher for the mixed-track sample compared to the intermediate track. So while the lower edge of the gap is populated by the more massive bare cores resulting from stars with high activity, planets with slightly smaller or similar core masses around stars with low or intermediate activity tracks can hold on to envelopes of 1-2\% of the total planet mass and populate the upper edge of the gap.

\subsubsection{The fuzziness of the 2D gap induced by stellar activity tracks}

\begin{figure*}
\centering
\begin{subfigure}{0.49\textwidth}
\includegraphics[width=\linewidth]{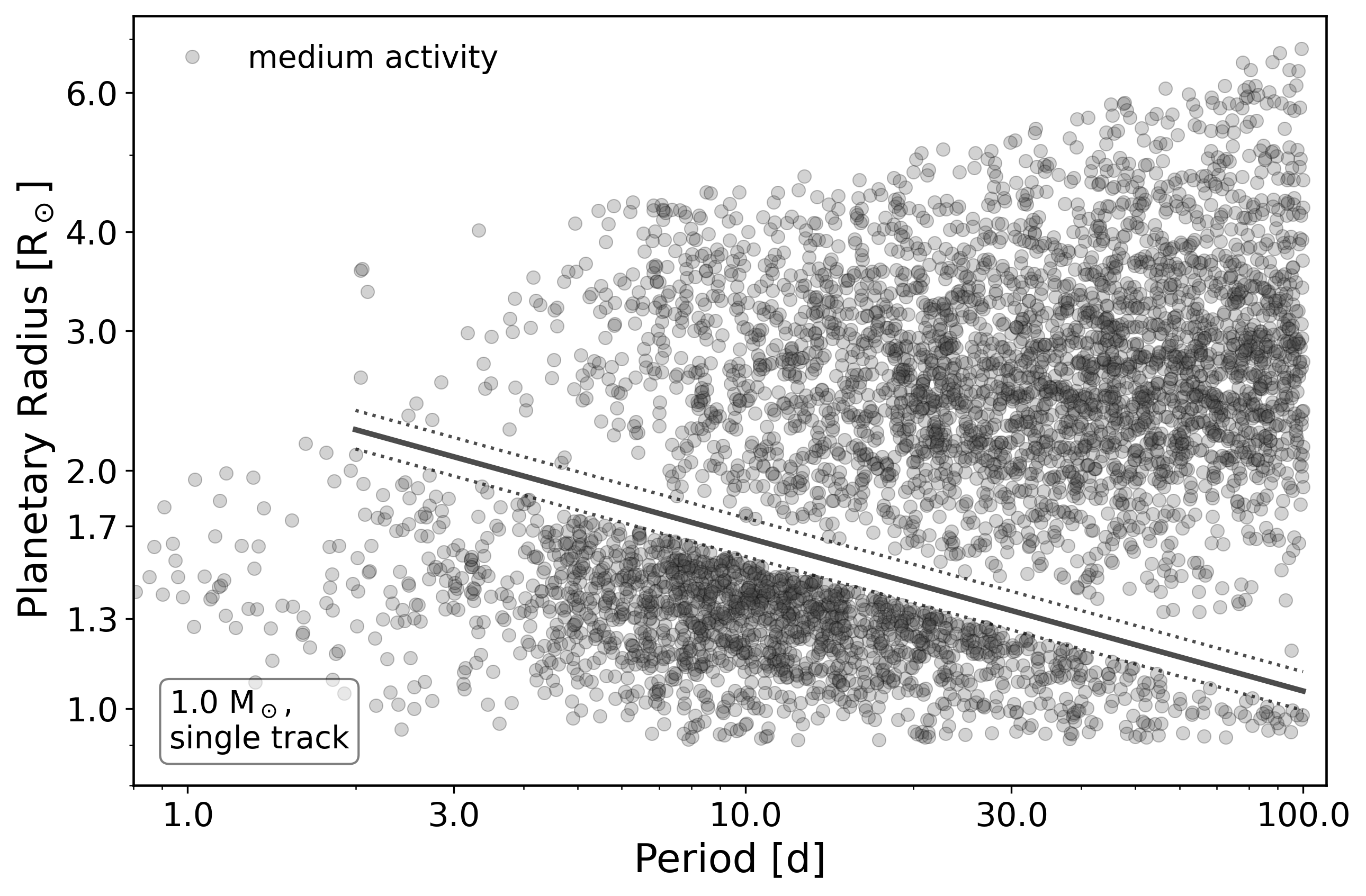}
\caption{}\label{fig:a}
\end{subfigure}\hspace*{\fill}
\begin{subfigure}{0.49\textwidth}
\includegraphics[width=\linewidth]{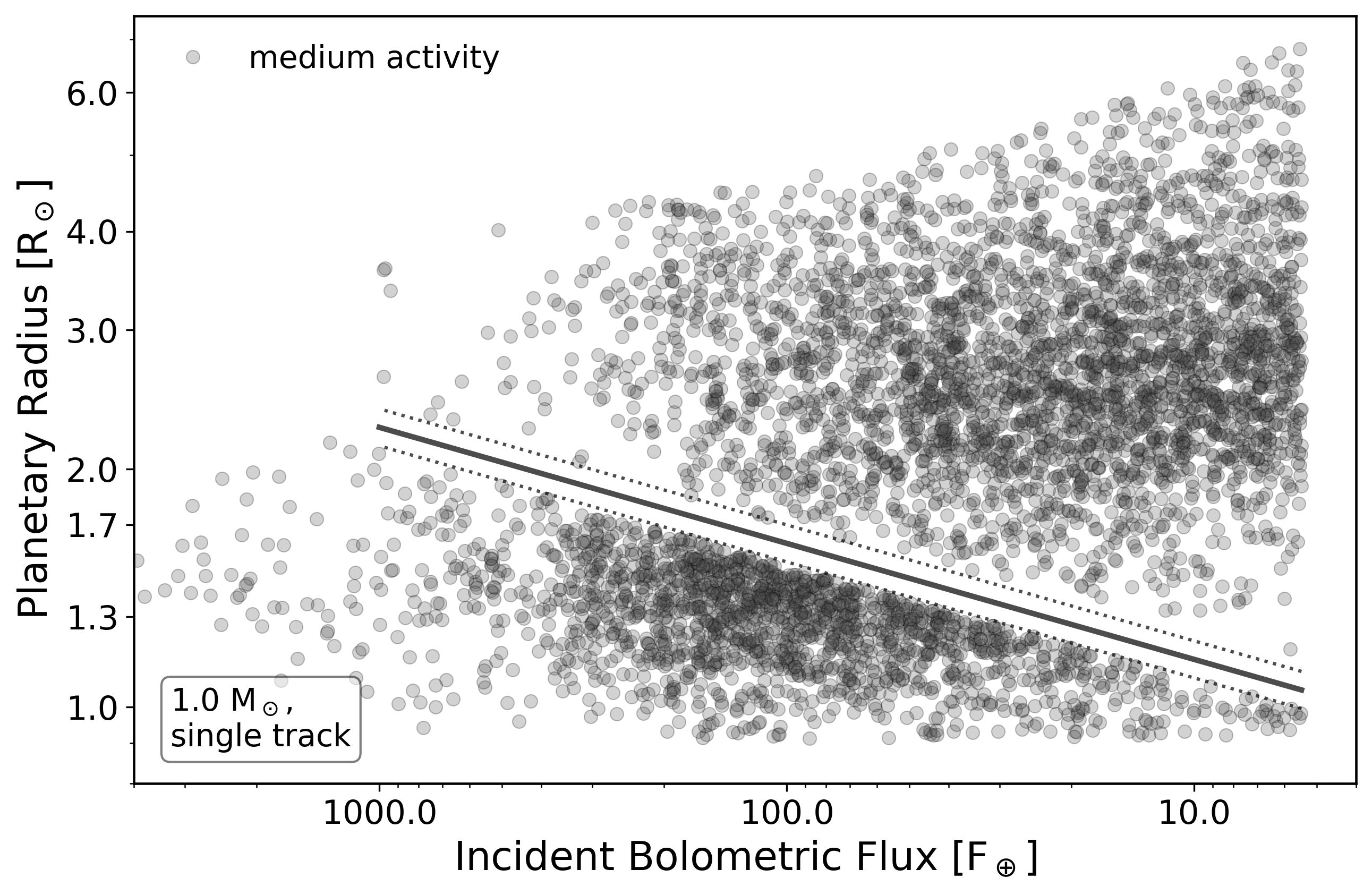}
\caption{} \label{fig:b}
\end{subfigure}

\medskip
\begin{subfigure}{0.49\textwidth}
\includegraphics[width=\linewidth]{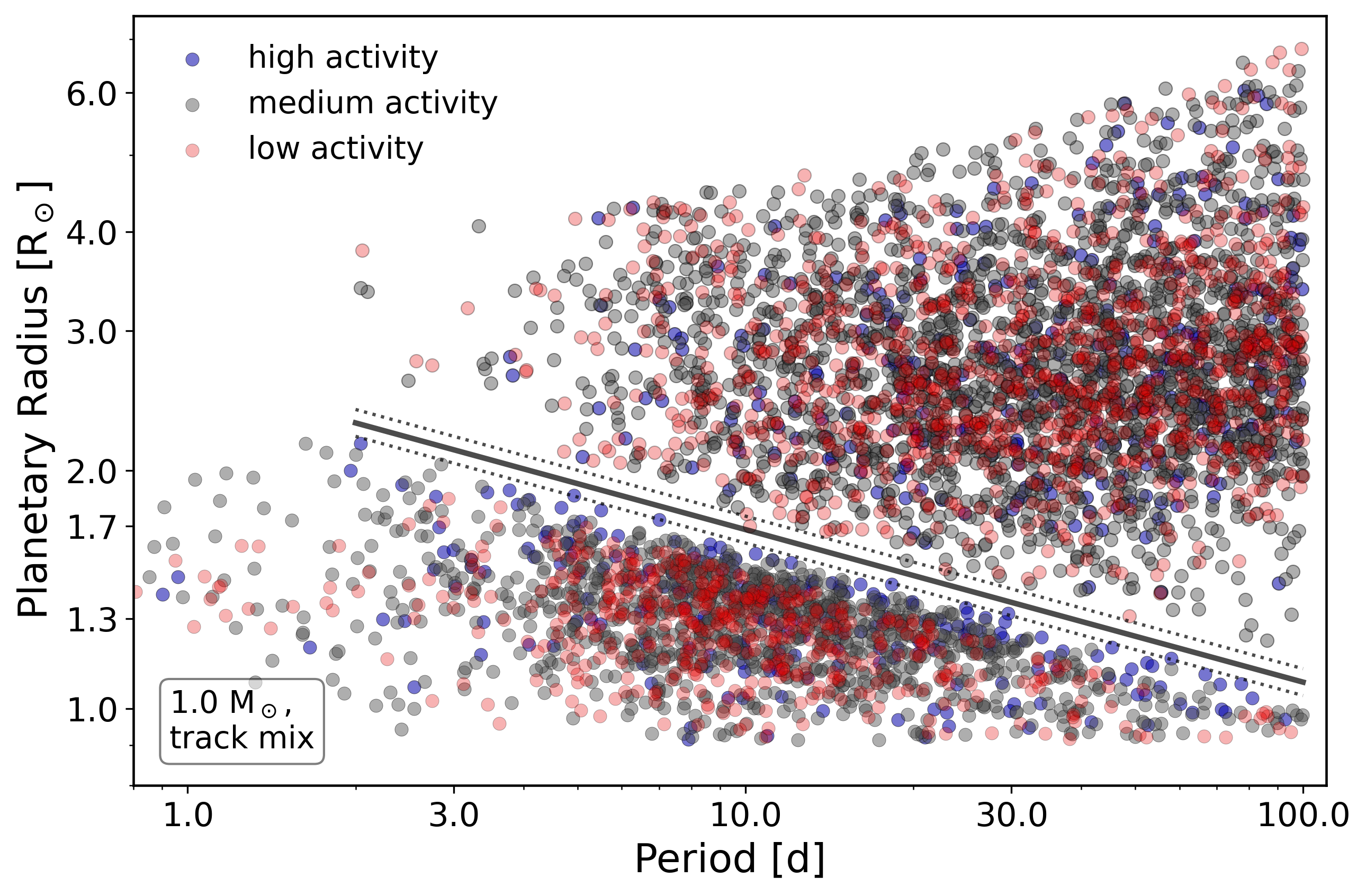}
\caption{} \label{fig:c}
\end{subfigure}\hspace*{\fill}
\begin{subfigure}{0.49\textwidth}
\includegraphics[width=\linewidth]{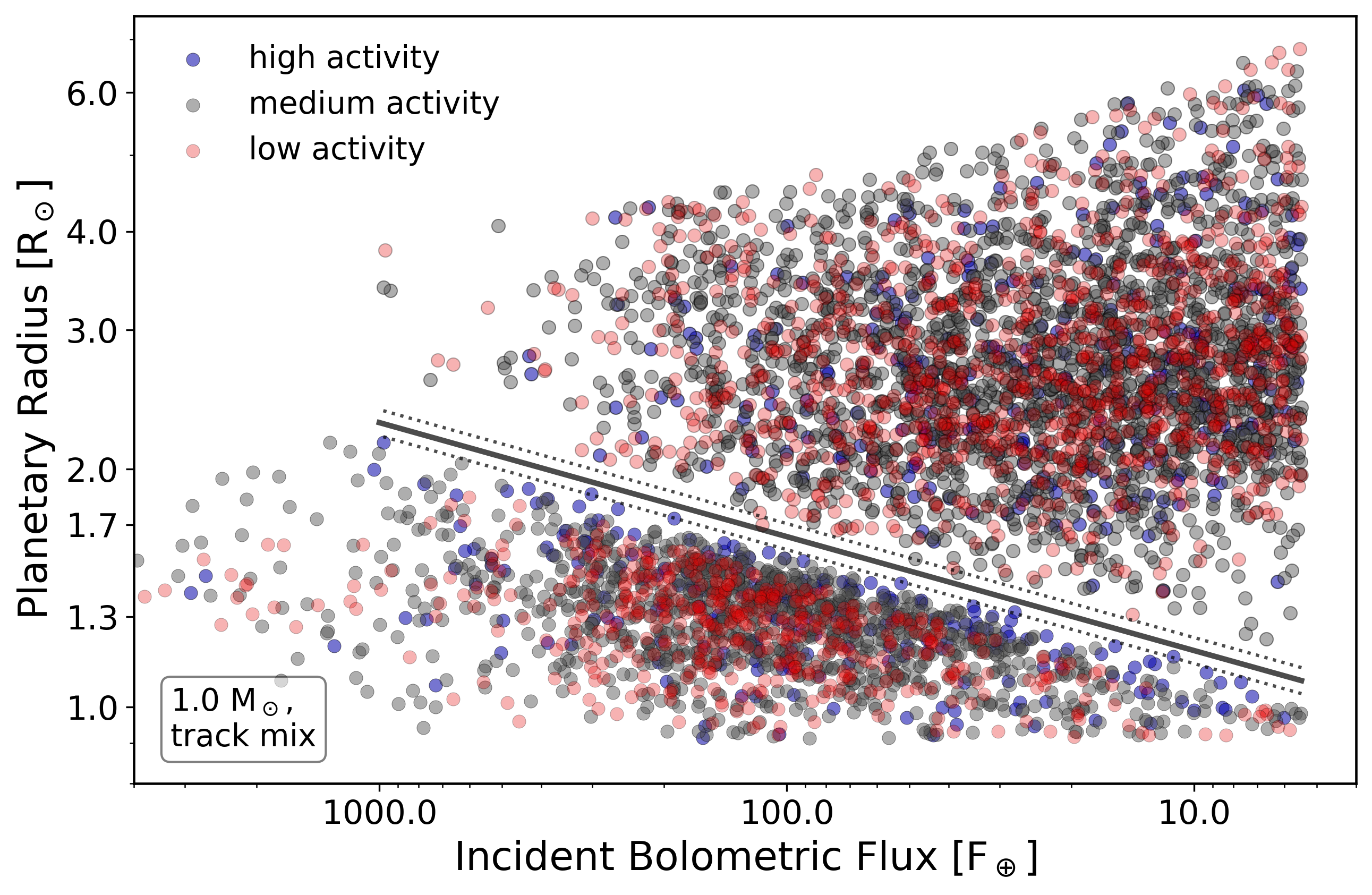}
\caption{} \label{fig:d}
\end{subfigure}

\medskip
\begin{subfigure}{0.49\textwidth}
\includegraphics[width=\linewidth]{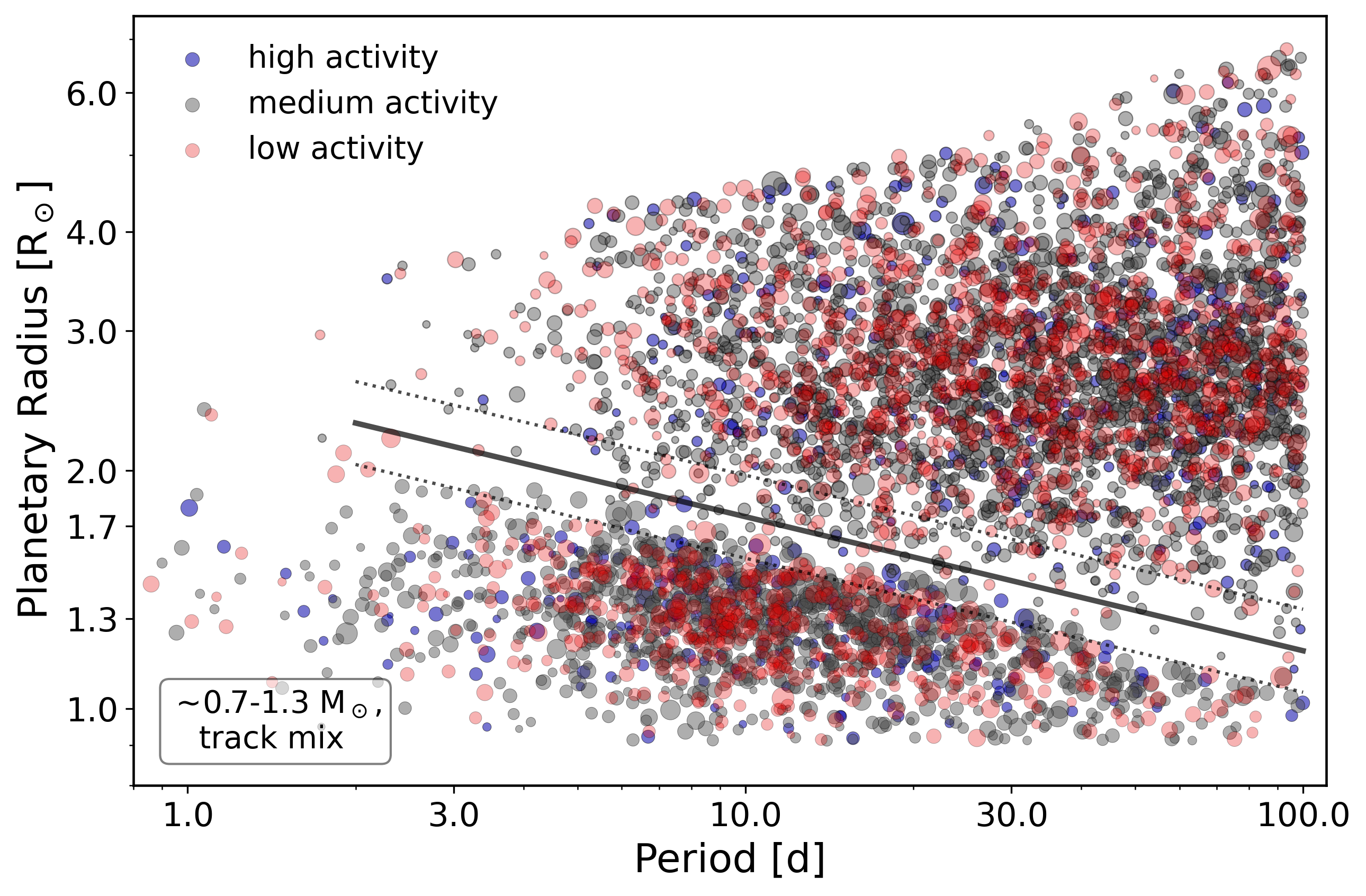}
\caption{} \label{fig:e}
\end{subfigure}\hspace*{\fill}
\begin{subfigure}{0.49\textwidth}
\includegraphics[width=\linewidth]{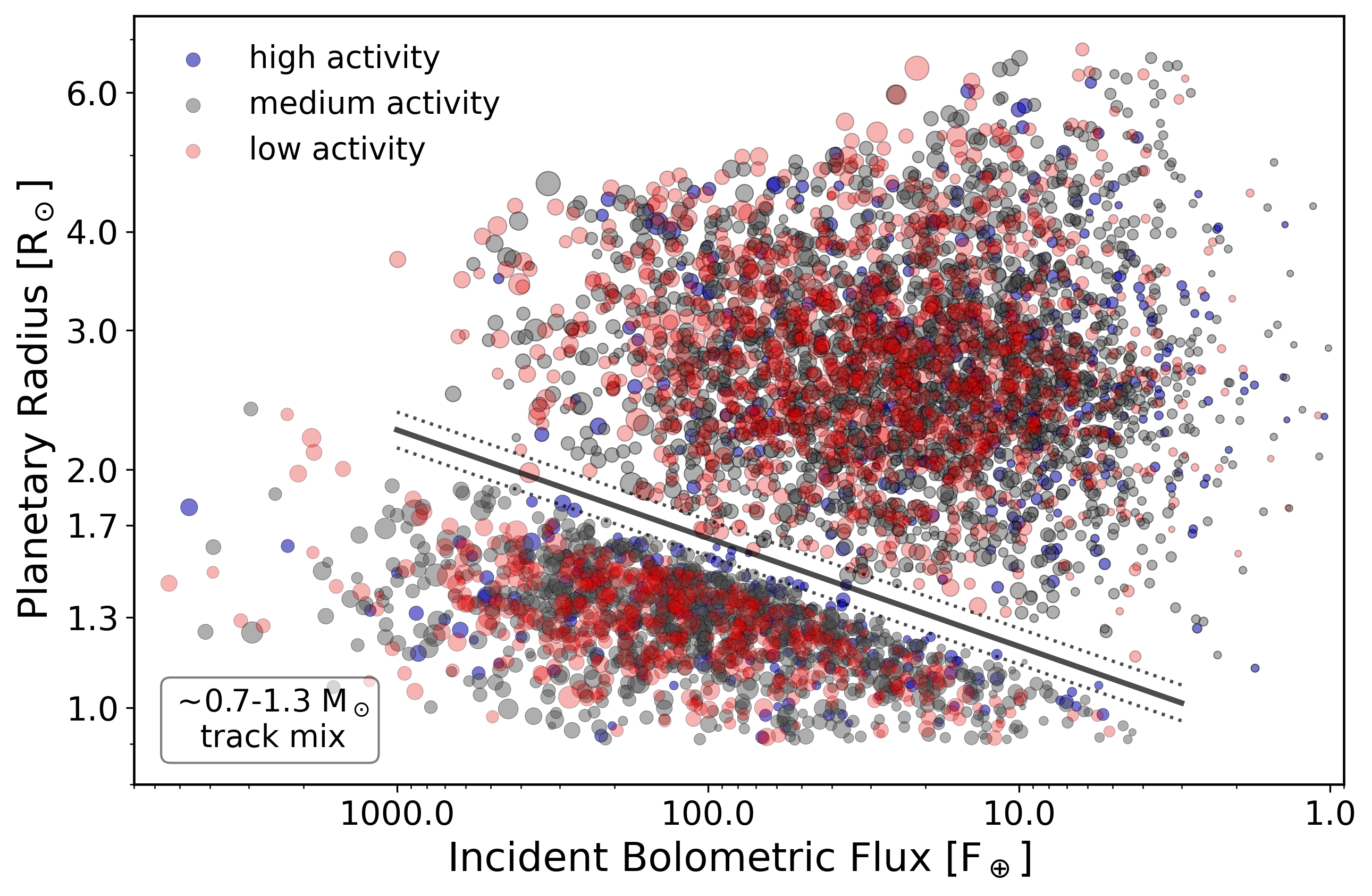}
\caption{} \label{fig:f}
\end{subfigure}

\raggedleft
\caption{Impact of stellar activity tracks on the exoplanet radius gap, for a single stellar mass and a distribution of K, G and F stars. In the left column we show planetary radius as a function of period, and on the right as a function of incident flux in Earth units. Going from panels (a) and (b) to (c) and (d) one can see the impact of a distribution of activity tracks on the simulated exoplanet population for one solar mass host stars. For a single stellar mass, the radius gap is still a relatively empty band, but the lower boarder is populated mainly by bare cores of planets around highly active stars. Going to a mixture of stellar masses ranging from K to F spectral types in panels (e) and (f) one can see that the gap is significantly less pronounced in period space, but stays relatively empty in flux space. In period, the lower boarder of the radius gap is mainly populated by planets around the more massive host stars in our sample, while in flux space the lower boarder is populated by planets around highly active stars. In flux space, stellar mass dependencies are reduced, which is why one can again see the impact of a distribution of stellar activity tracks on the gap edges, similar to the one solar mass case. Highly active stars emit large amounts of XUV flux longer compared to stars with low or intermediate activity, which causes planets of similar initial properties to end up as bare cores around highly active stars, and to retain thin envelopes around stars which drop out of the saturated regime earlier.} \label{fig:1}
\label{fig:2Dplots}
\end{figure*}

For a distribution of stellar masses ranging from K to F spectral type, going from an single activity track to a mix of tracks, our simulations show that the gap becomes wider and fuzzier in period space, but stays relatively empty in flux space. By fuzzier, we mean a gap that has a less sharp lower edge and a larger number of planets inside.
This is shown in Figure~\ref{fig:2Dplots}, panels (e) and (f), in comparison to a single stellar mass in panels (c) and (d). In period space, the gap is not completely empty anymore, even for a single activity track. This is due to the stellar mass dependence on the saturation XUV luminosity in our simulations. Planets around K dwarfs experience lower irradiation levels in the saturated phase compared to planets around F dwarfs, which impacts the critical core mass that can be fully evaporated and shifts the features of the bimodal radius distribution to lower radii for K dwarfs, and larger radii for F dwarfs. In period space, a planet at the same orbital period thus experiences different levels of evaporation around different host star spectral types. As a result, the lower boarder of the radius gap in period space is populated by planets around the more massive host stars in our sample ($\gtrsim$ 1 M$_\odot$), whereas the upper boarder with planets that retain thin envelopes is composed mostly of the G and K stars in our sample. In flux space, this stellar mass difference is effectively cancelled out and the gap appears cleaner. A stratification of planets with low, intermediate and high activity host stars is visible in the bare core population, with planets around highly active stars having slightly heavier evaporated cores and thus populate the lower edge of the radius gap. Comparing panels (d) and (f) in Figure~\ref{fig:2Dplots}, we find that in flux space the gap appears still quite empty, but with the inclusion of an activity track distribution a few massive planetary cores around highly active stars, or planets around low activity stars which can retain a thin envelope, fill the gap from below and above.

Regarding the planets in the vicinity of the gap, our simulations show that planets populating the lower border are stripped bare cores, or super-Earths, while planets inside and residing close to the top boarder are sub-Neptunes with thin envelopes. In the mixed stellar mass sample with the noisy gap in period space (panel (e) of Fig.~\ref{fig:2Dplots}), the planets inside the gap all have envelopes with less than 1\% in mass. At larger radii, just outside the gap, one enters a regime of relatively stable planets with envelope mass fractions of around 1-2\%. For the solar mass sample (panel (c) in  Fig.~\ref{fig:2Dplots}), the few non-bare planets inside the gap have very thin envelopes of less than 0.1\%. If we double our fitted gap width and investigate the planets inside a gap that has twice the width, the envelopes present extend to 0.5-1\%. The same behaviour is true for the solar-mass and mixed-mass samples when viewed in flux space (panels (d) and (f) in Fig.~\ref{fig:2Dplots}). The effect of stellar mass on the gap and its slope is further discussed in Section~\ref{subsec:smass_distro}

%%%%%%%%%%%%%%%%%%%%%%%%%%%%%%%%%%%%%%%%%%%%%%%%%%%%%%%%%%%%%%%%%%%
%%%%%%%%%%%%%%%%%%%%%%%%%%%%Discussion%%%%%%%%%%%%%%%%%%%%%%%%%%%%%
%%%%%%%%%%%%%%%%%%%%%%%%%%%%%%%%%%%%%%%%%%%%%%%%%%%%%%%%%%%%%%%%%%%
\section{Discussion}

We discuss now possibilities to compare structures seen in our simulated exoplanet populations to real populations that can actually be observed.
Observed exoplanet samples typically do not allow to infer the past activity history of the host star, with the possible exception of some multi-planet systems \citep[e.g.][]{2019Kubyshkina_a, 2019Kubyshkina_b, 2022Kubyshkina}. Therefore, observed samples will typically have a spread of stellar activity histories, whose effect we described in the results section. However, some other stellar parameters may be controlled for, namely stellar mass and stellar age.

%%%%%%%%%%%%%%%%%%%%%%%%%%%%%%%%%%%%%%%%%%%%%%%%%%%%%%%%%%%%%%%%%%%%%%%%%%%%%%%%%
%%%%%%%%%%%%%%%%%%%%%%%%%%%%%%%%%%%%%%%%%%%%%%%%%%%%%%%%%%%%%%%%%%%%%%%%%%%%%%%%%
\subsection{Influence of stellar mass}
\label{subsec:smass_distro}

The observed population of exoplanet host stars has a range of stellar masses. With a growing pool of known exoplanets, one can investigate the properties of the exoplanet radius gap in increasingly narrow bins of host star masses to explore the resulting relationship.

The stars in the detected Kepler exoplanet sample roughly follow a Gaussian distribution, which is peaked around one solar mass and covers an approximate mass range from 0.7 to 1.3 solar masses \citep{2017Petigura}, i.e.\ relatively few M dwarfs. We focus on F, G and K stars in this work, which all exhibit short time scales of activity saturation early in their lives and then follow various decreasing activity tracks, as described in Section~\ref{sec:activity_decay}. M dwarfs are observed and modelled to have much longer activity saturation timescales and possibly a narrower spread in activity tracks \citep[e.g.][]{2020Magaudda, 2021Johnstone}, but those stars are not at the focus of our present work here.

The main difference in terms of the exoplanet evolution comes from the different saturation X-ray luminosities of the young-age F, G and K stars, since their saturated phase is characterized by a constant level of $L_\mathrm{X}/L_\mathrm{bol}$, see Section~\ref{sec:activity_decay}; this translates to saturated X-ray luminosities of 0.2 and 3.5 times the X-ray luminosity of a $1\,\mathrm{M}_\odot$ star for stellar masses of $0.7\,\mathrm{M}_\odot$ and $1.3\,\mathrm{M}_\odot$, respectively. The XUV emission in the saturated phase crucially shapes the mass loss a planet undergoes early on, when the atmosphere might still be warm and expanded.

\begin{figure}
\includegraphics[width=0.45\textwidth]{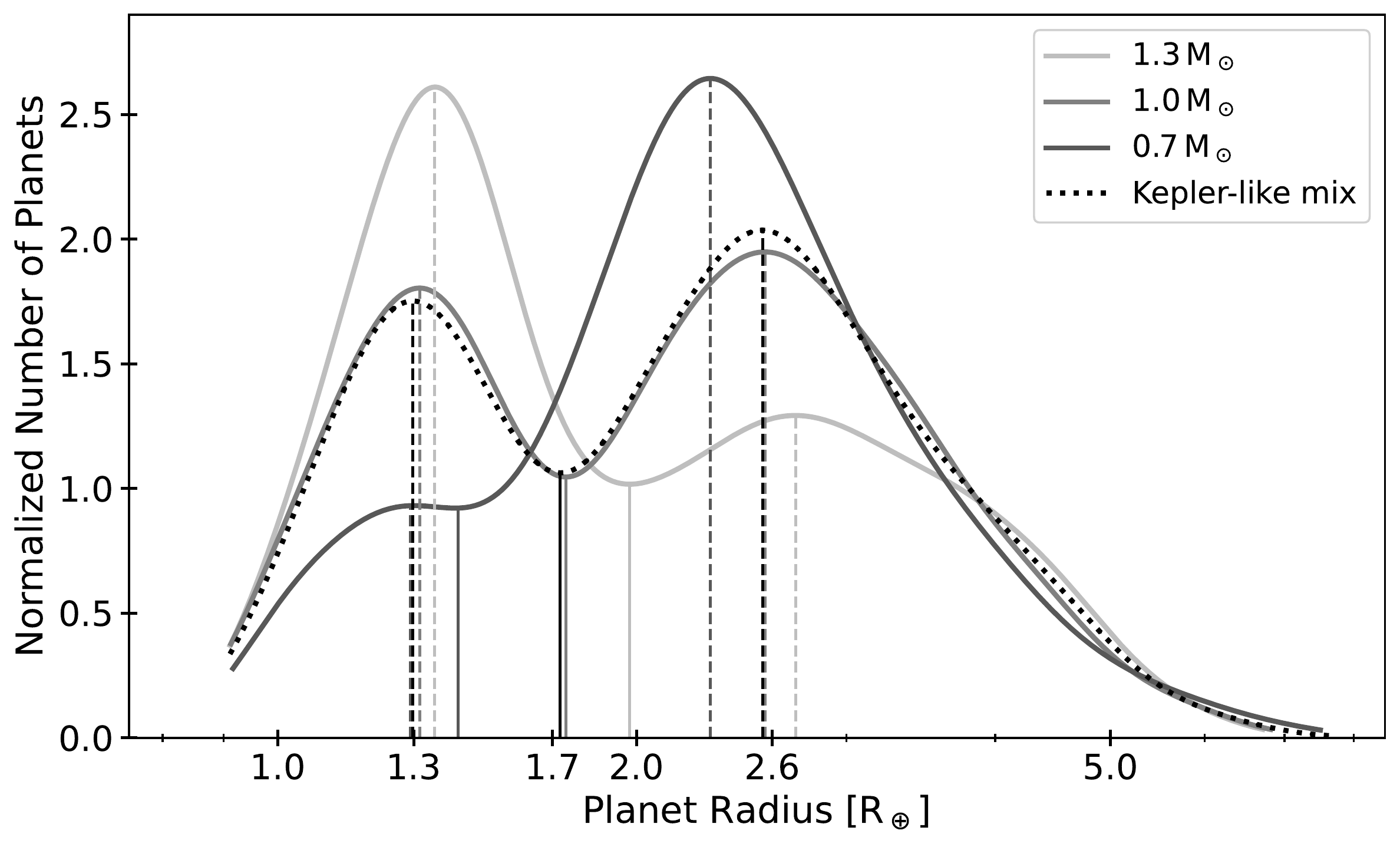}
\caption{Influence of stellar mass on the radius distribution. Comparison of the 1D-KDE estimation of the same exoplanet population but assuming that all planets orbit a star with mass 0.7 (light grey), 1.0 (medium grey) or 1.3 (dark grey) $M_\odot$. The radius distribution for a Gaussian stellar mass distribution peaked around $1.0\,M_\odot$ is shown as the dotted black line for comparison. In our simulation, the amount of evaporation around a $0.7\,M_\odot$ star is not large enough to populate the first radius peak, while for the $1.3\,M_\odot$ run, the second peak almost disappears due to the high mass loss. The gap locations extend from $\sim 1.4$ to $2.0\,R_\oplus$, which agrees with the observed gap extent as determined by \citet{2018Fulton}.}
\label{fig:07_10_13_track5}
\end{figure}

The result can be seen in Figure~\ref{fig:07_10_13_track5}, where we show our modelled radius distribution at 5 Gyrs for the same planet population around an $0.7\,\mathrm{M}_\odot$, $1.0\,\mathrm{M}_\odot$ and $1.3\,\mathrm{M}_\odot$ star, respectively. For the simulation with $0.7\,\mathrm{M}_\odot$ host stars (light grey), only $16\%$ of the planets lose their primordial atmosphere and as a consequence, the first peak is barely visible. For the $1.0\,\mathrm{M}_\odot$ simulation (medium grey) the bare core number increases to $35\%$ and the two peaks have comparable heights. For the highest host star mass of $1.3\,\mathrm{M}_\odot$ (dark grey), $51\%$ of planets in the sample fully evaporate, causing the second radius peak to almost disappear. The clearly different gap locations - $\sim\,1.4, 1.8, 2\,\mathrm{R}_\oplus$ for $0.7, 1.0, 1.3\,\mathrm{M}_\odot$, respectively - are governed by the distribution of core masses for the bare core planets and the planets with remaining envelope. On both sides of the gap, the location of the peak shifts to larger planet sizes for larger stellar masses: With higher XUV fluxes and thus larger amounts of evaporation, planets with more massive cores get stripped, leaving behind more massive bare cores below the gap. At the same time, planets with remaining envelope above the gap have more massive cores compared to the lower stellar mass simulations. So while the number of planets which can hold on to some envelope decreases, the location of the second peak shifts to larger radii as well.

If we introduce a stellar mass distribution like the one of the Kepler host stars, we find only small differences of the 1D radius distribution to the purely solar-mass case, since that sample happens to be centered on solar-mass stars. However, observationally, planet populations can be split into sub-samples by stellar mass, where the gap location could be observed to change similar to the light grey/medium grey/dark grey distributions in our Figure~\ref{fig:07_10_13_track5}.

In 2D, going from 0.7 to 1.3 M$_\odot$, the simulated planet populations also show a shift of the gap position upwards, as expected; this change is about 0.13 dex, which corresponds to a shift from 1.4 to 1.9 R$_\oplus$ at a period of 10 days. The gap slope is consistent across individual stellar masses, but turns out to be shallower for the mixed stellar population; we find a gap slope of typically -0.19(1) in period space for samples with a single stellar mass versus a slope of -0.16(1) for the mixed sample.  This is due to the fact that there is a larger number of planets below the gap in wider orbits for the F stars, which has a relatively high weight in the fit, since there are almost no bare-core planets in those orbits for the lower-mass stars. The slope of the gap in flux space is around 0.15(1) in our simulations, and in agreement for single stellar masses as well as a mixture of stellar masses.

We find that the best way to display the 2D planet population is in units of bolometric irradiation, not orbital period, consistent with e.g. \citet{2019Martinez, 2020Berger}. The fuzziness of the gap, i.e.\ how many planets populate the gap, is significant in period space, since the total accumulated XUV irradiation of a planet in a given orbit around an F, G or K star differs drastically depending on the host star, and therefore we observe an overlay of different gap positions in period space. However, in irradiation space, differences in the bolometric luminosity of the host stars are cancelled out and the difference in gap positions only comes from different $L_\mathrm{X}/L_\mathrm{bol}$ evolutions over time, which are much smaller than the differences in actual stellar X-ray (or XUV) luminosity. We therefore find a much emptier gap in our simulated planet populations in irradiation space, see Figure~\ref{fig:2Dplots}, panels (e) and (f).

The shifting of the radius gap to larger radii for larger stellar masses in the observed exoplanet population has been pointed out by several authors including \citet{2017Zeng, 2018Fulton, 2019Wu, 2020Berger}. One of the suggested reasons is a dependence of planetary mass on stellar mass, i.e. more massive host stars host more massive planets \citep{2018Fulton, 2019Wu, 2022Petigura}. In our simulations, planetary and host star mass are uncorrelated, and the shift in the radius gap is a result of the different XUV saturation luminosities. In period space, the lower edge of the radius gap is populated by the more massive host stars in our sample (see Figure~\ref{fig:2Dplots}, panel (e)), while in flux space the stellar mass-dependence is effectively scaled out. If such host star mass stratification in the upper boarder of the bare core population would be observed in period-space but not flux-space, this could indicate that the XUV dependence on stellar mass is shifting the radius gap rather than differences in the initial planet population around F, G, K stars.

%%%%%%%%%%%%%%%%%%%%%%%%%%%%%%%%%%%%%%%%%%%%%%%%%%%%%%%%%%%%%%%%%%%%%%%%%%%%%%%%%
\subsection{Influence of stellar age}
\label{subsec:age_distro}

The typical currently observed population of exoplanet host stars has a range of stellar ages. There is a bias towards older ages due to the easier detectability of exoplanets around quieter host stars, and the bulk of stars with exoplanets detected in era of the \textit{Kepler} mission has ages of around 3-7 Gyr \citep{2010Batalha}. Thanks to dedicated %observing (and analysis) 
campaigns, the number of young planets with ages less than 1-2 Gyr is gradually rising \citep[e.g.][]{2019Newton, David2019b, 2020Plavchan, 2022Mann}. In terms of exoplanet evolution, the age at which we observe a system can be decisive in the detection or non-detection of an atmosphere, in particular if the mass loss is not restricted to the first 100 Myr.

\begin{figure}
\includegraphics[width=0.45\textwidth]{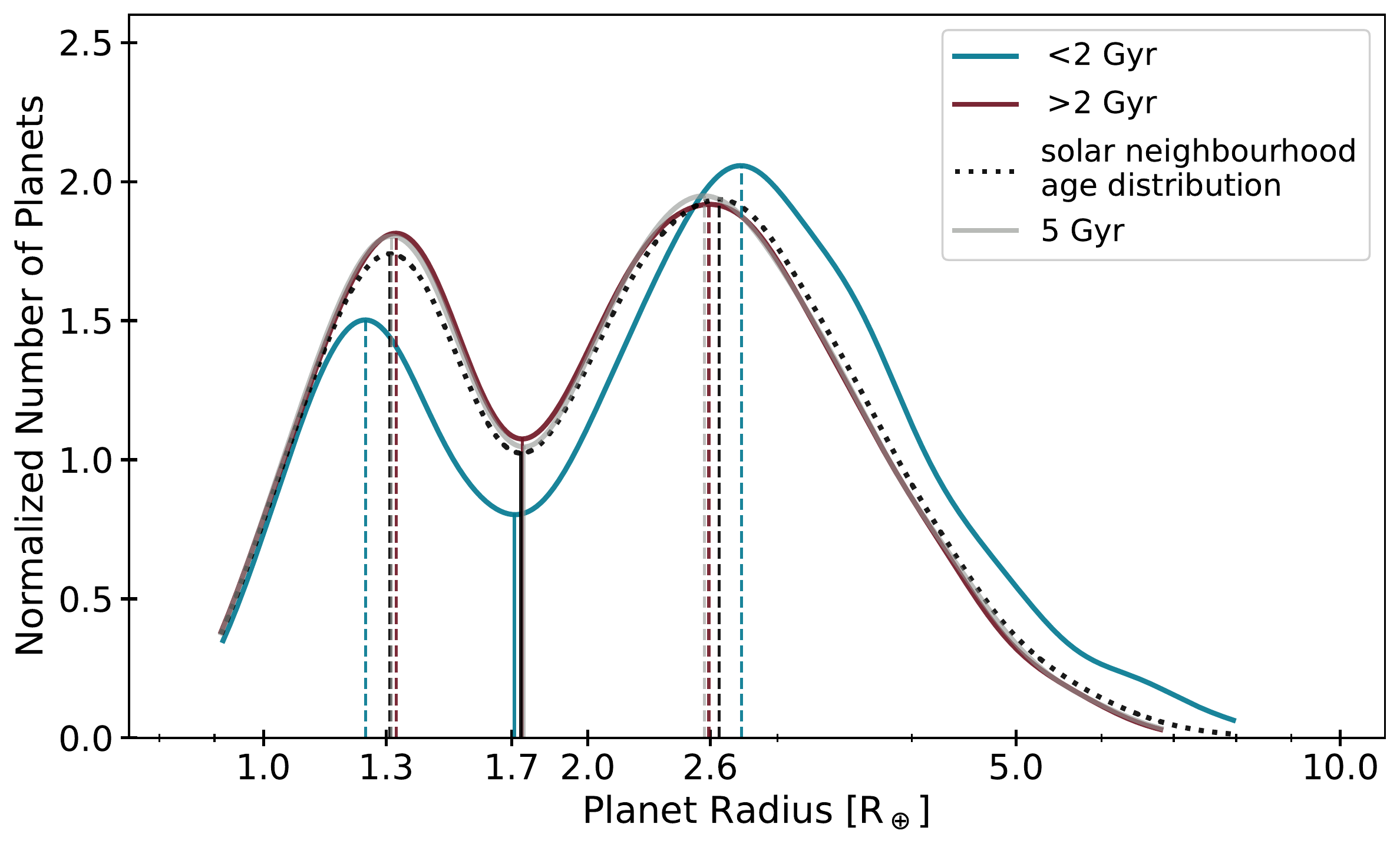}
\caption{Influence of a solar-neighborhood age spread on the radius distribution for one solar mass host stars. We compare the age-distributed sample in black (dotted line) to the single-age case at 5 Gyr (grey). The KDEs are qualitatively similar, but the small number of young planets in the age-distributed sample which still maintain an envelope cause the height of the first peak to be slightly smaller. These planets still reside above the gap in the age-distributed sample, but have ended up as bare cores below the gap in the 5 Gyr sample. We also show the age-distributed planet population, split up into two age bins: planets with ages less than 2 Gyr (blue) and planets with ages above 2 Gyr (red). The gap location is basically unchanged going from the younger to the older age bin, but the gap is much more filled in at later ages. The first radius peak is at smaller radii in the young sample, while the second peak is located at larger radii. At young ages, only planets with less massive cores have been fully evaporated, and planets with envelope  above the gap have had less time to cool and contract.}
\label{fig:age_evo}
\end{figure}

Photoevaporation population studies usually present the evolved radius distribution at one specific final age, ranging from $3$ to $10$ Gyr \citep[e.g.][]{2017Owen, 2017Lopez, 2018Jin, 2020Mordasini}. In Figure~\ref{fig:age_evo}, we show a population with an observationally-motivated age spread, similar to the one expected to be present in the solar neighborhood (see Sec.~\ref{subsec:population_details}), in comparison to the same sample at a single age of 5 Gyr. In the age-distributed sample, the bulk of planetary ages is around 3-7 Gyr, with a smaller number of younger and older planets. With the inclusion of a small fraction (20\%) of young planets (10 Myr to 2 Gyr), the radius gap is not influenced significantly compared to a single age (5 Gyr). A small difference in the peak height of the first peak and the location of the second peak are nonetheless visible. In the age-distributed sample, some of the young planets (less than 1.5\% of the whole sample) still retain an envelope and reside above the gap, while in the 5 Gyr sample, the same planets have completely lost their atmosphere and populate the first peak. These planets all sit close to the top border of the radius gap in the age-distributed population and have very thin atmospheres with envelope mass fractions less than 0.1-0.2\%. In the age-distributed sample, the opposite is true for a handful of close-in (<\,5 days) planets with ages above 5 Gyr (less than 0.5\% of the whole sample). They end up as bare cores in the age-distributed sample, but retain a thin envelope in the 5 Gyr sample. In general, about 1\% more planets reside above the gap in the age-distributed sample compared to the single age 5 Gyr population.

In the 2D view of the planet population, the slope and location of the radius gap are in agreement for a single stellar age sample of 5 Gyr and a solar neighborhood age composition. The width of the gap is also not significantly changed by the inclusion of an age distribution for a population which has evolved along a single intermediate activity track, and the gap remains almost as clean as in Fig.~\ref{fig:2Dplots} (a). For a mixture of tracks, we observe that the combination of young age and a low to intermediate activity track leads to a slightly larger number of planets to survive just above the gap with envelope mass factions less than 0.2\%. In the 5 Gyr sample, these planets will have lost their remaining atmosphere.
The lower border of the gap, which is populated by bare cores, is still sharp after the inclusion of an age-spread for a single acivity track, while this is no longer the case when we include a mixture of tracks. We therefore attribute the fuzziness of the lower border mostly to differences in the stellar activity history rather than an age spread.

Currently, most exoplanets have host stars whose ages are not known to high accuracy. This may change in the future through increased numbers of exoplanet detections in open clusters, where the member stars have very similar ages. In such a manner, one could investigate the effects of stellar age on the shape of the exoplanet radius distribution and the gap itself. This has recently been performed for the first time for a small number of age bins by e.g. \citet{2020Berger, 2021Sandoval, 2021David}.

Following these recent observational studies, we split our sample with the solar-neighborhood age distribution into two age bins; a young bin with planet ages less than 2 Gyr, and a second old bin with ages above. The corresponding KDEs are shown in Figure~\ref{fig:age_evo}. The results, as expected, show that when observing a sample of exoplanets with ages less than $1-2$ Gyr, the gap should be emptier and wider in the young sample, and more filled in a sample of older host stars. This is qualitatively similar to the observational findings by \citet{2021David}, who report that the radius valley appears much more filled in for planets older than $\sim 2-3$ Gyr.

\begin{figure}
\includegraphics[width=0.45\textwidth]{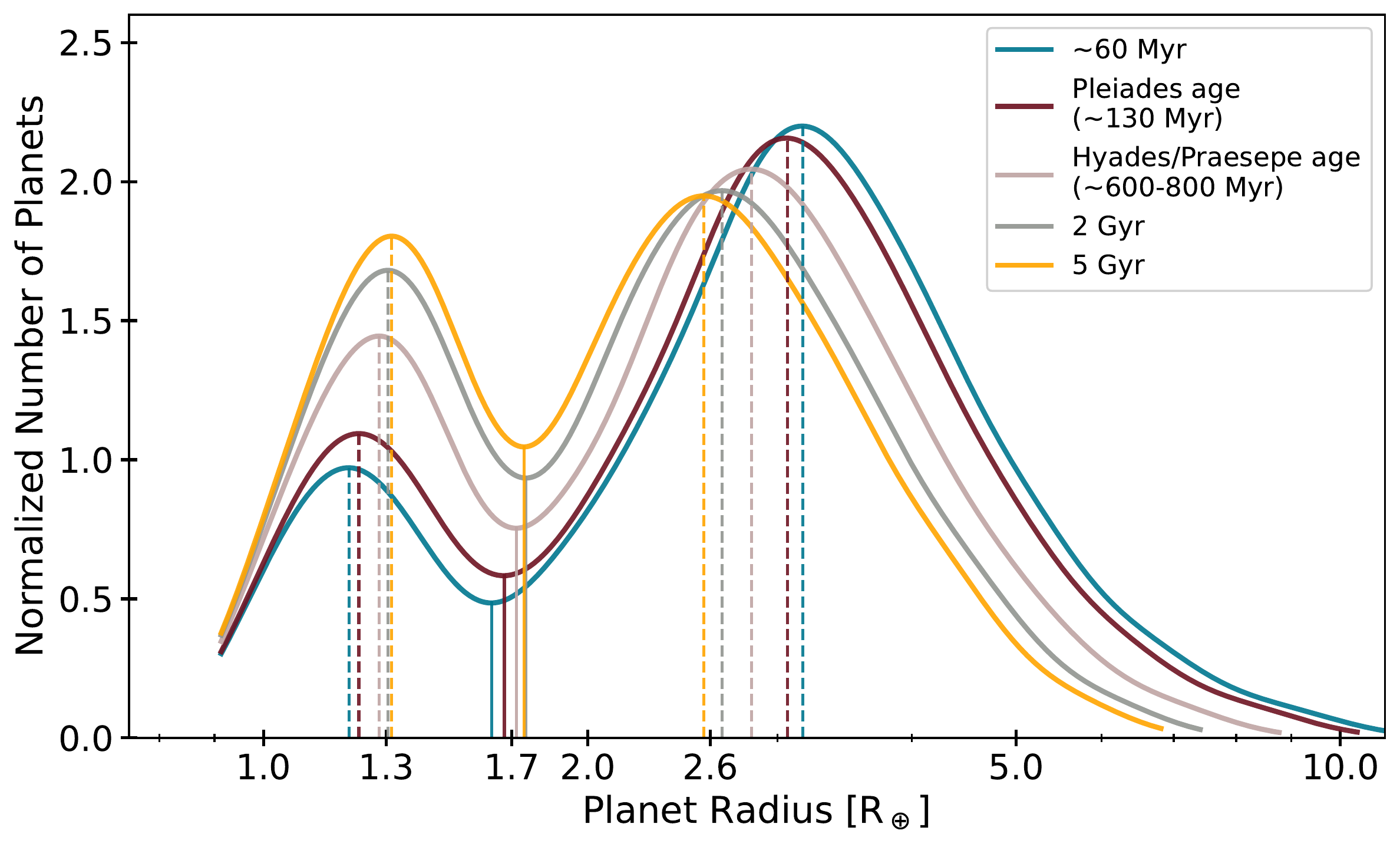}
\caption{Influence of age on the radius distribution for a single intermediate activity track. The extreme XUV exposure and high mass loss rates quickly strip the lightest, closer-in planets of their thin primordial atmospheres, giving rise to the bimodal structure early on. Over time, the first peak and gap minimum shift to larger radii due to the growing number and mass, or size, of the bare cores. The second peak drifts to smaller radii due to cooling and subsequent radius contraction. The height of the first peak increases while the second peak becomes smaller, and the gap region becomes narrower and progressively more filled.}
\label{fig:age_evo_cluster}
\end{figure}

While splitting the observed exoplanet sample into very thin age bins, similar to the ones presented in Figure~\ref{fig:age_evo_cluster}, is currently not feasible, the growing number of younger exoplanets has already allowed for a separation into a few coarse age bins. In the future, with more and more planet detections in young open clusters it might become possible to trace the radius distribution and gap evolution across narrow age bins. This will put important constraints on the strength of evaporation in the first several hundred million year. In Figure~\ref{fig:age_evo_cluster}, we show our modelled radius distribution for one solar mass and a single intermediate activity track at five different ages, motivated partly by known young clusters. Early on, when planets are exposed to extreme XUV fluxes and the mass loss is detrimental for close-in, low mass planets, the first peak and the radius gap shift visibly towards larger radii. This is explained by an increasing number of more massive bare cores. Going from ages 60, 130, $\sim$ 600 Myr, to 2 and 5 Gyr, the percentage of bare cores increases from 18, 21, 28, 33 to 35\%. So while the lightest close-in planets in our sample are stripped of their envelope within less than 100 Myr - giving rise to the bimodal structure early on - there still is significant mass loss afterwards. It is thus not just the first 100 Myr that play a role in sculpting the exoplanet population, but also the subsequent times up to Gyr ages. This is partly driven by the EUV contribution to the total XUV flux, with EUVs decaying less strongly than the X-rays, in agreement with the findings by \citet{2021King}.

After roughly 600-800 Myr, or the age of the Hyades or Praesepe, we find that while the location of the radius gap is mostly unaffected, we observe a gradual filling in of the gap up to Gyr ages. This is caused on the one hand by more massive/ larger bare cores gradually populating the lower boundary of the gap, and on the other hand by planets with core masses around 3-5 $\mathrm{M}_\oplus$ and envelopes of $1-2\%$ slowly losing mass and drifting towards the gap on Gyr time scales, ending up with either no envelope or very thin envelopes of less than $1\%$ in our simulation. Across all ages, the second peak shifts to smaller radii, which is driven mainly by planetary cooling and contraction, and only slightly by mass loss. As expected, and in agreement with \citet{2021David}, the gap is wider at younger ages and becomes narrower at later ages.

\begin{figure}
\includegraphics[width=0.45\textwidth]{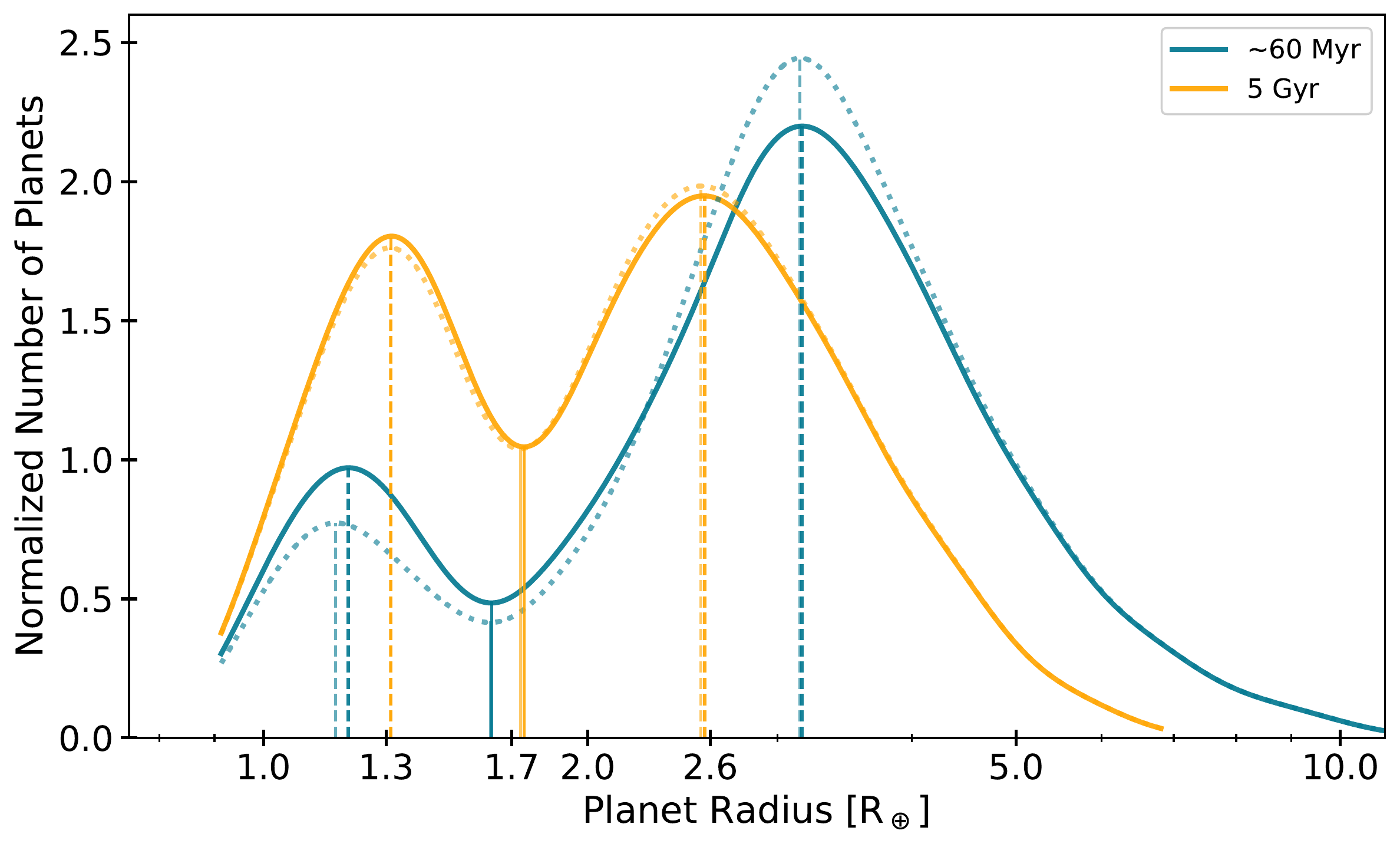}
\caption{Influence of a stellar activity spread on the 1D radius distribution at young and old age. The solid lines show the population for a single intermediate track with $t_{\mathrm{sat}} \approx 40$~Myr, and the dotted lines show the same population but with a distribution of low, intermediate and high activity tracks. At young age -- here we show $\sim 60$ Myr as an example -- there are fewer bare cores in the activity-distributed sample compared to the intermediate track. At later ages -- 5 Gyr shown here -- differences in the 1D distribution due to activity differences have been erased. Around low to intermediate activity stars, which drop out of the saturated regime before 40 Myr, the lower mass-loss rates stall the atmospheric stripping of their planets. However, the remaining XUV irradiation is still sufficient to eventually strip their remaining envelopes by 1-2 Gyr, erasing differences in the 1D distribution due to an acitivty spread at Gyr ages.}
\label{fig:age_trackmix}
\end{figure}

While we attribute the shifting of the gap to larger radii and gradual filling in over time first and foremost to age (and thus, the total amount of XUV exposure), our simulations show that the inclusion of an activity track, where some planets do drop out of the saturated regime very early on, while others stay saturated for prolonged times, slows down the growth of the first peak and filling in of the radius gap up to an age of about 1 Gyr.
In our one solar-mass simulation, a difference in the 1D radius distribution peak heights and gap depth/width between a single track with $t_{\mathrm{sat}} \approx 40$~Myr and a distribution of activity tracks is visible up to roughly 600-800 Myr. Going from ages 60, 130, $\sim$~600 Myr, to roughly 2 and 5 Gyr, the percentage of bare cores increases from 15, 18, 26, 32 to 34\% for an activity distribution, as opposed to 18, 21, 28, 33 to 35\% for a single intermediate track. Beyond about 1-2~Gyr, differences in the 1D distribution due to activity differences have been mostly erased; peak heights and radius gap depth are then roughly in agreement (see Fig.~\ref{fig:age_trackmix}). This behaviour is due to the overall larger number of tracks with intermediate and low activity (and saturation times less than $40$~Myr), than higher activity tracks in the sample with activity spread. A dropping out of the saturated regime early on alters the early phase of otherwise intense mass loss, allowing planets around such stars to hold on to their envelopes longer compared to planets around stars which stay in the saturated regime longer. However, due to the significant EUV contribution up to Gyr ages in our simulations, as time progresses, these planets still influence enough mass loss afterwards to eventually get stripped by an age of 5 Gyr. This erases major differences in the 1D distribution attributed to the spread in activity tracks. At older ages, only in 2D, e.g.\ the radius vs.\ period plane, the imprints of the activity spread remain, namely as the smoothed out, fuzzier borders of the gap -- a byproduct of planets with more massive cores undergoing complete envelope loss around stars with prolonged saturation levels, and thus larger cumulative XUV exposure. In summary, we find that the inclusion of an activity spread changes mainly the sharpness of the lower border of the gap, but the gap evolution over stellar age agrees with the trends found by works studying the observed planet populations (e.g.\ \citealt{2020Berger, 2021David}).

%%%%%%%%%%%%%%%%%%%%%%%%%%%%%%%%%%%%%%%%%%%%%%%%%%%%%%%%%%%%%%%%%%%%%%%%%%%%%%%%%
\subsection{Caveats and model limitations}
\label{subsec:limitations}

We briefly discuss some of the uncertainties introduced into our modelling from the underlying assumptions of the evaporation model and the input planet population. The main limitations of the energy-limited escape modelling, the role of magnetic fields and stellar wind interactions are given in \citet{2021Poppenhaeger} (Section 4.3). Some additional simplifying assumptions that we make here and that are explored in more realistic setups by other authors are: circular orbits and no planet migration after disk dispersal (see \citet{2021Attia} for eccentricity and migration effects on photoevaporative mass loss at later ages); lack of planetary compositional diversity (see \citet{2020Venturini} or \citet{2021Zeng} for the impact of compositional differences, like the inclusion of cosmic ices and water on the radius distribution); lack of a photon-limited escape regime \citep[e.g.][]{2016Owen, 2021Lampon}.

While we choose - if possible - observationally motivated stellar and planetary parameters for the initial exoplanet population, the exact details, like core mass distribution or post-formation envelope mass fraction, will influence the final predictions made by our simulations. Impacts of a range of simulation inputs on the evolved planet population or the radius gap have already been discussed in several works, including e.g. \citet{2020Modirrousta, 2020Mordasini, 2021Rogers_a, 2021Kubyshkina}. Further uncertainties arise from the estimation of X-ray and EUV fluxes in the saturated regime and beyond, due to the partly unknown high energy spectra of stars of different spectral type and age (see e.g. \citet{2021King}). The impact of the EUV estimation and the X-ray decay slope is further discussed in Appendix~\ref{sec:EUV_est}.

As we show in Appendix~\ref{sec:app_evaporation}, assuming energy-limited and radiation/recombination-limited mass loss yields quite different outcomes than assuming hydro-based mass loss. The mass loss predicted by the hydro-simulations, in particular in the first 100 Myr, is orders of magnitude greater compared to the energy-limited and/or radiation/recombination-limited simulations. This results in much more severe mass loss for planets across the whole range of orbital separations and core masses. Under our simulation assumptions, which includes mostly lower-mass cores below $\sim 10$ $\mathrm{M}_\oplus$, this leads to a large number of planets which fully evaporate, leaving behind almost no significant second radius peak in most simulations with hydro-based mass loss rates. These results suggest that for a planet population similar to the one used in this work, the predicted hydro-based mass loss rates under the assumed XUV irradiation levels are too high to reproduce the observationally known bimodal radius distribution. To preserve the second peak in the radius distribution, it would require significantly lower XUV fluxes or a population with a much larger number of cores above $\sim 10 \mathrm{M}_\oplus$ for there to be enough planets surviving with an envelope.

In general, our simulations yield a bimodal radius distribution, with a negative slope of the radius gap in radius-period space, qualitatively similar to observations. The exact values of the two peaks and the gap minimum, as well as the height of the peaks, depend sensitively on the choice of the input population, and the overall strength of the mass loss, which includes the evaporation scheme and its details, as well as the host star's XUV emission and its activity evolution. The main conclusion of this study, that the radius gap boarders become fuzzier and less pronounced by the inclusion of a stellar activity track distribution, hold true for a range of tested simulation inputs.

%%%%%%%%%%%%%%%%%%%%%%%%%%%%%%%%%%%%%%%%%%%%%%%%%%%
%%%%%%%%%%%%%%%%%%%% CONCLUSIONS %%%%%%%%%%%%%%%%%%
%%%%%%%%%%%%%%%%%%%%%%%%%%%%%%%%%%%%%%%%%%%%%%%%%%%

\section{Conclusions}

We investigate the effect of a distribution of host star activity evolutionary tracks on an observationally-motivated population of exoplanets shaped by photoevaporative mass loss. In our study, the host star activity track is associated with an X-ray saturation luminosity, a timescale for which the star stays at constant XUV flux at early ages, a set of XUV decay paths afterwards, as well as a method to approximate the contributing EUV radiation. The stellar activity track does have a mass dependence, with the more massive F stars being X-ray brighter, but having saturation times that are shorter than for G and K stars.  Within each spectral type, stars follow a distribution of low, intermediate and high stellar activity tracks.

Our qualitative comparison suggests that for the activity track distribution used in this work, the slope and location of the gap is not significantly changed compared to a single activity track for all host stars in the sample. However, compared to a single activity track for all host stars, the radius gap, which is a prediction of photoevaporative mass loss, does become fuzzier and less clean for a mixture of activity tracks, i.e.\ the lower edge of the gap is less sharp and the number of planets with radii inside the gap is larger. The inclusion of a small number of high and low activity tracks causes the gap to be narrower, because the lower gap boarder becomes filled in with more massive bare cores around stars with prolonged activity, while the upper gap becomes populated by planets around lower activity host stars which can hold on to thin envelopes of a few percent.

% outlook/future
A quantitative comparison of the exoplanetary radius gap with model predictions in principle can put constraints on the underlying core mass distribution, planetary composition, or the post-formation envelope mass fraction \citep[e.g][]{2021Rogers_a}. For individual multi-planet systems, even the rotational history of the host star can be deduced \citep[e.g.][]{2019Kubyshkina_a}. While this is a promising approach, it comes with a lot of challenges due to the large number of input parameters in the mass-loss modelling. Here we have explored the influence of the previously not well studied relaistic spread of stellar activity tracks. Going forward, there is also a large uncertainty in estimating the important EUV irradiation planets receive from their host stars. This calls for EUV observations of planet hosting stars from space, and more detailed observations of planetary mass loss are needed to put constraints on the strength of hydrodynamic escape or other mass loss processes, like core powered mass loss at ages below and above $\sim 1 \mathrm{Gyr}$. Models will benefit from decisive tests through observations, and will enable a better understanding of the fate of exoplanets over time.

%%%%%%%%%%%%%%%%%%%%%%%%%%%%%%%%%%%%%%%%%%%%%%%%%%
%%%%%%%%%%%%%%%% Acknowledgements %%%%%%%%%%%%%%%%
%%%%%%%%%%%%%%%%%%%%%%%%%%%%%%%%%%%%%%%%%%%%%%%%%%

\section*{Acknowledgements}

The authors thank the anonymous referee for providing insightful comments and suggestions on the paper. We also wish to thank the "2022 Summer School for Astrostatistics in Crete" for providing training on some of the statistical methods adopted in this work. This research made use of the Python packages \texttt{numpy} \citep{2020numpy}, \texttt{pandas} \citep{mckinney2010data}, \texttt{scipy} \citep{2020scipy}, and \texttt{matplotlib} \citep{hunter2007matplotlib}. Part of this work was supported by the German \emph{Leibniz-Gemeinschaft}, project number P67-2018.

%%%%%%%%%%%%%%%%%%%%%%%%%%%%%%%%%%%%%%%%%%%%%%%%%%
%%%%%%%%%%%%%% DATA AVAILABILITY %%%%%%%%%%%%%%%%%
%%%%%%%%%%%%%%%%%%%%%%%%%%%%%%%%%%%%%%%%%%%%%%%%%%

\section*{Data Availability}

This work is based on simulations with the publicly available code "Planetary Photoevaporation Simulator (PLATYPOS)" \citep{2022Ketzer}, which can be accessed on Github (\url{https://github.com/lketzer/platypos/}).


\bibliographystyle{mnras}
\bibliography{paper_bib} % if your bibtex file is called example.bib

%%%%%%%%%%%%%%%%%%%%%%%%%%%%%%%%%%%%%%%%%%%%%%%%%%
%%%%%%%%%%%%%%%%%%%%%%%%%%%%%%%%%%%%%%%%%%%%%%%%%%
%%%%%%%%%%%%%%%%% APPENDICES %%%%%%%%%%%%%%%%%%%%%
%%%%%%%%%%%%%%%%%%%%%%%%%%%%%%%%%%%%%%%%%%%%%%%%%%
%%%%%%%%%%%%%%%%%%%%%%%%%%%%%%%%%%%%%%%%%%%%%%%%%%

\appendix

\section*{Appendix}

We show here the effects of different simulation inputs and assumptions on an exoplanet population that has been shaped by photoevaporation. This includes the planetary structure model, the effective absorption radius, the evaporation model and efficiency, the primordial envelope mass estimation, the core mass distribution and the EUV estimation method. This section aims to visualize and give the reader a feel for the impact of the different simulation inputs on the 1D and 2D radius distribution. To better disentangle these different effects, we only show a one solar mass population at an age of 5 Gyr. Unless stated otherwise, we assume the energy- and radiation-recombination-limited mass loss model, the Lopez-$\beta$, an evaporation efficiency of 10\,\%, and only show the results for a single intermediate activtiy track ($t_{\mathrm{sat}} \approx$ 40 Myr). The details on the exoplanet populations can be reviewed in Section~\ref{subsec:population_details}.

%%%%%%%%%%%%%%%%%%%%%%%%%%%%%%%%%%%%%%%%%%%%%%%%%%
\section{Planetary structure model}
\label{sec:app_planet_model}

\begin{figure}
\includegraphics[width=0.45\textwidth]{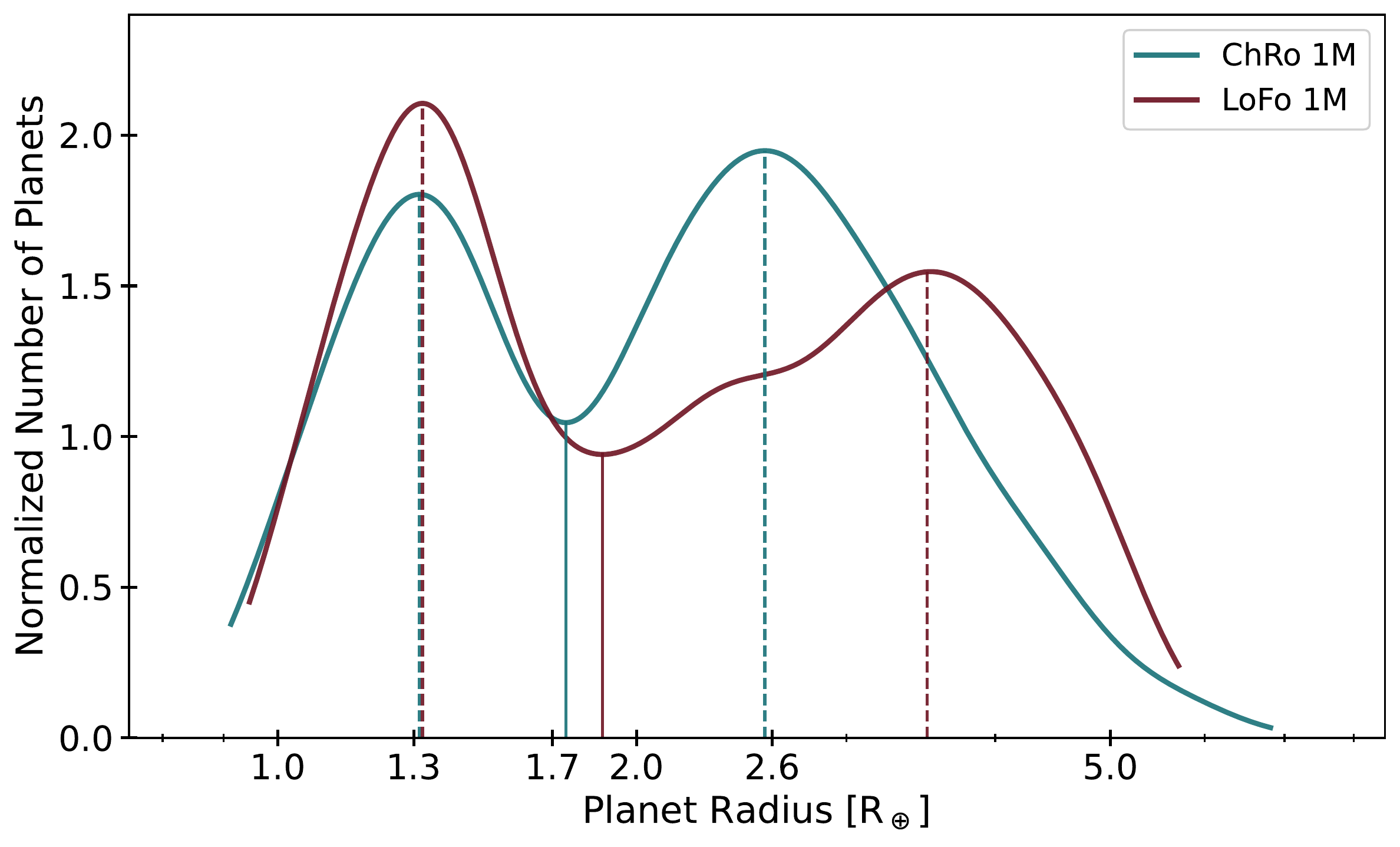}
\caption{Influence of the planetary model on the 1D radius distribution for one solar mass host stars. We compare the 1D KDEs for two comparable planet populations using the ChRo (red) and the LoFo (blue) planet models. The main difference can be seen in the second peak, which is broader and at much larger radii for the LoFo models.}
\label{fig:LoFo_ChRo}
\end{figure}

The exact details of the planetary structure model impact the final radius distribution. The most striking difference between the LoFo14 and the ChRo15 models is the location and width of the second peak in the 1D radius distribution. Using the LoFo models, the evolved population shows a relatively broad second peak that is located at larger radii compared to the ChRO models (see Figure~\ref{fig:LoFo_ChRo}). The difference arises from differing planetary radius dependencies on planetary mass, envelope mass fraction, stellar insolation and age between the models. This leads to slightly different initial radii at young age, before photoevaporation shapes the envelopes, and a different radius evolution due to cooling and contraction.

\begin{figure}
\includegraphics[trim={0.3cm 0.5cm 0cm 0.3cm},clip,width=0.48\textwidth]{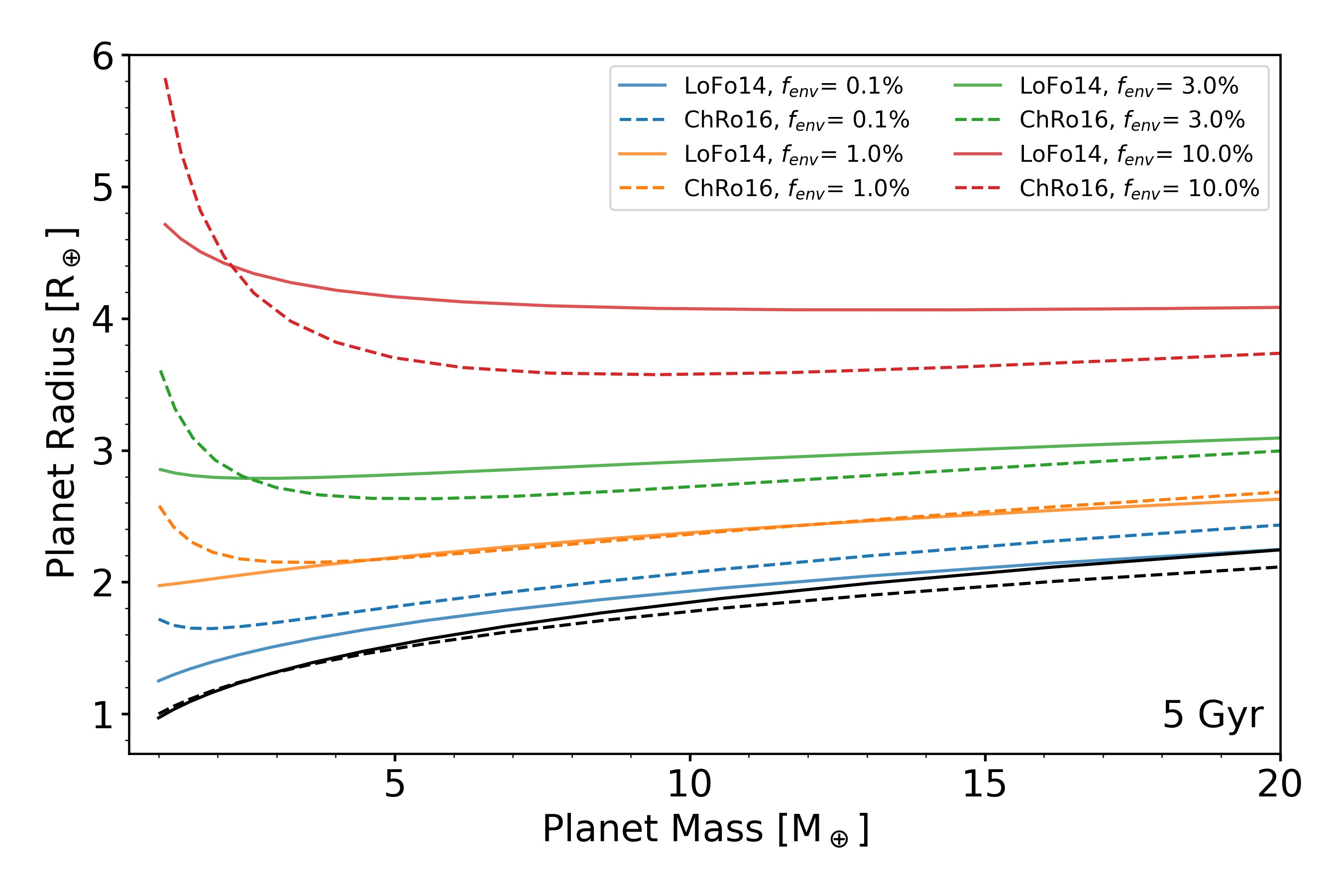}
\caption{Mass-radius relation for four different envelope mass fractions ($f_{env}=0.1, 1.0, 3.0, 10.0\,\%$) at 5 Gyr for LoFo14 (solid) and ChRo16 (dashed) planetary models. The black lines show the mass-radius relation for an Earth-like rocky core, as implemented in the LoFo14 and ChRo16 models. There are two notable differences between the models that affect the planetary mass loss calculations. The LoFo14 models predict larger radii for planets with envelope mass fractions greater than a few percent, while the ChRo16 models predict larger radii for thin envelopes less than 1\,\%.}
\label{fig:MR_rel}
\end{figure}

The planets under consideration in this paper ($M_\mathrm{core} \leq 25\,\mathrm{M}_{\oplus}$) show enhanced radii at early ages, with lower-mass planets showing the most extreme radius enhancement due to lower surface gravities, but similar internal energies compared to their more massive counterparts. As planets cool over time (with the puffy planets cooling the fastest), this radius difference for different planetary masses becomes less pronounced, resulting in a gradual flattening of the mass-radius curve.
In Figure~\ref{fig:MR_rel}, we show the mass-radius relation for the ChRo16 and LoFo14 models at 5 Gyr. It is evident that even at more mature ages, the predicted radii can vary due to slight differences in the modeling.

The LoFo models predict, in general, larger planetary radii at any given age for the planets in our sample with masses between $\sim 3-15\,\mathrm{M}_\oplus$, stellar insolation less than a few $100\,\mathrm{F}_\oplus$, and envelope masses greater than $\sim 2\,\%$. As a consequence, the predicted sizes for such planets are in most cases larger than the ChRo models. This gives rise to the second radius peak in the 1D distribution being shifted to larger radii (see Fig.~\ref{fig:LoFo_ChRo}). The observed radius distribution of close-in exoplanets indicates that the planets above the gap peak between 2.4-2.7 R$_\oplus$ \citep{Fulton2017, VanEylen2018b, 2019Martinez}. This is most closely matched by our ChRo simulations, which is why we only show the results using the ChRo models in the main body of the paper.

\begin{figure*}
\centering
\begin{subfigure}{0.49\textwidth}
\includegraphics[width=\linewidth]{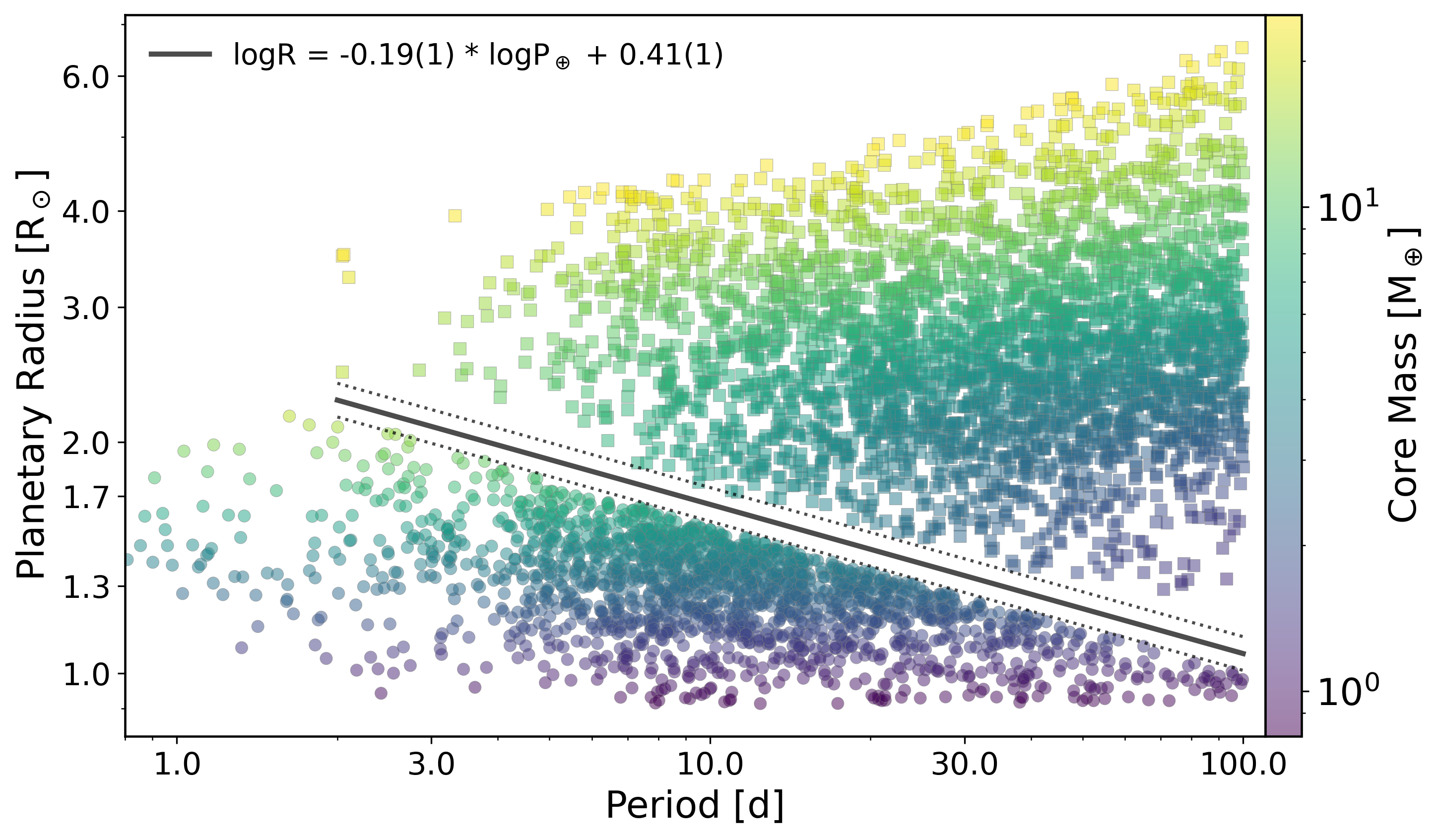}
\caption{}\label{fig:a}
\end{subfigure}\hspace*{\fill}
\begin{subfigure}{0.49\textwidth}
\includegraphics[width=\linewidth]{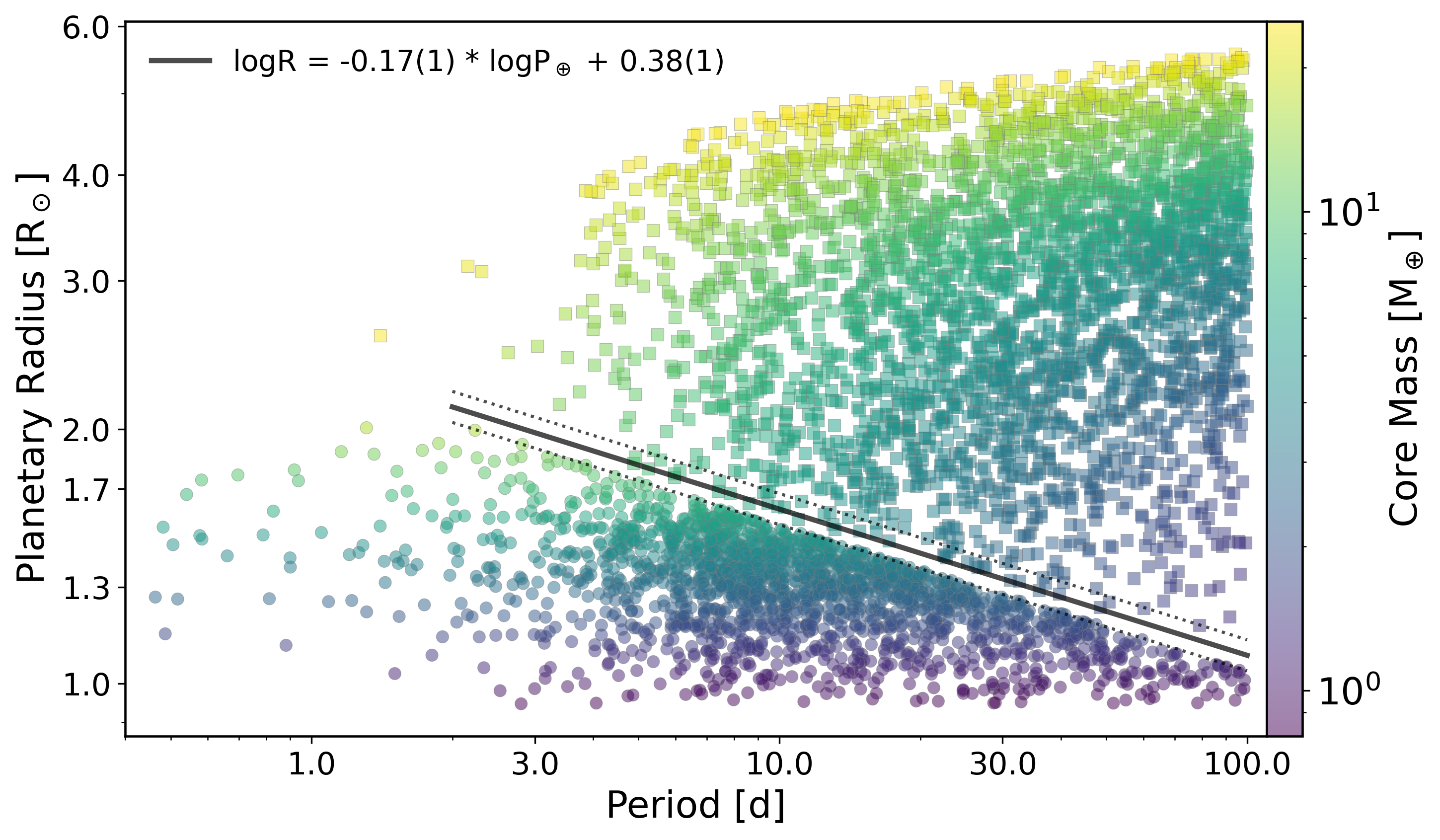}
\caption{} \label{fig:b}
\end{subfigure}
\raggedleft
\caption{Influence of the planetary model on the 2D exoplanet radius gap for one solar mass host stars. We show the radius vs. period distribution for the ChRo and LoFo population (panel (a) and (b), respectively), together with the gap fits. For the ChRo models, a thin, empty gap with negative slope is clearly visible. The LoFo models on the other hand produce a gap region that is much more filled in compared to the ChRo population. The slope of the gap is slightly shallower for the LoFo models compared to the ChRo models (dlog$R_\mathrm{p}$/dlog$P\approx-0.17\pm0.01$ vs. $\approx-0.19\pm0.01$).}
\label{fig:pl_models}
\end{figure*}

The location of the first radius peak, which is mainly set by the peak of the core mass distribution (see also Sec.~\ref{sec:app_Mcore}) and amount of evaporation, is similar for both planetary models and only differs in height. This is due to the LoFo simulation ending up with slightly more low mass bare cores, causing the first peak to be more pronounced. When we compare the slope of the gap fit in Figure~\ref{fig:pl_models} panels (a) and (b), we see that the simulation run with the LoFo planets produces a gap with a shallower slope. Compared to the ChRo models under the same simulation assumptions, more LoFo planets with slightly more massive and thus larger cores at large distances from their host star end up fully stripped, influencing the slope of the gap. Additionally, it is evident that the gap is more pronounced and the gap region much emptier for the ChRo models. For the LoFo models, the gap is not as well-defined, and the fit is mostly driven by the bare core boundary. The ChRo16 models predict larger radii for low mass planets ($\lesssim 3\,\mathrm{M}_\oplus$) and/or planets with very thin envelopes ($\lesssim 1-2\,\%$) (see Fig.~\ref{fig:MR_rel}), and as a consequence of this, many ChRo planets, once they reach an envelope mass small enough, still undergo sufficient mass loss to have this last bit of atmosphere removed. A comparable LoFo planet on the other hand might still be able to hold on to a very thin atmosphere ($\lesssim 1\,\%$) in our simulations, ending up with a radius "inside" the radius gap between $\sim 1.4-2.0\,\mathrm{R}_\oplus$. This causes the radius gap to be more filled in for the LoFo simulation, and the gap to be more spread out in the 1D radius histogram.

%%%%%%%%%%%%%%%%%%%%%%%%%%%%%%%%%%%%%%%%%%%%%%%%%%
\section{Effective absorption radius}
\label{sec:app_beta}

The effective absorption radius, or $\beta$-parameter, is another ill-constrained parameter in photoevaporation simulations, and thus relies on being estimated. In addition to the "Lopez"-$\beta$ calcualtion introduced in \ref{subsec:Lopezbeta}, PLATYPOS has a second estimation method for the important XUV absorption radius built-in. We call this the "Salz"-$\beta$ calculation.

Based on detailed upper planetary atmosphere simulations, \citet{Salz2016b} derived a scaling law for the planetary XUV radius as a function of planetary gravitational potential and their XUV irradiation. They find that super-Earth-sized planets with hydrogen atmospheres host extended atmospheres, which increases the effective surface area exposed to the high-energy stellar irradiation, and thus causes them to experience enhanced photoevaporative mass loss. In general, their findings indicate that the effective XUV absorption radius increases with decreasing gravitational potential, and to a lesser degree, with higher irradiation levels due to the resulting expansion of the thermosphere. We take Eq.\,4 from \citet{Salz2016b} to estimate $\beta$ for all planets in our sample at any point in time during the mass-loss calculation. We note that the aforementioned study only included planets with a gravitational potential of $\mathrm{log}_{10}(-\Phi_{\mathrm{G}}) > 12.0$. Since many of the lower-mass planets in our sample fall below the lower limit of the Salz-relation, we impose a cutoff by setting $\beta$ to the value predicted by the relation for the smallest and largest valid gravitational potential and XUV flux, respectively.

\begin{figure}
\includegraphics[width=0.45\textwidth]{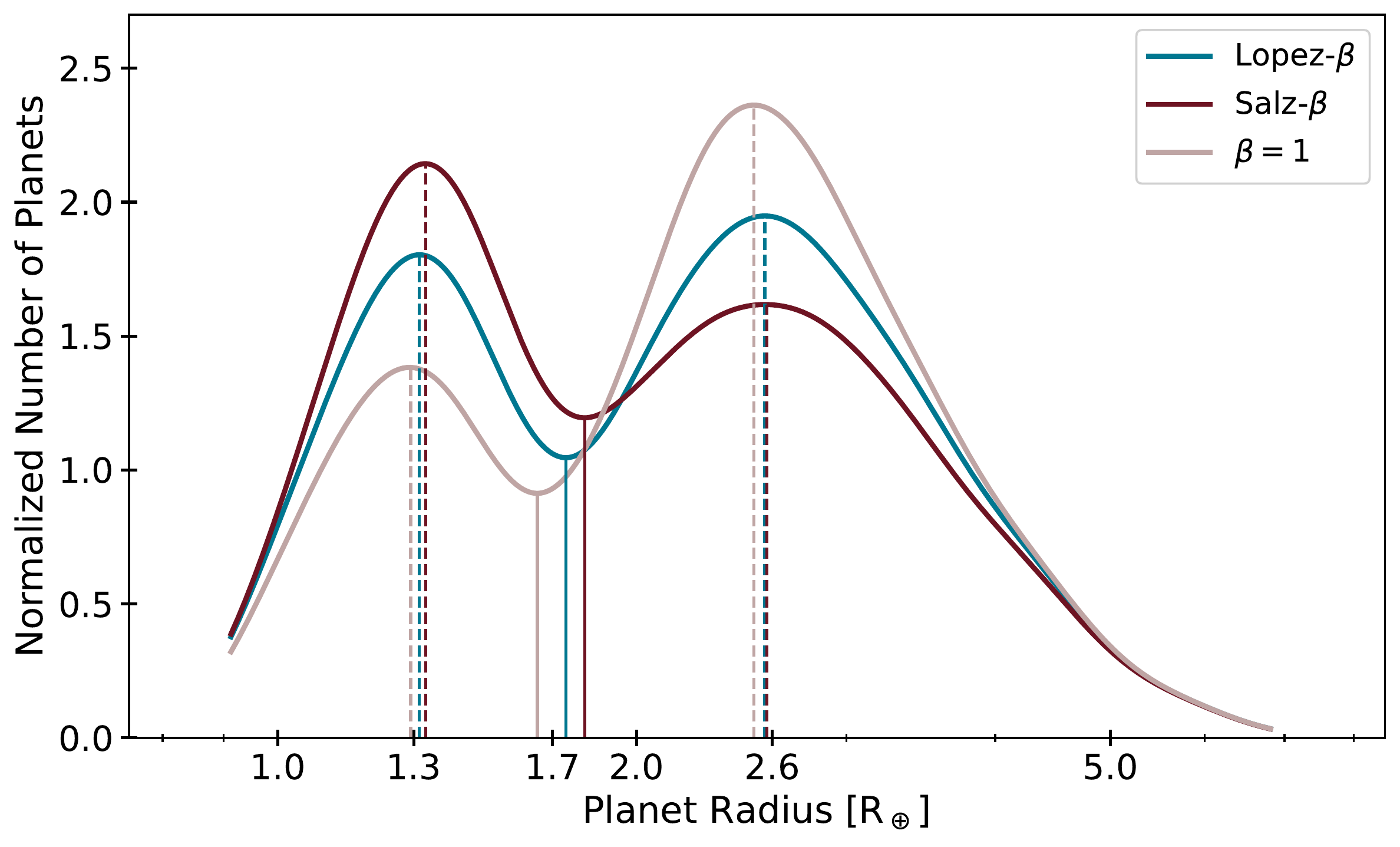}
\caption{Influence of the effective absorption cross-section, $\beta$, on the 1D radius distribution for one solar mass host stars. We show the 1D KDE estimation of the same planet population having estimated $\beta$ according to the "Salz"- or "Lopez"-method (blue and red, respectively). For comparison, we also show the corresponding run where $\beta$ is set to unity (rose). In this simulation, the Salz-$\beta$ values are on average 25\,\% larger than the Lopez ones, causing more planets to fully evaporate, fill in the first "bare core"-peak, and shift the gap to slightly larger radii.}
\label{fig:Salz_vs_Lopez}
\end{figure}

We show the effect of the effective absorption radius on the same planet population in Fig.~\ref{fig:Salz_vs_Lopez}. The blue and red 1D radius distributions show the evolved population having estimated $\beta$ according to the Lopez- and Salz-approximation described in Sec.~\ref{subsec:Lopezbeta} and \ref{subsec:Salzbeta}. For comparison, we also show the same run but with $\beta = 1$ (rose), which means that the effective XUV absorption radius is equal to the optical radius for any given age step in the simulation.

The most notable difference between the 1D radius distributions is the height of the two peaks. In general, a larger $\beta$, or larger effective absorption cross-section, leads to a greater amount of photoevaporation. This causes more planets in the sample to completely lose their atmosphere, decreasing the height of the second peak around 2.6 R$_\oplus$, and subsequently filling in the first peak around 1.3 R$_\oplus$. More massive evaporated cores explain the shift of the first peak and with it the gap to slightly larger radii. We tested this behavior for both the ChRo and LoFo planet models and found the behaviour to be qualitatively similar.

For the simulation run described here, the Salz-$\beta$ ranges from $\sim$ 1.5 to 2.8 at the start of the simulation (mean of around 2.1), and decreases to $\sim$ 1.5 to 2.3 (mean of around 1.7) at an age of 5 Gyr. The Lopez-$\beta$ covers a larger range of values at early ages, ranging from 1.1 to 4.5, but has a lower overall mean of 1.6 compared to the Salz-estimation. At 5 Gyr, the Lopez-$\beta$ is predicted to range from 1.1 to 2.4, with a mean around 1.3, again lower than the Salz-$\beta$. On average, the Salz-$\beta$ is about 25\% larger than the Lopez-$\beta$, leading to more mass loss and more evaporated bare cores in the Salz-simulation.

Since many of the lower-mass planets in our sample fall below the lower limit of the Salz-relation, we also tested the impact of imposing a cutoff to the Salz-beta to prevent a potential overestimation for low-gravity planets. Once, we allow for the extrapolation of the relation to smaller gravitational potential values, and once we impose a cutoff to the Salz-$\beta$ when a planet falls below the minimum range of applicability. In this case, $\beta$ is set to the value predicted by the relation for the smallest and largest valid gravitational potential and XUV flux, respectively. For the run under consideration, at early ages, where we find the largest number of low-gravity planets due to their puffy envelopes, the cutoff leads to a slightly lower mean of 1.9, compared to the beta's estimated without the cut. This difference is too small to change the the location of the gap and the result is almost indifferent to the Salz-run without a cutoff.

\begin{figure*}
\centering
\begin{subfigure}{0.49\textwidth}
\includegraphics[width=\linewidth]{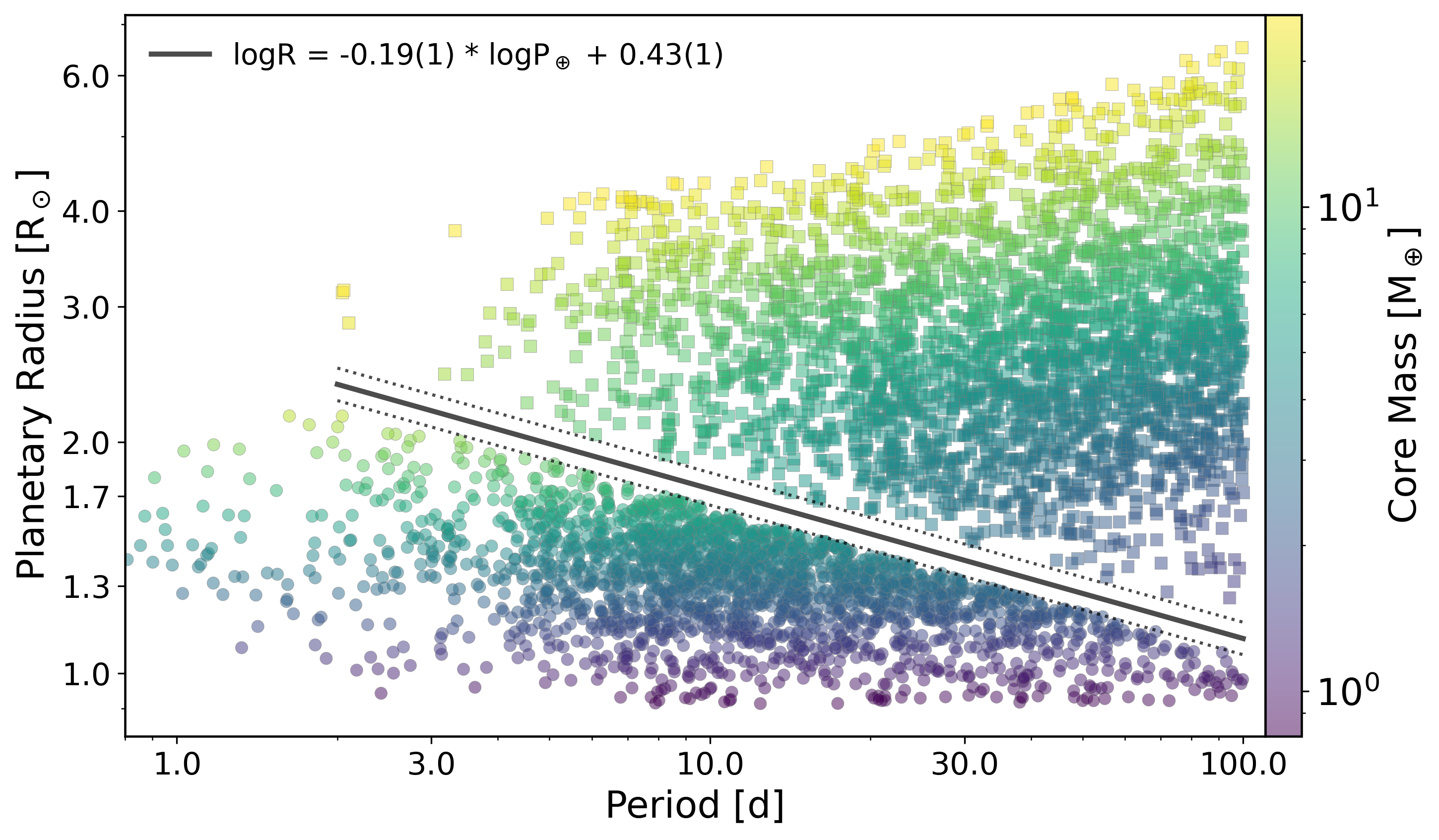}
\caption{}\label{fig:a}
\end{subfigure}\hspace*{\fill}
\begin{subfigure}{0.49\textwidth}
\includegraphics[width=\linewidth]{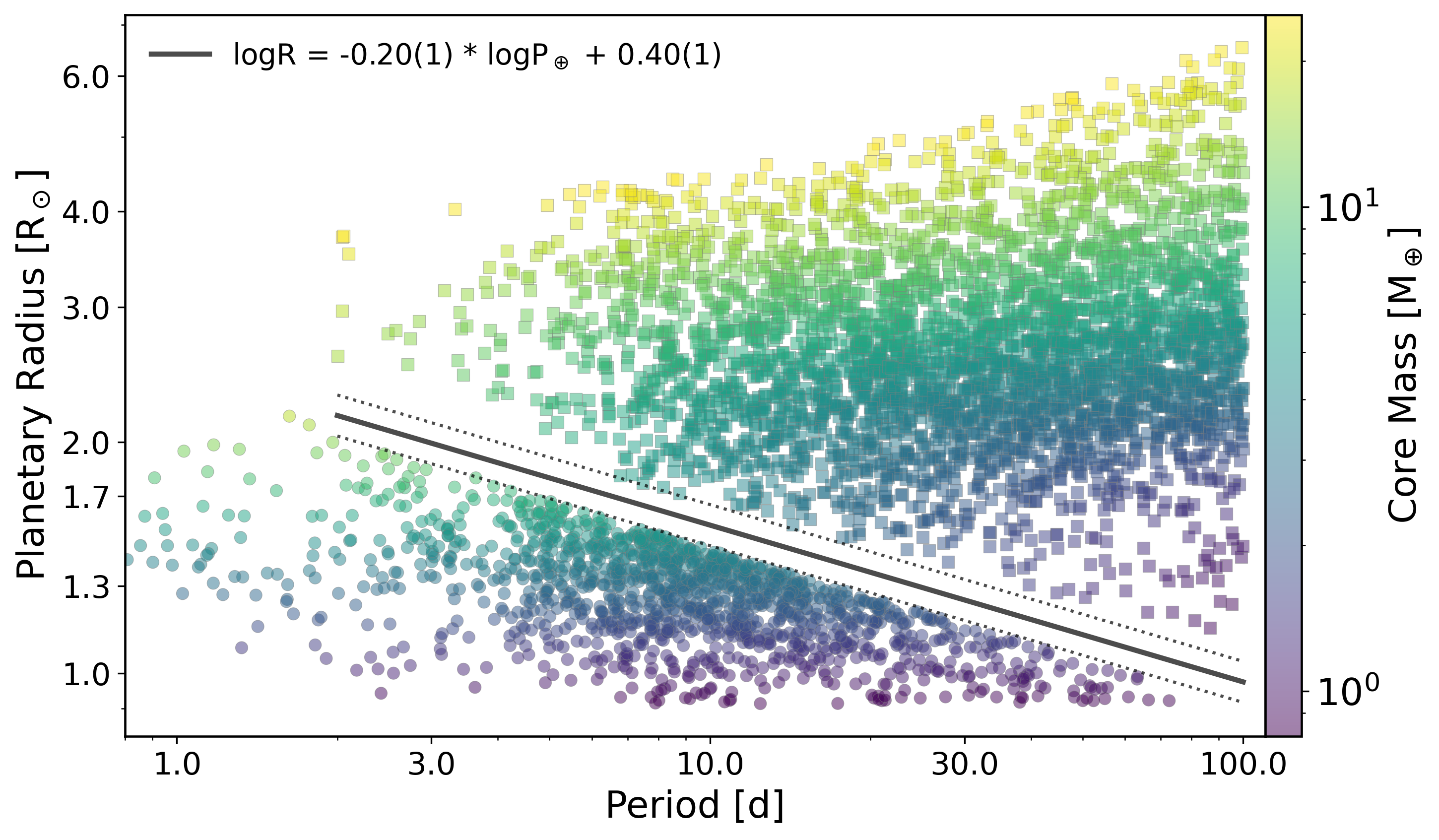}
\caption{} \label{fig:b}
\end{subfigure}
\raggedleft
\caption{Influence of the effective absorption cross-section on the 2D exoplanet radius gap for one solar mass host stars. We show the radius vs. period distribution for the ChRo population with the Salz-$\beta$ and $\beta = 1$ in panels (a) and (b), respectively, together with our fit of the gap. These plots can be compared with the Lopez-$\beta$ run in panel (a) of Fig.~\ref{fig:pl_models}. When $\beta$ is set to unity, the gap is lower compared to both the Salz- and the Lopez-$\beta$ runs. The mass-loss rates at any given age are lower, which means fewer planets end up rocky and less massive cores can be stripped. The gap appears to be a bit wider with less evaporation.}
\label{fig:ChRo_betas}
\end{figure*}

When investigating differences in two dimensions, we find that the slope of all three runs is consistent within one sigma. Since the Salz-$\beta$ is on average larger, more massive planets can be fully stripped, ending up below the gap. This shifts the gap slightly upwards to larger radii compared to the run with the Lopez-$\beta$. As expected, in the $\beta=1$ run fewer planets and planets with lower core masses turn into rocky cores, causing the gap to be shifted downwards to smaller radii. The radius vs. period plots are shown in Fig.~\ref{fig:pl_models} panel (a), and Fig.~\ref{fig:ChRo_betas} panels (a) and (b).

While the choice of $\beta$ does not significantly change the properties of the gap itself, it strongly affects the number of planets propagating downwards across the gap over time and is therefore an important parameter in the atmospheric evolution of planets, as shown in Fig.~\ref{fig:Salz_vs_Lopez}.

%%%%%%%%%%%%%%%%%%%%%%%%%%%%%%%%%%%%%%%%%%%%%%%%%%
\section{Evaporation model and efficiency}
\label{sec:app_evaporation}

The calculation of the mass loss rates is one of the crucial assumptions in photoevaporation studies. First, we compare energy-limited mass loss (short: E) only, with a combination of energy- and radiation\slash recombination-limited mass loss (short: ER). At early ages, when the XUV flux is highest, the radiation\slash recombination-limited mass loss rates apply to close in and\slash or lower-mass planets in the population, when calculated according to Sec.~\ref{subsec:RRlim}. For the planets in our simulation this, however, has no significant effect on their final radius, as almost all planets with recombination-limited mass loss end up as bare cores. For the one-solar-mass run with the intermediate stellar activity track, less than 4\% of all planets in the sample, which retain an envelope under the Lopez-beta assumption, have final radii which differ by less than 1\%. For the Salz-beta the same is true for about 7\% of the planets. All the planets which are affected by the recombination-limited mass loss rates have remaining envelopes of $\lesssim 1\%$ and reside just above the gap. Only in the Salz-case a handful of planets cannot retain a thin envelope and instead end up as bare rocky cores in the case of energy-limited mass loss only. Due to the negligible difference on the planets in our sample, we chose to show the simulation runs with a combination of energy-limited and radiation\slash recombination-limited mass loss rates only.

The third mass loss rate estimation that is implemented in PLATYPOS are the hydro-based mass loss rates (short: HBA) by \citet{Kubyshkina2018a}. Planets are predicted to be particularly susceptible to mass loss at early ages due to their warm, puffy, and low-gravity atmospheres. For comparison, for the one solar mass population, the HBA mass-loss rates at 10 Myr are on average a factor 15 higher than for the simulation using the ER mass loss rates and the Lopez- or Salz-beta.
This early phase of intense mass loss leads to more planets losing their envelope completely and ending up as bare cores in the HBA run. At Gyr ages, all three mass-loss rates are comparable within a factor 2, and the Salz  and HBA rates are almost identical across the planetary sample.

\begin{figure}
\includegraphics[width=0.45\textwidth]{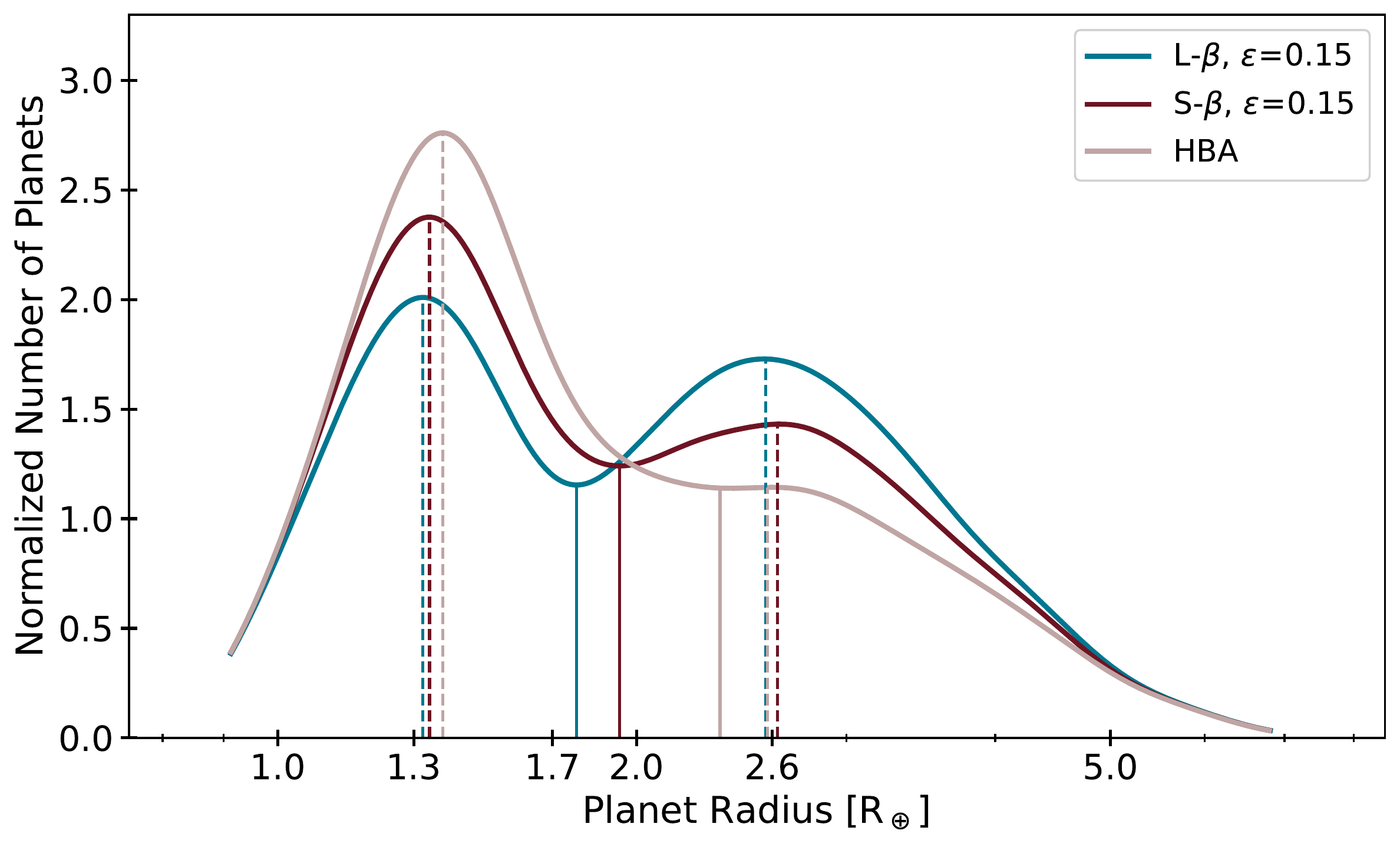}
\caption{Influence of the mass loss rate calculation on the 1D radius distribution for one solar mass host stars. We show in blue and red the 1D KDEs of the same planet population using the ER mass loss rates with the Lopez- and Salz-beta, respectively. In rose, we plot the final distribution of planetary radii having used the hydro-based mass loss rates. This radius distribution appears significantly different to the ER distributions. With the relatively large number of low-mass cores in our population, the majority of planets cannot retain any of its primordial envelope and end up as bare cores, leaving behind no pronounced second radius peak. An additional result of the higher HBA mass loss rates is that more massive cores can be stripped, shifting the radius gap to larger radii. Note that for the ER runs shown here, we used an evaporation efficiency of 0.15 for better comparison with the HBA mass-loss rates.}
\label{fig:ER_HBA}
\end{figure}

Figure~\ref{fig:ER_HBA} clearly shows that, compared to the ER simulation, in the HBA population the majority of planets fully evaporate, leaving behind almost no planets above the gap populating the second peak in 1D radius distribution. This makes the radius gap to disappear almost completely. Another consequence of the initally high HBA mass-loss rates is that more massive cores can be fully stripped compared the ER runs. This can be seen in the first radius peak slightly shifting to larger radii due to the larger number of more massive, and thus larger cores, as well as in a shift of the radius gap above 2 $R_\oplus$. Even for a slightly larger core mass distribution as used in Sec.~\ref{sec:app_Mcore}, the same conclusions for the HBA run hold true and almost no second peak remains. This means that a planet population similar to ours, which undergoes photoevaporation on the same order as predicted by the HBA mass loss rates using our XUV irradiation levels, cannot reproduce the observed bimodal radius distribution. The population would need to contain many more massive cores and/or envelopes for there to be enough planets able to retain an envelope to preserve the second peak in the radius distribution.

\begin{figure}
\includegraphics[width=0.49\textwidth]{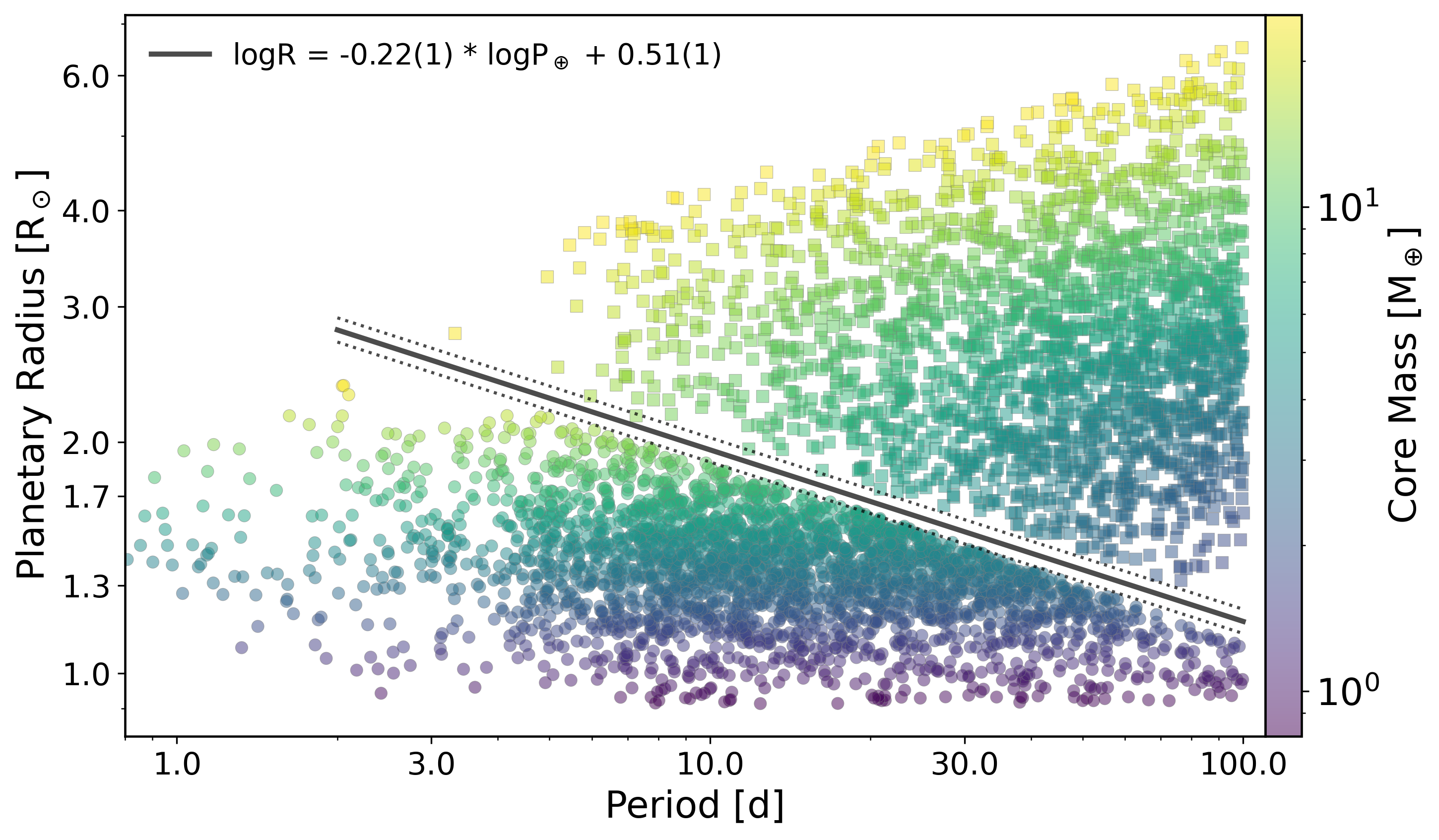}
\caption{Radius vs. period distribution for the planet population which has evolved using the hydro-based mass loss rates. Our fit to the radius gap is shown in grey. One can easily see that, compared to the ER runs (e.g. panel (a) in Fig.~\ref{fig:pl_models} and Fig.~\ref{fig:ChRo_betas}), more massive cores end up as rocky cores below the gap, shifting the gap upwards. Compared to the ER runs, the gap is slightly steeper and narrower.}
\label{fig:ChRo_HBA}
\end{figure}

When comparing the slope of the radius gap in 2D, we find that the gap, although almost non-visible in 1D, is still clearly visible in two dimensions. When comparing the HBA gap in Fig.~\ref{fig:ChRo_HBA} with the ER gap in panel (a) of Fig.~\ref{fig:pl_models} or Fig.~\ref{fig:ChRo_betas}, we see that although empty, it appears to be narrower and shifted upwards to larger radii. Interestingly, the slope of the gap is also slightly steeper for the HBA mass loss rates, most likely because of the larger number of massive cores in close-in orbits (dlog$R_\mathrm{p}$/dlog$P\approx-0.22\pm0.01$ vs. $\approx-0.19\pm0.01$).

\begin{figure}
\includegraphics[width=0.45\textwidth]{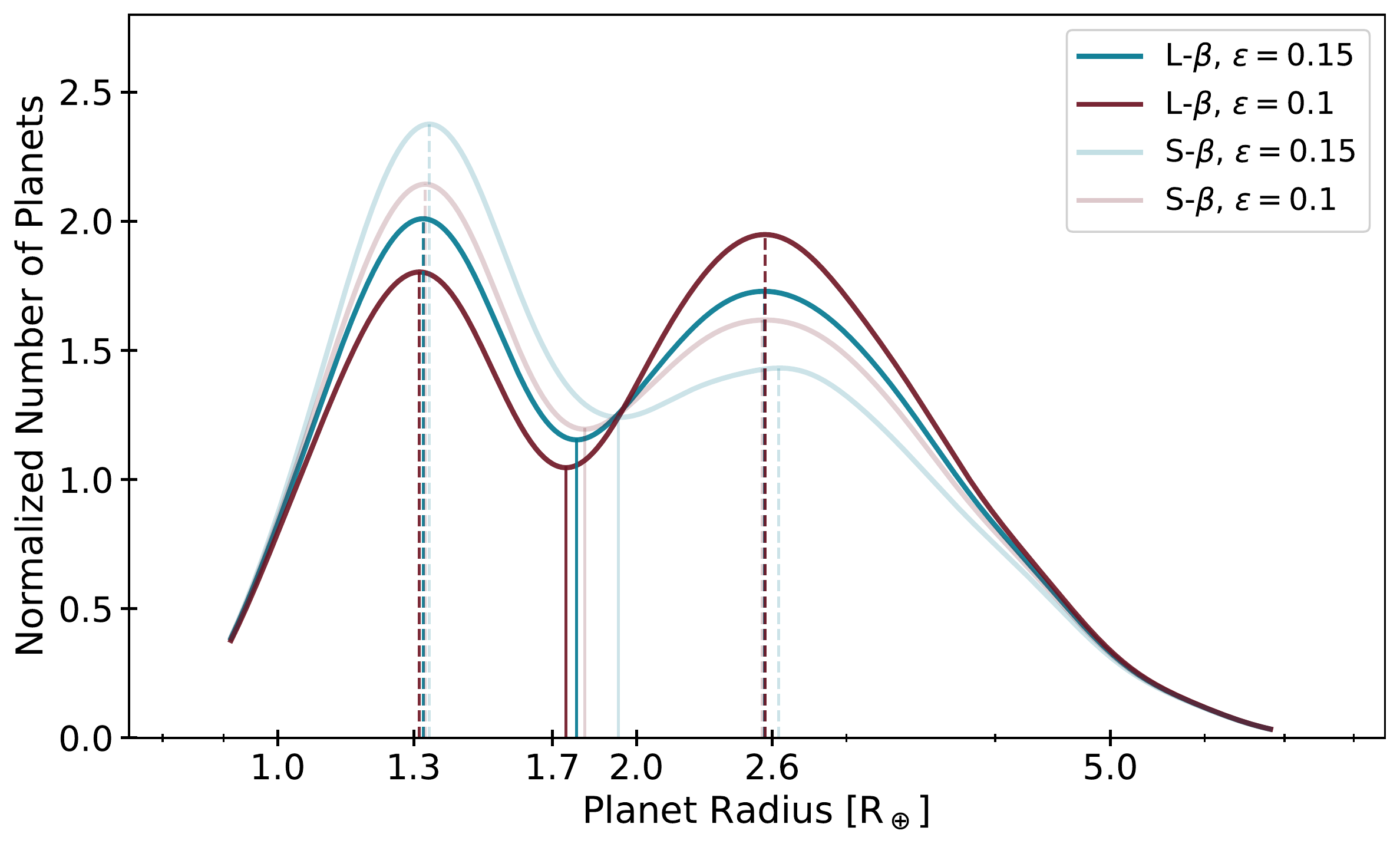}
\caption{Influence of the evaporation efficiency on the 1D radius distribution for one solar mass host stars. In red and blue we show the KDE of the evolved planet population for the Lopez-$\beta$ and $\epsilon = 0.1$ and $0.15$, respectively. The slightly transparent KDEs show the same for the Salz-$\beta$. A small change in the evaporation efficiency can change the peak heights but not their location. The radius gap is almost unaffected.}
\label{fig:eps}
\end{figure}

The evaporation efficiency is another parameter which affects the strength of the energy-limited mass loss rates. While we make a simplification and set the efficiency parameter as a constant, its exact value will still linearly increase or decrease the mass loss rates at any given time in the simulation. We show, in Fig.~\ref{fig:eps}, how a small increase of the evaporation efficiency from 0.1 to 0.15 impacts the 1D distribution features. The location of the two peaks still agrees within one sigma even with a small difference in evaporation strength, but of course the height of the peaks changes, with more planets evaporating and filling in the first peak when $\epsilon = 0.15$.

When it comes to the location of the gap, more massive bare cores tend to cause a shift to larger radii. Assuming an intermediate activity track for the evolution, the shift is small for the Lopez-beta simulation, but for the Salz run, which already predicts higher mass-loss rates compared to the Lopez run, this shift in the gap minimum becomes larger, as the gap shifts from 1.8 to 1.9 R$_\oplus$. We also tested this behavior for a lower and higher  stellar activity track and found that the gap shift increases going from low to high activity, or XUV flux. In general this indicates that the larger the mass loss, as set by the evaporation model, the beta-parameter and the evaporation efficiency, the larger is the shift on the gap location in 1D. In 2D, the slope of the gap is unchanged by a small change in the evaporation efficiency.

%%%%%%%%%%%%%%%%%%%%%%%%%%%%%%%%%%%%%%%%%%%%%%%%%%
\section{Primordial gas-envelope mass}
\label{sec:app_fenv_init}

\begin{figure}
\includegraphics[width=0.45\textwidth]{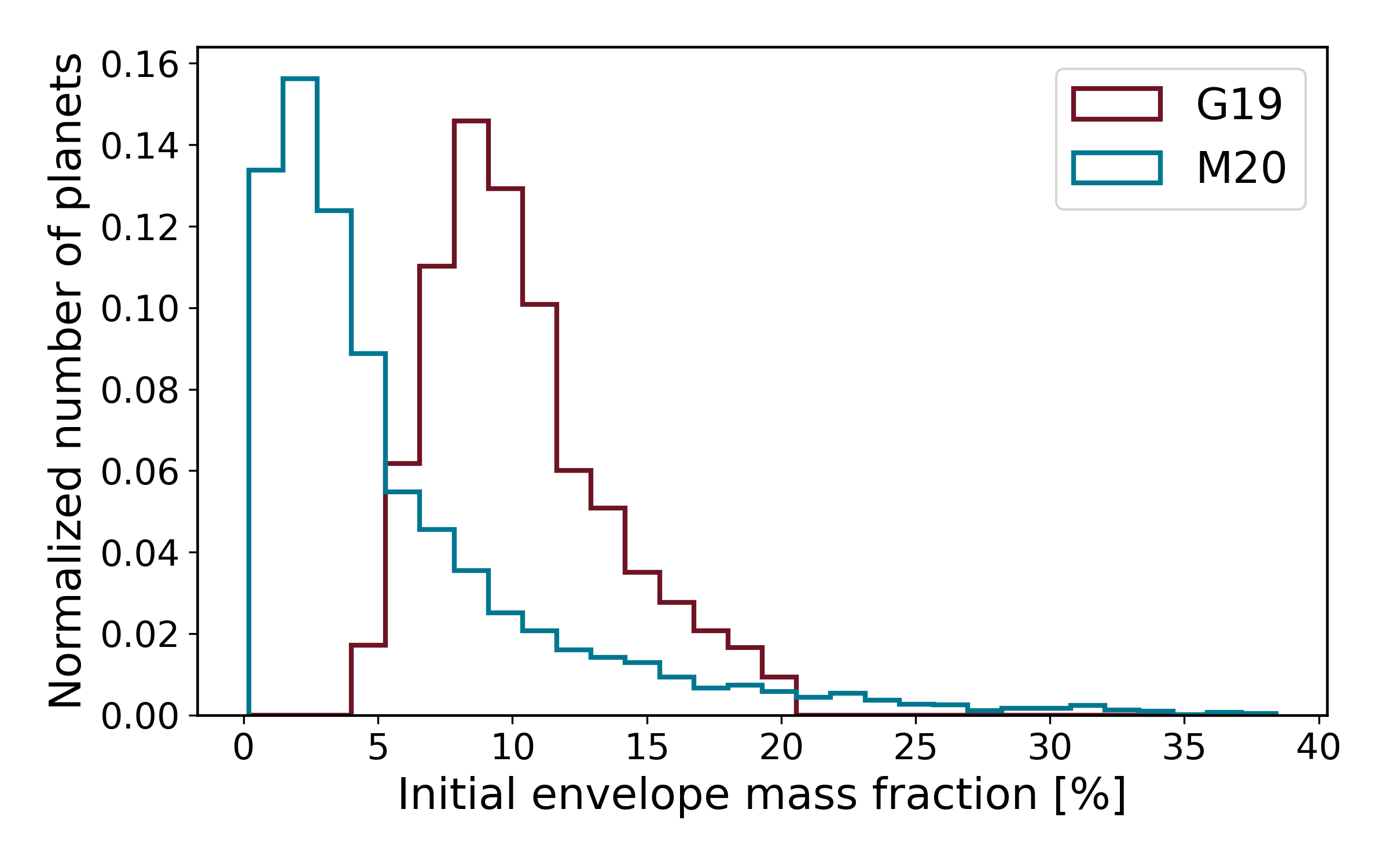}
\caption{Distribution of the gas envelope mass fractions at 10~Myr, the starting age of the simulation. In blue we show the M20 distribution, where the initial envelope mass fraction is dependent on planetary core mass and envelope mass fraction, and in red we show the G19 distribution with a dependence on core mass only. Planet with primordial atmospheres according to G19 are significantly higher in mass.}
\label{fig:fenv_init}
\end{figure}

We check the impact of varying the initial envelope mass fraction of the planetary sample at the starting age of our simulation on the final radius distribution and the radius gap. We either calculate the primordial gas-envelope mass according to the simulations by \citet{2020Mordasini} (short: M20), or based on the results by \citet{2016Ginzburg, 2019Gupta} (short: G19). As a reminder, the envelope masses predicted by M20 have a positive dependence on core mass and a negative dependence on orbital separation, while the G19 primordial envelopes have a dependence on core mass only. For the one-solar-mass run considered here, the M20 envelope mass fractions cover a larger range ($\sim 0.2-38\,\%$) but have a lower median ($4\,\%$) compared to the G19 ones, which range from ($\sim 4-20\,\%$) with a median of $\sim 10\,\%$. This is illustrated in Figure~\ref{fig:fenv_init}.

\begin{figure}
\includegraphics[width=0.45\textwidth]{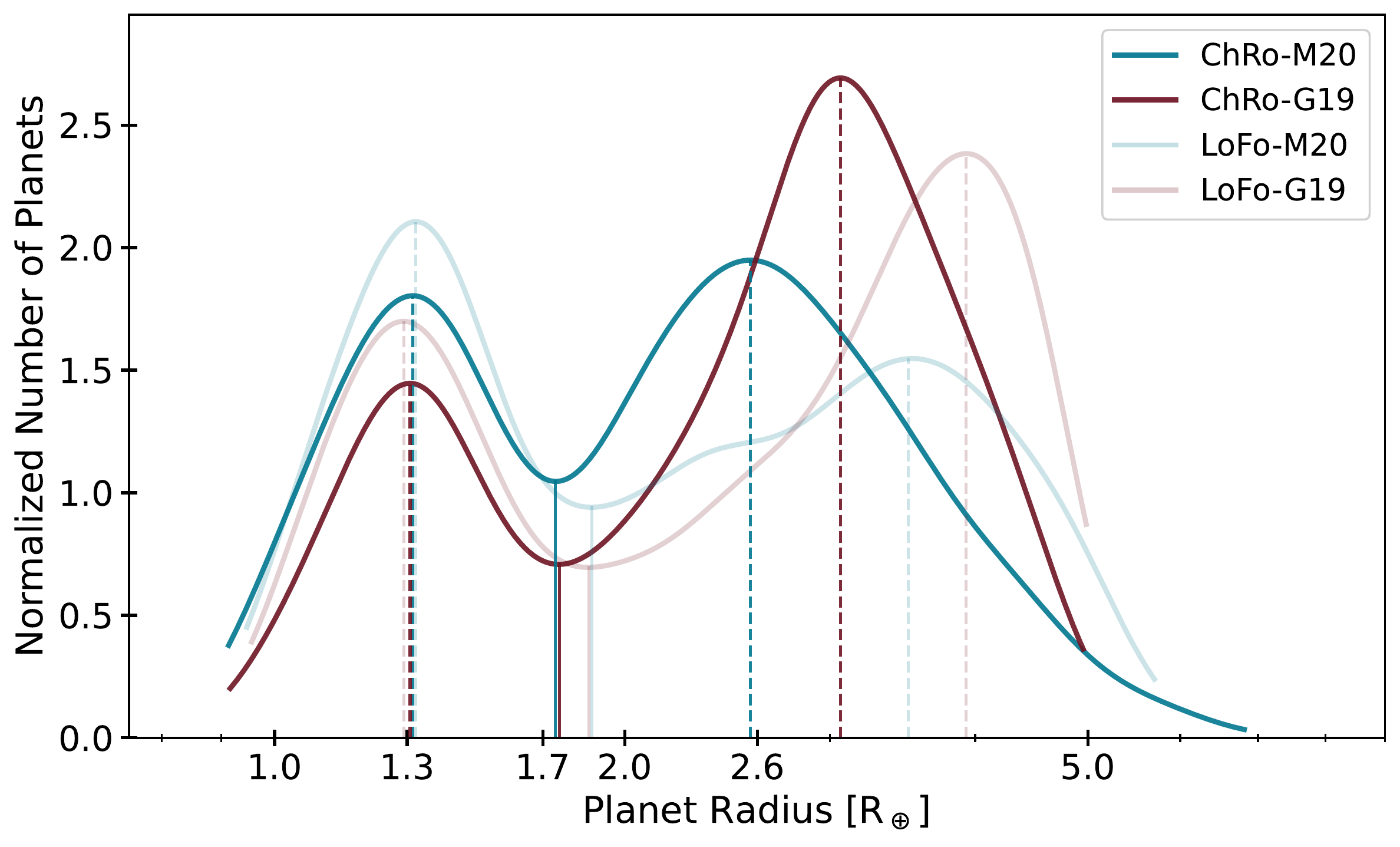}
\caption{Influence of the primordial gas-envelope mass on the 1D radius distribution for one solar mass host stars. We show the KDEs for a population with primordial envelope masses calculated according to M20 (blue) and G19 (red). The transparent blue and red plots the results when using the LoFo models and are just shown for comparison. The main difference between the M20 and the G19 radius distributions is the location and height of the second peak. The majority of G19 planets has a larger primordial envelope. Under the same simulation assumptions, more planets can retain envelope in the G19 run and also more planets end up with heavier and thus larger envelopes compared to the M20 population. The location of the first peak and the gap are unchanged by the choice of primordial envelope mass estimation.}
\label{fig:M20_vs_G19}
\end{figure}

The 1D radius distributions in Figure~\ref{fig:M20_vs_G19} show that the second peak is shifted to larger radii for the G19 run. The reason for this is that, on average, the G19 primordial envelopes are more massive than the M20 ones, which makes them harder for the planets to lose under the same simulation assumptions, and as a result the planets end up with more massive final envelopes and thus are larger in size. In the G19 population, fewer planets are able to completely lose their envelope, which is why the first peak is less pronounced. Interestingly, the location of the gap in 1D is not significantly affected by the choice of the primordial envelope mass fraction.

\begin{figure}
\includegraphics[width=0.49\textwidth]{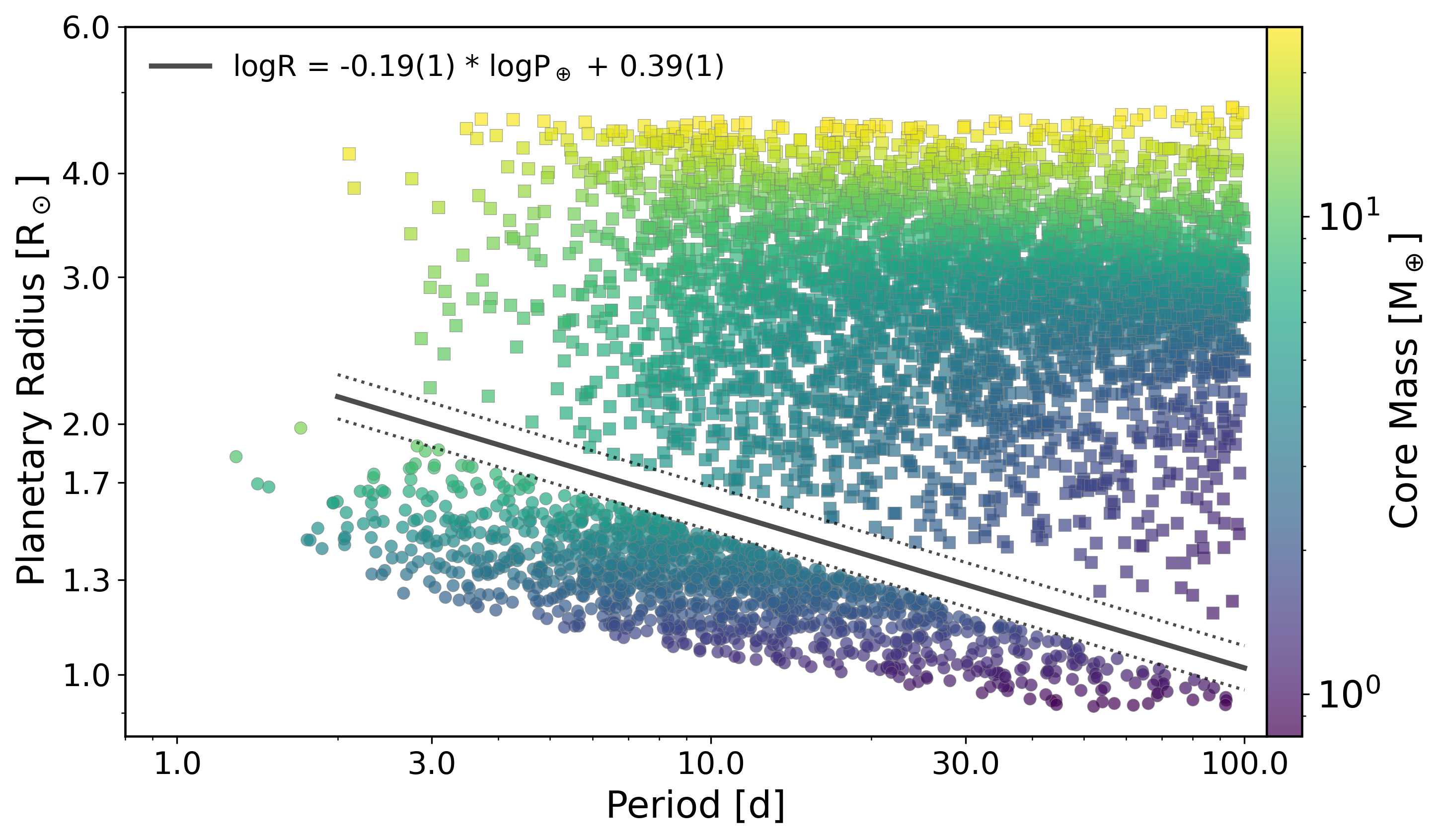}
\caption{We show the radius vs. period distribution for the planet population for which the primordial envelope mass fractions were calculated according to G19. Planets which can hold on to envelopes tend to cluster around $3\,\mathrm{M}_\oplus$, and an empty gap is clearly visible. Our fit to the gap is shown in grey. This plot can be compared with the M20 simulation in panel (a) of Fig.~\ref{fig:pl_models}.}
\label{fig:ChRo_GGG}
\end{figure}

In 2D, when looking at the G19 radius vs. period distribution in Fig.~\ref{fig:ChRo_GGG}, one can see that the bulk of the planets above the gap, which are able to retain envelope, reside at larger radii, clustered around $\sim 3\,\mathrm{M}_\oplus$. The region underneath until the gap is sparsely populated compared to the M20 run (see Fig.~\ref{fig:pl_models} panel (a) for comparison), but an empty gap is still clearly visible. The slope of the G19 gap is consistent with the slope of the M20 gap, but the gap is slightly shifted downwards because there are fewer, more massive bare cores in this simulation. The G19 and M20 gaps do however still overlap, telling us that the impact of the primordial envelope mass fraction only has a minor impact on the location of the gap. This conclusion is unchanged for the choice of $\beta$ and/or the evaporation scheme. As a caution, the two starting populations are not exactly identical due to the imposed "physicality"-check (see Sec.~\ref{sec:planetary_sample}). With the G19 primordial envelopes having no dependence on the distance to the host star, planets with core masses below $2\,\mathrm{M}_\oplus$ have envelopes massive and large enough to fulfill our low-density and Roche lobe cutoff, and thus are excluded. For this reason, the G19 population has about 100 fewer planets with core masses below $2\,\mathrm{M}_\oplus$, which is why there are fewer planets with radii around $1\,\mathrm{R}_\oplus$ compared to the M20 population.

%%%%%%%%%%%%%%%%%%%%%%%%%%%%%%%%%%%%%%%%%%%%%%%%%%
\section{Core-mass distribution}
\label{sec:app_Mcore}

Previous studies have shown that the core mass distribution has a significant influence on the radius distribution in photoevaporation studies. More massive cores can hold on to their envelopes more easily, but once a more massive core does get stripped, it's bare core radius will also be larger. For reference, a $3\,\mathrm{M}_\oplus$  bare rocky core has a radius of $1.3\,\mathrm{R}_\oplus$, whereas a $5\,\mathrm{M}_\oplus$ core has a radius of $1.5\,\mathrm{R}_\oplus$. This can impact the location of the first and second peak in the 1D radius distribution, and with it the gap.

\begin{figure}
\includegraphics[width=0.45\textwidth]{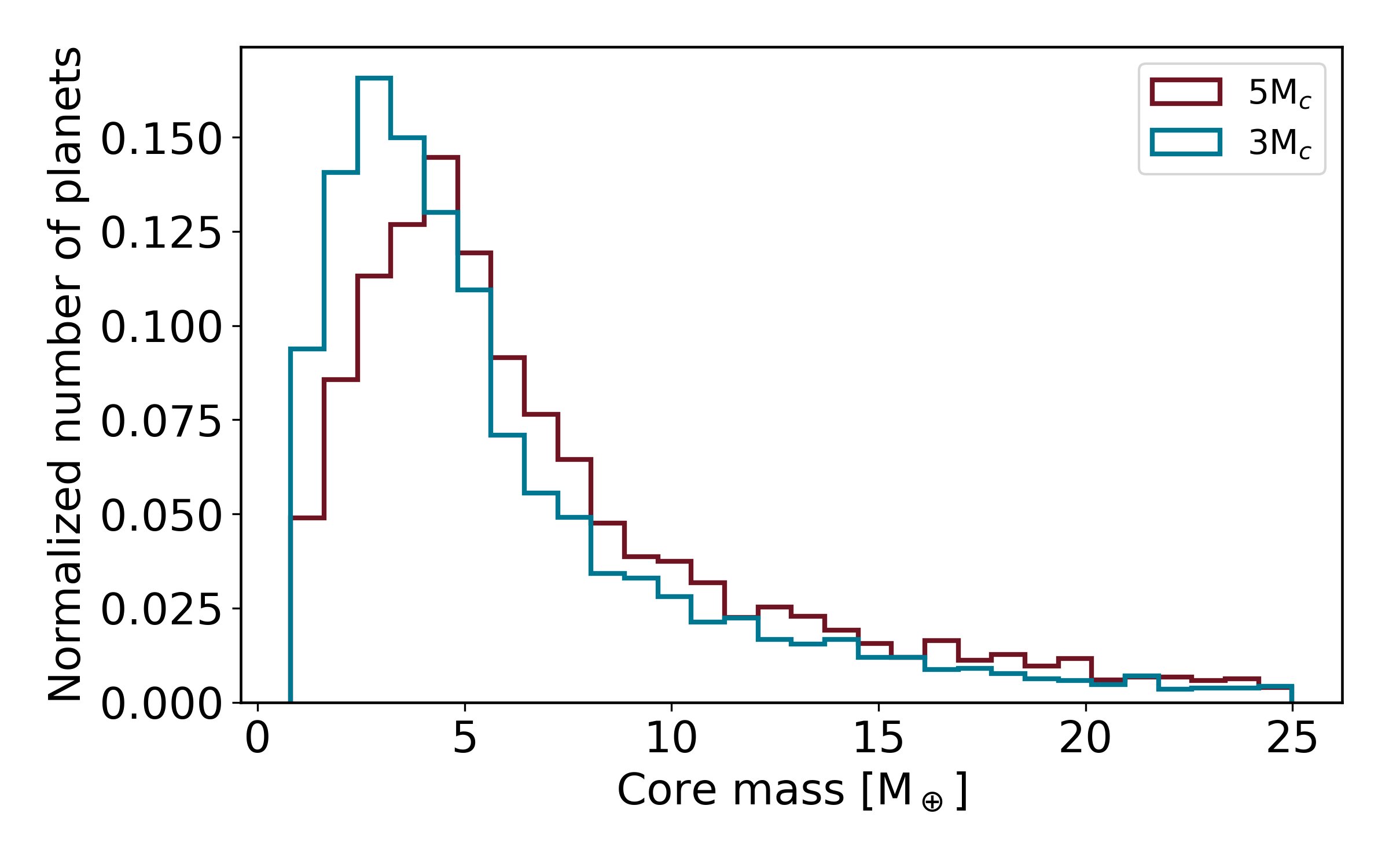}
\caption{Histogram of the core mass distributions for the $3M_c$ and $5M_c$ simulation, which peak at $3$ and $5\,\mathrm{M}_\oplus$, respectively. The $5M_c$ population has fewer planets below $3\,\mathrm{M}_\oplus$ and more planets at higher masses above $5\,\mathrm{M}_\oplus$.}
\label{fig:Mcore_hist}
\end{figure}

To illustrate how the radius distribution is affected by the core masses of the sample, we compare the population where the core masses peak at $3\,\mathrm{M}_\oplus$ (short: $3M_c$) with a slightly heavier population which peaks at $5\,\mathrm{M}_\oplus$ (short: $5M_c$). As can be seen in Fig.~\ref{fig:Mcore_hist}, the $5M_c$ simulation thus has fewer planets below $\sim 3\,\mathrm{M}_\oplus$, and more planets with intermediate masses between $5$ and $20\,\mathrm{M}_\oplus$.

\begin{figure}
\includegraphics[width=0.45\textwidth]{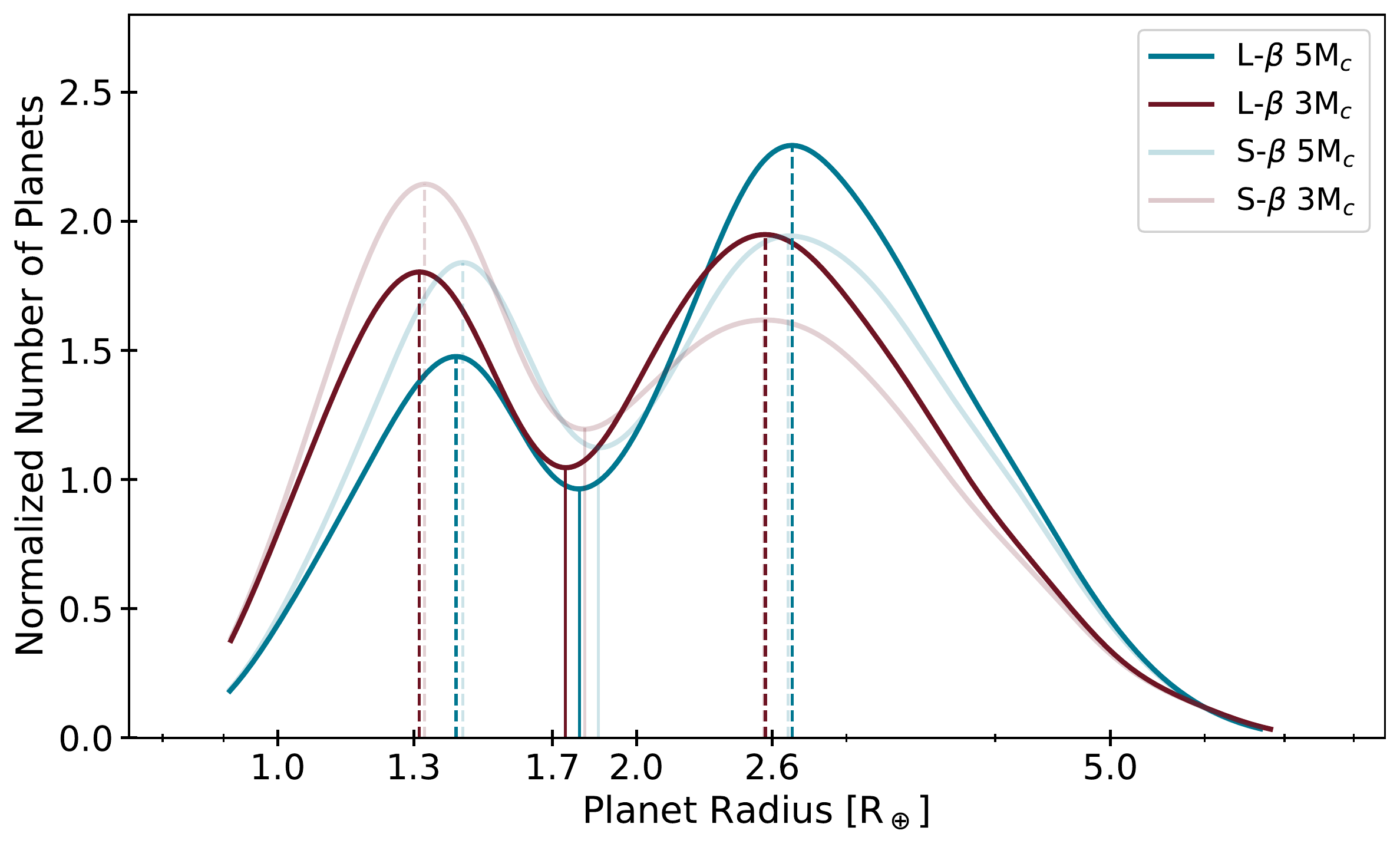}
\caption{Influence of the core mass distribution on the 1D radius distribution for one solar mass host stars. The planet population with core masses peaking at $5\,\mathrm{M}_\oplus$ is shown in blue and the population with a $3\,\mathrm{M}_\oplus$ peak in red. For comparison, the transparent lines show the same populations but using the Salz-$\beta$. The shift of the first peak from $1.3$ to $1.5\,\mathrm{R}_{\oplus}$ is clearly visible, as well as a shift of the second peak to larger radii. The change in the location of the gap is almost negligible.}
\label{fig:3vs5Mcore}
\end{figure}

In the 1D radius distribution, one can most easily identify the effect of a heavier planet sample. Figure~\ref{fig:3vs5Mcore} clearly shows that the first peak is shifted from $1.3$ to $1.5\,\mathrm{R}_\oplus$ for the $5M_c$ simulation due to the larger number of close-in, more massive, and thus larger in size cores, which end up fully stripped. In general, more massive stripped cores shift the radius gap to larger radii, but as a result of having more massive cores in the sample, the number of planets which can fully evaporate in the $5M_c$ case decrease overall. These two effects combined lead to the location of the radius gap being almost unchanged between the $3M_c$ and the $5M_c$ run. The second radius peak is also shifted to slightly larger radii because more massive planets have larger primordial envelopes to begin with and can better hold on to their envelope, reducing the total mass loss. As a result, there are more planets with larger sizes in the $5M_c$ simulation.

In 2D, there are noticeably fewer planets below the gap, but both the slope and height of the $5M_c$ gap are in good agreement with the $3M_c$ one. This shows that at least small changes in the core mass distribution, while being noticeable in 1D, have no significant impact on the radius gap in 2D.

\begin{figure}
\includegraphics[width=0.49\textwidth]{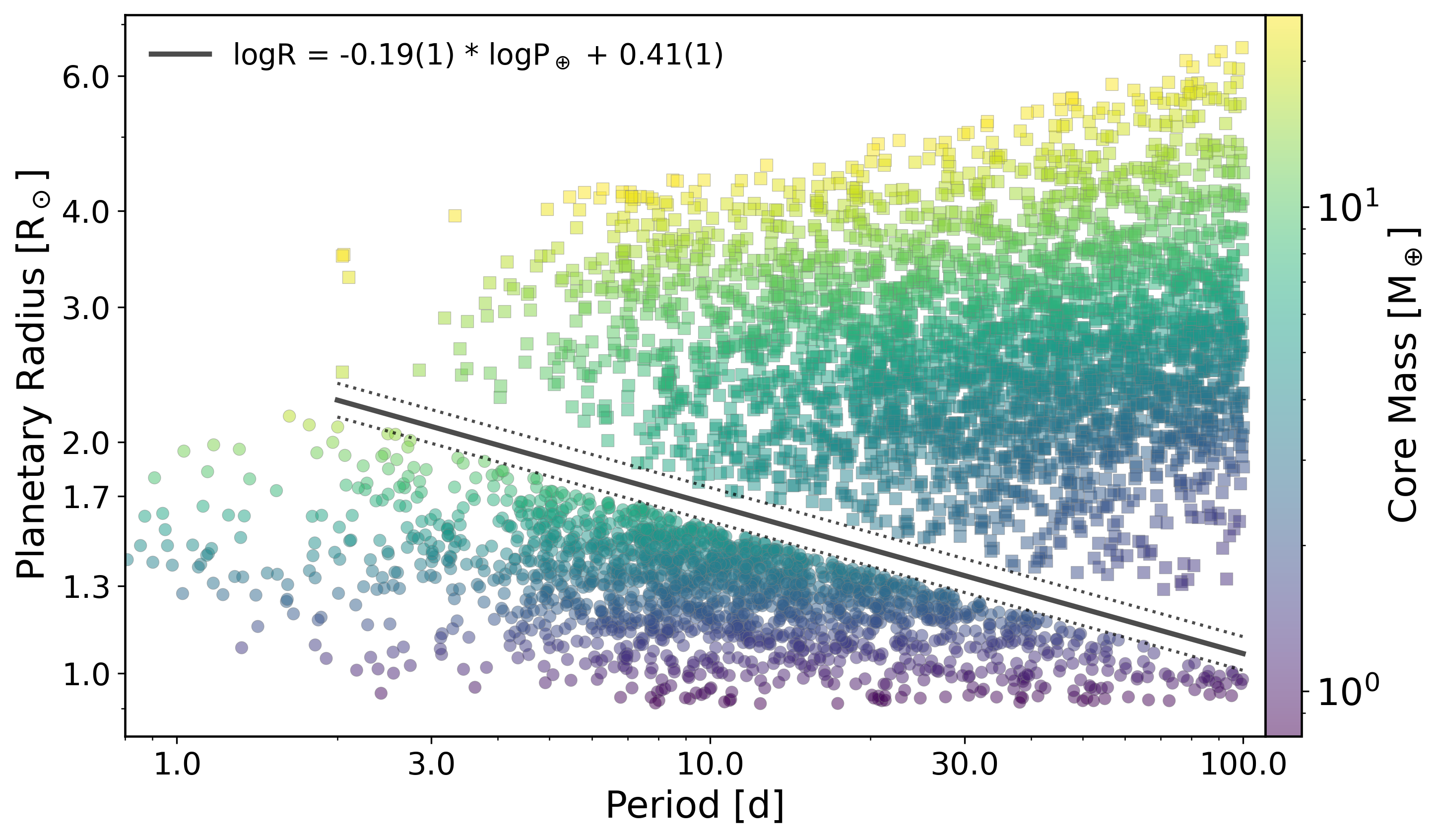}
\caption{Radius vs. period distribution for the planet population with more massive cores ($5M_c$), for comparison with the $3M_c$ simulation in panel (a) of Fig.~\ref{fig:pl_models}. Our fit to the gap is shown in grey.}
\label{fig:5Mcore}
\end{figure}

%%%%%%%%%%%%%%%%%%%%%%%%%%%%%%%%%%%%%%%%%%%%%%%%%%%%%%%%%%%%%%%%%%%%%%%%%%%%%%%%%
\section{Influence of EUV estimation method and X-ray power-law slope}
\label{sec:EUV_est}

Stellar EUV emission ionizes hydrogen and drives the mass loss of planetary atmospheres, in particular at later ages when the stellar X-ray emission has significantly died down. Understanding the decay details of both X-ray and EUV emission over time and across the whole range of stellar spectral types is crucial for accurately predicting the XUV irradiation of an exoplanet.

As described in Sections~\ref{subsec:Xray_decay} and \ref{sec:EUV_estimation}, we use the works by \citet{2015Tu, 2012Jackson, 2021Johnstone} for solar-mass stars to approximate our X-ray evolutionary tracks. In \citet{2021Poppenhaeger}, we used the X-ray evolution model by \citet{2015Tu} only, and estimated the EUV flux using the SF11 relation. The difference to the current work is a slightly lower X-ray luminosity around $1$ Gyr, where all the activity tracks are set to converge, as well as a steeper decay beyond this age. We now show how the assumptions about the X-ray decay slope and\slash or the EUV estimation method impact the evolved 1D and 2D radius distribution.

\begin{figure}
\includegraphics[width=0.45\textwidth]{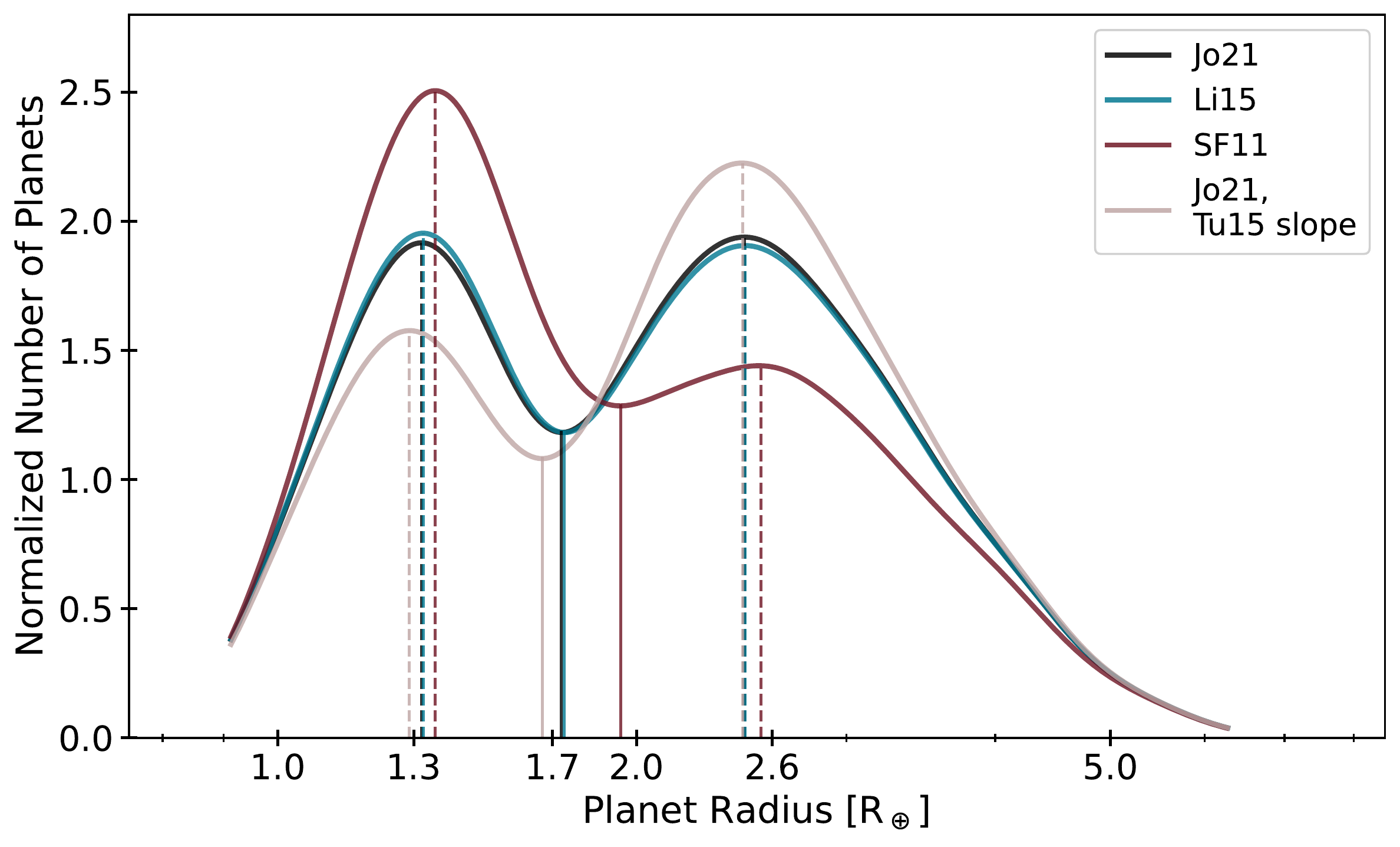}
\caption{Influence of EUV estimation and power-law slope of the X-ray decay. Comparison of the 1D-KDE estimation of the same planet population which have evolved using the Jo21 (black), Li14 (blue) and SF11 (red) method to estimate the stellar EUV output. The Jo21 and Li14 simulations are almost identical, but the SF11 simulation predicts many more evaporated cores due to the high EUV fluxes at young ages. In rose we also show the same population with EUVs calculated according to Jo21, but with a steeper power-law slope of the X-day decay of $-1.58$ instead of $-1.13$. The steepness of the slope impacts the total amount of XUV irradiation the planets receive, with a steeper slope leading to less XUV exposure and fewer evaporated cores.}
\label{fig:Xslope_EUVs}
\end{figure}

Figure~\ref{fig:Xslope_EUVs} presents the 1D radius distribution for different EUV estimation methods and two different power-law slopes for the X-ray decay. We show the distribution at 10 Gyr, because only at this old age, a small difference between the Jo21 (black) and Li14 (blue) EUV estimation is visible by eye. Li14 predicts EUV luminosities which are up to a factor of 10 higher than the Jo21 ones at X-ray luminosities below $\sim 10^{28}\,\mathrm{erg/s}$, which the stars in our sample reach roughly after one to three Gyrs. For this reason, the EUV estimation at Gyr ages, although the overall XUV fluxes are low, can still have an influence on the radius distribution, with higher EUV luminosities leading to some of the planets at the top edge of the radius gap to become fully stripped of their remaining thin atmosphere. This changes the height of the peaks slightly, but has no impact on the location of the peaks or the gap. In 2D, the location, slope and width of the gap are unchanged.

By using the SF11 method (red) for estimating the important stellar EUV output, however, the radius distribution at Gyr ages looks significantly different. Due to the EUV luminosity being about 10 times larger than the X-ray luminosity for the first few Gyr, the total XUV flux a planet receives is immense compared to all the other EUV estimation methods. This leads to almost 50\% of the planets in our sample to end up as bare rocky cores, compared to around 36\% for the Jo21 and Li14 simulations. Due to more planets and planets with more massive cores being completly stripped, the second peak becomes very weakly pronounced and the radius gap shifts to larger radii. In 2D, the gap slope is still consistent within one sigma, but the gap is shifted to larger radii. We observe that in our simulations, the gap becomes slightly narrower as the number of evaporated bare cores increases and the number of planets with envelopes above the gap increases.
This comparison shows that the EUV estimation method, as in the case of SF11, can significantly alter the radius distribution. The results support the notion that the SF11 method overestimates the crucial EUV output and should be used with caution for young and\slash active stars. However, methods adopted to young and active stars like Jo21 (or Ch15) and Li14 lead to matching radius distributions which agree in peak and gap heights and locations, and preserve the second radius peak much like the one present in the observational data.

Not only the EUV estimation method can play a role, but also the steepness of the X-ray decay. In Figure~\ref{fig:Xslope_EUVs} we also show the radius distribution for the same population, but having used a steeper power-law slope ($-1.58$ compared to $-1.13$) for the X-ray decay - and thus lower EUV and XUV predictions - in Figure~\ref{fig:Xslope_EUVs} (rose). It is evident that with less XUV exposure, fewer planets fully evaporate. The second peak is still well populated and due to the smaller number of heavier evaporated cores, the gap minimum is located at sightly smaller radii compared to the simulation runs with a shallower power-law slope. This is just to highlight that the amount and decay shape of X-ray and EUV radiation from the host star influences the location and height of the peaks and the gap in the 1D radius distribution. When trying to model the observed radius distribution, similar to \citet[e.g.][]{2020Modirrousta, 2021Rogers_a}, the choice of the slope for the X-ray and EUV decay can substantially impact the conclusions of such studies.

Ultimately, the amount of XUV flux a planet receives is set by the value of the saturation X-ray luminosity, the time a star spends in the saturated phase, the steepness of the X-ray decay and the amount of EUV emission at early and later ages. A general outcome for the lowest mass planets in our sample, or planets really close to their host star, is, that the exact details do not change the fate of these planets - they will lose their atmosphere. For planets with intermediate masses and/or larger distances, the details of the XUV evolution determine if a planet gets fully stripped or can hold on to an envelope. In addition, depending on the strength of the X-ray emission in the saturated phase and the steepness of the decay, the EUV estimation method can have a bigger or smaller impact on the total amount of evaporation a planet undergoes. When the X-ray fluxes are lower, the EUV contribution to the total XUV output becomes more important. As a consequence, small differences in the EUV estimation become more pronounced. On the other hand, if the X-ray fluxes are high, small differences between, for example, the Jo21 and Li14 EUV estimation do not change the total XUV flux significantly, shrinking any differences between the two EUV calculations.

To summarize, our simulations show that the choice of the power-law slope for the X-ray decay can have a much larger impact on the final radius distribution than the EUV estimation method - when using the methods adopted for young and active stars.

%%%%%%%%%%%%%%%%%%%%%%%%%%%%%%%%%%%%%%%%%%%%%%%%%%
%%%%%%%%%%%%%%%%%%%%%%%%%%%%%%%%%%%%%%%%%%%%%%%%%%

% Don't change these lines
\bsp	% typesetting comment
\label{lastpage}
\end{document}